\documentclass[reprint,superscriptaddress,amsmath,amssymb,aps,prd,twocolumn,floatfix,nofootinbib]{revtex4-1}
\usepackage{amsmath, amssymb}
\usepackage{listings}
\usepackage{graphicx}
\usepackage{dcolumn}
\usepackage{bm}
\usepackage{float,color}
\usepackage{xcolor}
\usepackage[normalem]{ulem}
\usepackage{tabularx}
\usepackage{mathtools} 
\usepackage{multirow}
\usepackage{amssymb}
\usepackage{scrextend}
\usepackage{hyperref}

\begin{document}

\title{
Possible binary neutron star merger history of the primary of GW230529
}
\author{Parthapratim Mahapatra}\email{MahapatraP@cardiff.ac.uk}
\affiliation{Chennai Mathematical Institute, Siruseri, 603103, India}
\affiliation{School of Physics and Astronomy, Cardiff University, Cardiff, CF24 3AA, United Kingdom}
\author{Debatri Chattopadhyay}
\affiliation{School of Physics and Astronomy, Cardiff University, Cardiff, CF24 3AA, United Kingdom}
\affiliation{Center for Interdisciplinary Exploration and Research in Astrophysics (CIERA) and Department of Physics \& Astronomy, Northwestern University, 1800 Sherman Ave, Evanston, Illinois 60201, USA}
\author{Anuradha Gupta}
\affiliation{Department of Physics and Astronomy, The University of Mississippi, University, Mississippi 38677, USA}
\author{Fabio Antonini}
\affiliation{School of Physics and Astronomy, Cardiff University, Cardiff, CF24 3AA, United Kingdom}
\author{Marc Favata}
\affiliation{Department of Physics \& Astronomy, Montclair State University, 1 Normal Avenue, Montclair, New Jersey 07043, USA}
\author{B. S. Sathyaprakash}
\affiliation{Institute for Gravitation and the Cosmos, Department of Physics, Penn State University, University Park, Pennsylvania 16802, USA}
\affiliation{Department of Astronomy and Astrophysics, Penn State University, University Park, Pennsylvania 16802, USA}
\author{K. G. Arun}
\affiliation{Chennai Mathematical Institute, Siruseri, 603103, India}
\date{\today}

\begin{abstract}
Black holes (BHs) with masses between $\sim 3-5M_{\odot}$, produced by a binary neutron star (BNS) merger, can further pair up with a neutron star or BH and merge again within a Hubble time. However, the astrophysical environments in which this can happen and the rate of such mergers are open questions in astrophysics. Gravitational waves may play an important role in answering these questions. In this context, we discuss the possibility that the primary of the recent LIGO-Virgo-KAGRA binary GW230529\_181500 (GW230529, in short) is the product of a previous BNS merger. Invoking numerical relativity (NR)-based fitting formulas that map the binary constituents' masses and tidal deformabilities to the mass, spin, and kick velocity of the remnant BH, we investigate the potential parents of GW230529's primary. Our calculations using NR fits based on BNS simulations reveal that the remnant of a high-mass BNS merger similar to GW190425 is consistent with the primary of GW230529. This argument is further strengthened by the gravitational wave-based merger rate estimation of GW190425-like and GW230529-like populations. We show that around 18\% (median) of the GW190425-like remnants could become the primary component in GW230529-like mergers. The dimensionless tidal deformability parameter of the heavier neutron star in the parent binary is constrained to $67^{+163}_{-61}$ at 90\% credibility. Using estimates of the gravitational-wave kick imparted to the remnant, we also discuss the astrophysical environments in which these types of mergers can take place and the implications for their future observations.
\end{abstract}
\maketitle
\section{Introduction}\label{sec:intro}
Dynamical mass measurements of low-mass x-ray binaries in our Galaxy have reported a gap in the black hole mass spectrum between $\sim 3$--$5M_{\odot}$~\cite{Bailyn:1997xt,Ozel2010ApJ,Farr2011ApJ}, often referred to as the ``low-mass gap.'' The lower edge of this gap depends on the maximum mass of a neutron star (NS) formed via a supernova explosion. This maximum mass is set by the NS equation of state~\cite{RhoadesPRL1974, Friedman1987ApJ, Cook1994ApJ, Mueller:1996pm, Kalogera:1996ci}, as well as supernova mechanisms ~\cite{Fryer2012ApJ} and the rotation of the progenitor star~\cite{Olejak2022MNRAS, Muhammed2024arXiv, Zuraiq2024PhRvD}.
By definition, the upper edge of this gap depends on the minimum possible black hole (BH) mass formed from stellar core collapse.
As these x-ray binaries are accreting from stellar companions in the Galactic field, this observed upper edge could imply that stellar processes cannot produce BHs with masses less than $5M_{\odot}$~\cite{Bailyn:1997xt,Ozel2010ApJ,Farr2011ApJ,Belczynski2012ApJ}.

On the contrary, there have been some recent observations of noninteracting binaries in our Galaxy~\cite{Thompson2019Sci,vandenHeuvel:2020chh,Thompson:2020nbd,Jayasinghe:2021uqb,Kareem2022MNRAS} that suggest there are candidate BHs that could potentially lie in the low-mass gap. It was argued that the radio observation of a probable NS--BH binary~\cite{Barr:2024wwl} in a globular cluster (GC) may also have a BH whose mass may lie in this gap.
Recently, Ref.~\cite{Song:2024tqr} also reported a possible low-mass gap BH candidate using data from the Large Aperture Multi-Object Spectroscopic Telescope and astrometric observations in Gaia Data Release 2 and 3 catalogs.
Asymmetric compact binary mergers GW190814~\cite{GW190814} and GW200210\_092254~\cite{GWTC-3}, observed through gravitational waves (GWs) in the third observing run of LIGO/Virgo, are also found to harbor compact objects that fall in the low-mass gap.
The population of NS binary candidates in wide orbits, as identified from Gaia astrometry, supports the existence of the low-mass gap~\cite{Kareem2024OJAp}.
Moreover, gravitational microlensing searches do not support the existence of the low-mass gap, but cannot exclude it either~\cite{Wyrzykowski2020A&A,Wyrzykowski:2015ppa}. Hence, understanding different ways of populating the purported low-mass gap can shed light on the astrophysical sites where these objects are spotted.

Recently, the LIGO-Virgo-KAGRA (LVK) Collaboration reported the detection of a compact binary merger GW230529$\_$181500 (hereafter GW230529) whose primary has a mass between $\sim2.4$--$4.4M_{\odot}$ at 90\% credibility; this has significant support in the low-mass gap~\cite{GW230529}. The secondary has a mass between $\sim 1.2$--$2.0M_{\odot}$ at 90\% credibility.
As the tidal deformability posteriors for both the primary and secondary of this binary are uninformative,
it cannot be determined if either object was an NS or a BH (see Fig.~14 of Ref.~\cite{GW230529}). However, based on our current understanding of the populations of BHs and NSs from GW observations, this event is argued to be consistent with a NSBH merger with the BH lying in the gap (see Table~3 of Ref.~\cite{GW230529} and Tables~III and VI of Ref.~\cite{Koehn:2024ape}).

Assuming the primary of GW230529 to be a BH, the goal of this paper is to determine the \emph{parents} of that primary, operating under the assumption that it was formed from the merger of two NSs. This analysis will make use of a formalism previously developed by the authors to study the parents of the LVK binary BHs (BBHs)~\cite{Mahapatra:2024qsy}.

In the subsections below, we review the literature on mechanisms that could produce BHs in the low mass gap and the caveats associated with those predictions.

\subsection{Insights from core-collapse supernova simulations}
Some core-collapse supernova models do produce BHs in the low-mass gap~\cite{Fryer:1999ht,Kushnir:2015mca}, allowing the creation of NSBH binaries with component masses $\approx$1.5 and $\approx$3.5\,M$_\odot$~\cite[see Table\,3 of Ref.][]{Chattopadhyay:2022cnp}.
Other models allow for the possibility of rapid explosion mechanisms creating no BH in the gap~\cite{Fryer2012ApJ} or delayed explosion mechanisms resulting in a continuous mass distribution between NSs and BHs~\cite{,Belczynski2012ApJ}. References~\cite{Zhu:2024cvt,Xing:2024ydg,Qin:2024ojw,Chattopadhyay:2024hsf,Chandra:2024ila} proposed an isolated binary evolution channel for the formation of GW230529 by adopting the delayed supernova prescription; this is also consistent with the Galactic pulsar mass and spin distributions~\cite{Chattopadhyay:2024hsf}. However, some models of core-collapse supernovae suggest that masses as low as that of the primary of GW230529 cannot be formed via the direct collapse of massive stars~\cite{OConnor:2010moj,Janka:2012wk,Sukhbold:2015wba,Muller:2016ujh}. 
On the other hand, the failed explosions of certain lower-mass progenitor stars may form low-mass gap BHs~\cite{Couch:2019mrd}. Likewise, substantial fallback, albeit rare, following a successful explosion could lead to the formation of a BH in the gap~\cite{Sukhbold:2015wba,Ertl:2019zks}. Ref.~\cite{Ertl:2019zks} suggests that there may not be a ``gap'' but only a less populated mass range between $\sim3$--$5M_{\odot}$.
Significant accretion onto a newborn NS in a binary may trigger an accretion-induced collapse producing a low-mass gap BH~\cite{Siegel:2022gwc,Zhu:2023nhy}. 
Therefore, unless substantial fallback or significant accretion is common in binaries, it is difficult for the isolated binary evolution channel to explain the merger rate of $55^{+127}_{-47}\,{\rm Gpc^{-3}\, yr^{-1}}$~\cite{GW230529} for GW230529-like binaries.

\subsection{Low mass gap BHs in dense star clusters}
It has been argued that binary NS (BNS) mergers provide a natural mechanism for populating the low-mass gap (see, for instance, Ref.~\cite{Gupta:2019nwj}). For a BNS merger remnant to pair up with another compact object, one would need an astrophysical environment that is dense enough and rich in compact objects. Astrophysical environments such as GCs, young star clusters, nuclear star clusters, and discs of active galactic nuclei are expected to satisfy these conditions.

\subsubsection{Globular clusters}
There is an ongoing debate about estimates of NS binary\footnote{{The term ``NS binary'' is used throughout this paper to mean a compact object binary with one NS and a companion that is either another NS or a BH; i.e., either a BNS or a NSBH system.}} mergers in GCs and whether they can constitute a significant fraction of the NS binaries detected by LIGO/Virgo (see Table~2 of Ref.~\cite{Ye:2019xvf} for a comparison of rate estimates from different studies). Several past studies have reported that the NS binary merger rate in GCs is small compared to that from the galactic fields~\cite{Phinney1991ApJL,Bae:2013fna,Clausen2013MNRAS,Belczynski:2017mqx,Ye:2019xvf}.
Since BHs are more massive than NS, they dominate the dynamical interactions at the cluster cores due to mass segregation. It is only after the ejection of most of the BHs from the cluster that NSs can segregate to the cluster core and interact to form binaries~\cite{Sigurdsson:1994ju,Fragione:2018jxd,Ye:2019luh}. 

On the other hand, merger rate estimates in massive and core-collapsed GCs that account for direct collisions and tidal captures of NSs suggest that NS binaries in GCs make a significant contribution to the overall merger rate in the Universe~\cite{Grindlay:2005ym,Lee2010ApJ}. Further, Ref.~\cite{Guetta2009A&A} inferred a very high NS binary merger rate in GCs from the luminosity function and observed redshift distribution of short gamma-ray bursts. Moreover, Ref.~\cite{Andrews:2019vou} argued that the four observed galactic double NS pulsars with short orbital periods ($<$1 day) and high eccentricities ($>0.5$) may all originate from GCs because their formation is difficult to explain within isolated binary evolution. (Accounting for radio selection effects of eccentric pulsars, however, may provide an alternative explanation for these detections~\citep{Bagchi:2013wga, Chattopadhyay:2020lff}.)
Recent observations also report a growing catalog of low-mass x-ray binaries with an NS as the compact object in star clusters~\citep{Fortin:2024siz}.
These findings contrast with studies suggesting that GCs contribute negligibly to the overall NS binary merger rate in the local Universe~\cite{Ye:2019xvf}. Thus, while the most conservative estimates of merger rates involving NSs from dynamical channels in star clusters remain low, this channel still merits further exploration.

A prior BNS merger scenario was proposed to form a compact object discovered in a binary with a pulsar PSR J0514-4002E in the GC NGC 1851~\cite{Barr:2024wwl}. Recently, Ref.~\cite{Ye2024} studied the formation of low-mass gap BHs in dense stellar clusters, similar to the companion of PSR J0514-4002E. They found that both massive star evolution and dynamical interactions can contribute to forming low-mass gap BHs. It was previously estimated that high natal kicks of NSs would almost always eject them from GCs~\citep{Hobbs2005MNRAS}. However, later studies indicate that the binary population of NSs exhibits evidence of much lower natal kicks~\citep{Igoshev2021,Fortin:2022ukx, Disberg:2024yrd}. Further, more massive GCs, such as Terzan 5 and 47 Tucanae, harbor nearly a third of the millisecond pulsar population~\citep{Manchester:2004bp, Harris1996, pulsars_gc_link}, demonstrating that a combination of high cluster escape speeds and low NS kicks can retain a significant number of NSs. Additionally, high-metallicity clusters tend to have a higher NS-to-BH fraction because stronger stellar winds produce lighter BHs, which receive higher natal kicks due to their smaller fallback mass. This, in turn, results in a smaller fraction of low-mass BHs~\citep{Fryer2012ApJ, Belczynski2010, Belczynski2012ApJ}.
In BH-poor clusters, NSs naturally dominate the mass segregation dynamics, significantly influencing overall cluster dynamics.

\subsubsection{Young star clusters, nuclear star clusters, and active galactic nuclei discs}
Metal-rich (close to solar metallicity) young star clusters may contribute significantly to the overall merger rate of NS binaries, although most of the contributions come from primordial binaries (i.e., binary systems that form during the initial star formation process within a cluster)~\cite{Ziosi:2014sra}.
References~\cite{Rastello:2020sru,ArcaSedda:2021zmm} suggested the dynamical formation scenario of NS binaries in metal-rich young star clusters to explain GW190814-like systems.  
Reference~\cite{Petrovich:2017otm} studied the merger rates of NS binaries in nonspherical nuclear star clusters with a massive BH at the center; they found a small contribution to the overall NS binary merger rate from nuclear star clusters. Our understanding of nuclear star clusters is not on par with that of GCs, and the feasibility of NSBH mergers with low mass-gap BHs is yet to be studied in detail. Active galactic nuclei discs can also contribute to the overall merger rate of NS binaries~\cite{McKernan:2020lgr}. They can also provide sites for producing low-mass gap BHs via both hierarchical NS mergers and significant accretion into NSs~\cite{Yang:2020xyi,Tagawa:2020qll}. This scenario was also proposed to interpret the secondary in GW190814-like binaries~\cite{Yang:2020xyi,Tagawa:2020qll}.

\subsection{Multiple star systems}
Hierarchical triples assembled either in the field or in dense clusters, with an inner binary containing NSs and an outer component that is also an NS, may provide yet another potential mechanism to produce a NSBH with a low mass gap BH~\cite{Lu:2020gfh,Liu:2020gif,Bartos:2023lfu,Gayathri:2023met}. This mechanism was put forward to explain GW events containing low-mass gap compact objects~\cite{Lu:2020gfh,Liu:2020gif,Gayathri:2023met}, such as GW190814~\cite{GW190814} and GW200210\_092254~\cite{GWTC-3}. Again, much less is known about the abundances of such triples from both observations and $N$-body simulations. Moreover, multiples such as hierarchical quadruple systems (2+2 and 3+1) containing NSs could also explain low-mass gap BHs~\cite{Safarzadeh:2019qkk,Fragione:2020aki,Vigna-Gomez:2020fvw,Hamers:2021olp,Vynatheya:2021mgl}. Ref.~\cite{Safarzadeh:2019qkk} proposed this channel to elucidate the formation pathway of GW190814~\cite{GW190814} and GW190924\_021846~\cite{GWTC2}. It remains to be explored if conditions for forming quadruple systems are commonly realized in the Universe.

\subsection{Primordial origins}
Recently, a study by Ref.~\cite{Afroz:2024fzp} revealed that GW events GW190425~\cite{GW190425} and GW230529 may originate from the primordial BH formation channel~\cite{Clesse:2020ghq}, where a subsolar mass BH grows to a higher mass by accretion.
However, recent results from microlensing surveys suggest that primordial BH formation channels involving dark matter may not be the dominant contributors to LVK sources in the local Universe (i.e., redshift $z\le0.1$)~\cite{Mroz:2024mse, Garcia-Bellido:2024yaz}.

\subsection{Our approach and key results}
The tension in the literature about the contribution of dense star clusters to the NS merger rate makes the problem even more interesting, and suggests further observational inputs are  necessary to settle this debate. Our approach is to turn the problem around and ask how GW observations can help us gain insights into the dynamical formation of NS binaries. A natural pathway for the primary of GW230529 is that it is formed from a BNS merger. Here, we investigate this scenario in detail, independent of the astrophysical details affecting binary formation. Such an agnostic approach could help constrain astrophysical scenarios that might produce GW230529-like systems through hierarchical mergers.

We extend the Bayesian framework of Ref.~\cite{Mahapatra:2024qsy}---developed in the case of BBHs---to the case of BNS, making use of the corresponding numerical relativity (NR) fitting formulas~\cite{Coughlin:2018fis} to infer the parameters of the parent binaries of the observed binary constituents. More specifically, we ask and answer the following questions:
\begin{enumerate}
    \item {\it If the primary of GW230529 originated from a prior BNS merger, what are the most likely parameters of the parent binary system?
    \item How do the inferred properties of these potential parent neutron stars compare to other neutron star binaries already observed by LIGO/Virgo?
    \item Assuming a previous merger led to the formation of the primary of GW230529, what proportion of such remnants eventually pair with another compact object and merge within a Hubble time?}
\end{enumerate}

\begin{figure}[ht]
\centering
\includegraphics[width=\columnwidth]{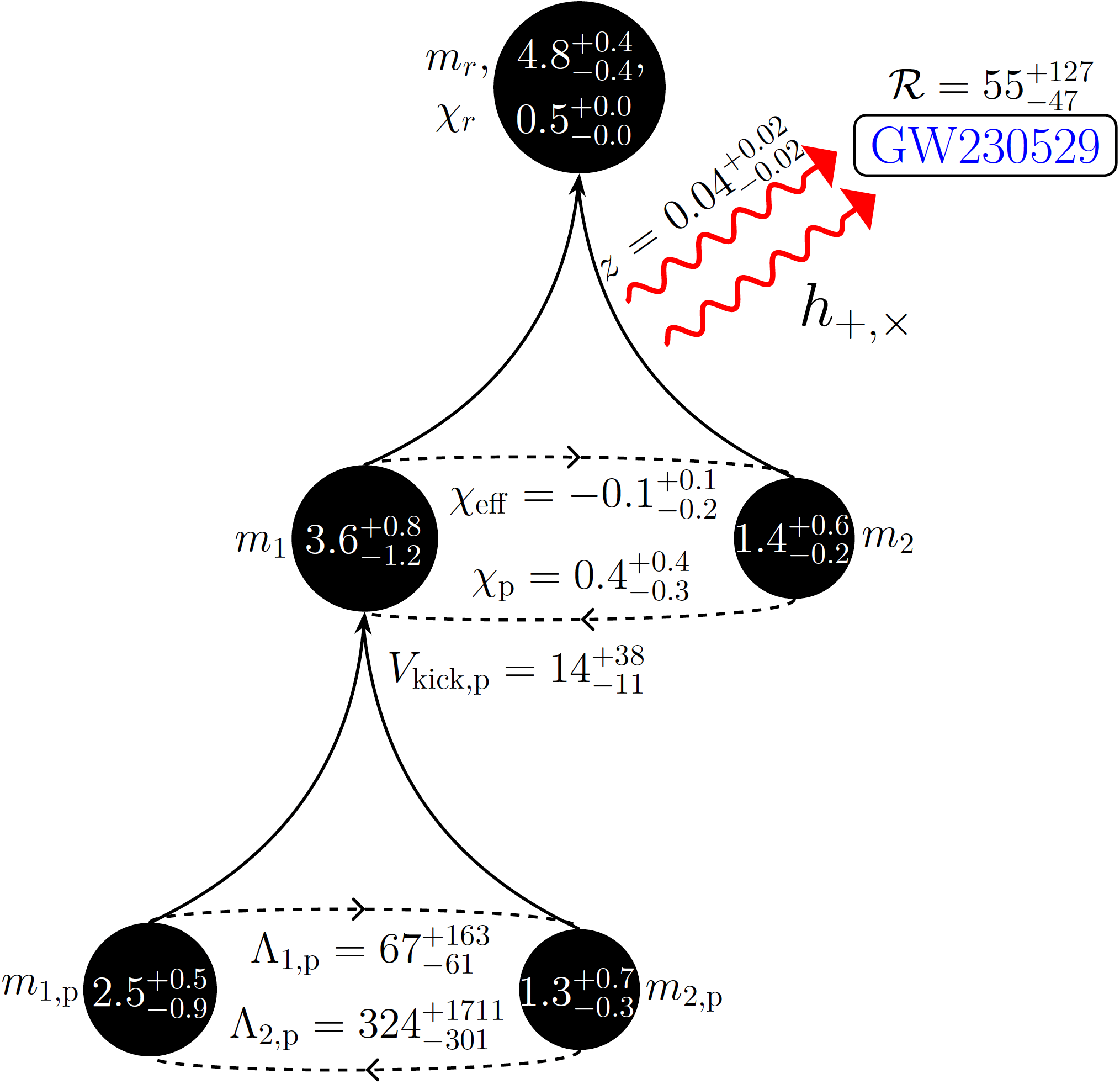}
    \caption{
    A schematic depiction of the possible merger history of GW230529 inferred by the method proposed in Ref.~\cite{Mahapatra:2024qsy} and using the NR fits for BNS mergers in Ref.~\cite{Coughlin:2018fis}. The middle of the figure depicts the observed binary components of GW230529, indicating the masses (in units of solar masses, $M_{\odot}$) as well as the effective dimensionless spin parameters $(\chi_{\rm eff}, \chi_{p})$ as inferred from the LVK Collaboration analysis~\cite{GW230529}. The remnant mass and spin of GW230529 are obtained using the NR fit for NSBH systems in Ref.~\cite{Zappa:2019ntl}. While estimating the remnant mass and spin, we have assumed a uniform distribution between 0 and 3000 for the dimensionless tidal deformability of the secondary component.
    The lower part of the figure shows the parameters $(m_{1, \rm p}, m_{2, \rm p}, \Lambda_{\rm 1, p}, \Lambda_{2, \rm p})$ for the parents of the primary component of GW230529. The kick magnitude $V_{\rm kick,p}$ (in km/s) imparted to the primary of GW230529 is also shown. Those values are inferred via the method described in Sec.~\ref{sec:method} and are among the main results of this paper. (We make use of the {\tt Flat} prior as discussed in Sec.~\ref{sec:prior}.) The numbers shown here quote the median parameter values and the upper and lower limits of the 90\% credibility interval of the inferred posteriors. We also show the redshift and merger rate (in units of $\rm Gpc^{-3}yr^{-1}$) for GW230529.
    } 
    \label{fig:gw230529}
\end{figure}

The inferred merger history of the primary of GW230529 is shown in Fig.~\ref{fig:gw230529}.
We find that the primary of GW230529 is consistent with the merger remnant of a BNS with median component masses of about $2.46M_{\odot}$ and $1.28M_{\odot}$. The tidal deformability of the heavier NS is predicted to be $\Lambda_{1,p}=67^{+163}_{-61}$; this is consistent with NS equation of state bounds from GW170817~\cite{GW170817}. Intriguingly, the masses of the parent BNS closely resemble those of GW190425~\cite{GW190425}, the second BNS merger observed by LIGO/Virgo during the third observation run. We discuss the astrophysical implications of these findings in Sec.~\ref{sec:results} and Sec.~\ref{sec:branching_fraction}.

The rest of the paper is organized as follows. In Sec.~\ref{sec:method}, we briefly summarize the adopted Bayesian inference framework from Ref.~\cite{Mahapatra:2024qsy} to estimate the parameters of the parent BNS. We describe our choice of priors on various parameters in Sec.~\ref{sec:prior}. The results from the analysis of the primary of GW230529 are reported in Secs.~\ref{sec:results} and \ref{sec:branching_fraction}. Finally, Sec.~\ref{sec:summary} summarizes our study.

\begingroup
\renewcommand{\arraystretch}{2.4} 
\begin{center}
	\begin{table*}[ht]
		\begin{tabular}{|| c | c | c ||} 
			\hline
			Prior name  & Distribution & Range for mass ratio\\ [0.5ex] 
			\hline\hline 
			{\tt Flat} & $\pi(m)\propto \mathcal{H} \, ({m - 1M_{\odot}})  \, \mathcal{H} ({ 3M_{\odot} - m})$ & [$\tfrac{1}{3}$, 1]\\ 
			\hline
            {\tt Powerlaw} & $\pi(m)\propto m^{-2} \, \mathcal{H} \, ({m - 1.2M_{\odot}})  \, \mathcal{H} ({ 2.8M_{\odot} - m})$ & [$\tfrac{1.2}{2.8}$, 1]\\
			\hline
            {\tt Truncated Gaussian} & $\pi(m)\propto e^{-\frac{(m-1.5M_{\odot})^{2}}{2\times(1.1M_{\odot})^{2}}} \, \mathcal{H} \, ({m - 1.2M_{\odot}})  \, \mathcal{H} ({ 2.7M_{\odot} - m})$ & [$\tfrac{1.2}{2.7}$, 1]\\
            \hline
		\end{tabular}
		\caption{Choice of different prior distributions for NS masses. Here, $\mathcal{H}$ is the Heaviside step function. For each prior distribution, the allowed ranges of mass ratio are shown in the last column.}
		\label{tab:mass-priors}
	\end{table*}
\end{center}
\endgroup

\section{Method}\label{sec:method}
The mapping of the properties of the binary constituents to the final remnants of BNS, NSBH, and BBH systems have been well studied using numerical relativity simulations and are described by phenomenological fitting functions. (see, for instance, Refs.~\cite{Pretorius:2005gq,Campanelli:2005dd,Baker:2005vv,Lousto:2009mf,Barausse:2012qz, Hofmann:2016yih,Varma:2019csw} for BBHs, Refs.~\cite{Zappa:2019ntl,Gonzalez:2022prs} for NSBHs, and Ref.~\cite{Coughlin:2018fis} for BNSs).  In the case of hierarchically formed BBH candidates, Refs.~\cite{Baibhav:2021qzw, Barrera:2022yfj, Barrera:2023qde, Alvarez:2024dpd, Mahapatra:2024qsy} used the NR fitting formulas to reconstruct the properties of the parents of selected BBHs in GWTC-3 following different methodologies.
Here, we closely follow the  Bayesian framework put forward in our earlier work~\cite{Mahapatra:2024qsy} but applied in the context of BNS fits. The BNS fitting formulas used in our study are summarized in the Supplemental Material.
  
We will assume that the primary of GW230529 is a low-mass gap BH formed through a previous generation BNS merger. Our goal is to estimate the properties of the parent BNS system. Let us denote the source-frame mass and the dimensionless spin parameter of the primary of GW230529 by $m_{\rm obs}$ and  $\chi_{\rm obs}$. We define $\vec{\theta}_{\rm hc}\equiv\{m_{\rm obs}, \chi_{\rm obs}\}$, where ``hc'' stands for ``hierarchical candidate.'' Let $\vec{\theta}_{\rm p}$ be the set of all parameters that describe the parent BNS.
Therefore, $\vec{\theta}_{\rm p}$ is given as\footnote{In general, we can have $\vec{\theta}_{\rm p} \equiv\{ m_{1,{\rm p}}, \vec{\chi}_{1,{\rm p}}, \Lambda_{1,{\rm p}}, m_{2,{\rm p}}, \vec{\chi}_{2,{\rm p}}, \Lambda_{2,{\rm p}} \} \,$, where $\vec{\chi}_{1,{\rm p}}$ and $\vec{\chi}_{2,{\rm p}}$ are the dimensionless spin angular momentum vectors of the primary and secondary, respectively. As we are using the NR fits for nonspinning BNS, we are considering the parameter set of Eq.~(\ref{eq:theta_parent}) for $\vec{\theta}_{\rm p}$, ignoring the NS spins. Once the NR fits for spinning BNS are available, we can use the above-mentioned parameter set for $\vec{\theta}_{\rm p}$. NSs are generally expected to have lower spin magnitudes compared to BHs~\cite{Lorimer:2008se, Chakrabarty:2008gz}. Among the known BNSs that will merge within a Hubble time, PSR J0737-3039A~\cite{Burgay:2003jj} has the highest observed spin magnitude, with an extrapolated spin magnitude of less than $\sim0.04$ at the time of merger.
 }
\begin{equation}\label{eq:theta_parent}
    \vec{\theta}_{\rm p}\equiv\{ m_{1,{\rm p}}, \Lambda_{1,{\rm p}}, m_{2,{\rm p}}, \Lambda_{2,{\rm p}} \} \,,
\end{equation} 
where, $m_{1,{\rm p}}$, $\Lambda_{1,{\rm p}}$, $m_{2,{\rm p}}$, and $\Lambda_{2,{\rm p}}$ are the masses and dimensionless tidal deformability (or polarizability) parameters\footnote{When the orbital separation in a BNS coalescence approaches the size of the NS, each NS is tidally distorted by its companion. For the $i$-th NS in the binary, its companion's quadrupolar electric-type tidal field $\mathcal{E}^{(i)}_{jl}$ induces a mass-quadrupole moment $\mathcal{I}^{(i)}_{jl}$. These are related via $\mathcal{I}^{(i)}_{jl}=-\Lambda_{i} m_{i}^{5} \mathcal{E}^{(i)}_{jl}$, where $\Lambda_{i}=\tfrac{2}{3}k_{2}(\tfrac{R_{i}}{m_{i}})^{5}$ is the dimensionless tidal deformability parameter, $k_2$ is the quadrupolar Love number, and $m_{i}$ and $R_i$, are the NS mass and radius.} of the primary and secondary of the parent BNS, respectively. We denote the more massive compact object as the ``primary''.
Here, we will use NR fits for the remnant mass and spin of nonspinning BNSs given in Ref.~\cite{Coughlin:2018fis} to relate $\vec{\theta}_{\rm hc}$ to $\vec{\theta}_{\rm p}$.

To compute the posterior distribution $p(\vec{\theta}_{\rm p}|d)$ on $\vec{\theta}_{\rm p}$ given the GW data $d$, we adopt Eq.~(11) of Ref.~\cite{Mahapatra:2024qsy}:
\begin{equation}\label{eq:Bayes-ancestral}
    p( \vec{\theta}_{\rm p} | d) = \frac{\pi(\vec{\theta}_{p})}{{\mathcal{Z}_{\rm p}(d)}} \, \frac{p(\vec{\theta}_{\rm hc}|d)}{\pi(\vec{\theta}_{\rm hc})} \bigg\rvert_{\vec{\theta}_{\rm hc}=\vec{F}(\vec{\theta}_{\rm p})}\,.
\end{equation}
In the above equation, $\pi(\vec{\theta}_{\rm p} )$ is the prior probability density function for $\vec{\theta}_{\rm p}$, $p(\vec{\theta}_{\rm hc}|d)$ and $\pi(\vec{\theta}_{\rm hc})$ are the posterior and the prior distributions of $\vec{\theta}_{\rm hc}$, and 
$$\mathcal{Z}_{\rm p}(d)\equiv\int \pi(\vec{\theta}_{p})\, \frac{p(\vec{\theta}_{\rm hc}|d)}{\pi(\vec{\theta}_{\rm hc})} \bigg\rvert_{\vec{\theta}_{\rm hc}=\vec{F}(\vec{\theta}_{\rm p})}\, d\vec{\theta}_{p}$$
is the evidence. The NR fits for the remnant mass and spin of BNS mergers are denoted by 
$$\vec{F}(\vec{\theta}_{\rm p})\equiv\{m^{\rm NR}_{ f} (\vec{\theta}_{\rm p}), \,  \chi^{\rm NR}_{ f} (\vec{\theta}_{\rm p})\}.$$ 
The reader may refer to Sec.~2 of Ref.~\cite{Mahapatra:2024qsy} for a detailed derivation of Eq.~(\ref{eq:Bayes-ancestral}).

We have taken the posterior samples of $\vec{\theta}_{\rm hc}$ from Ref.~\cite{GW230529}.\footnote{See Refs.~\cite{GW230529,Chandra:2024ila,Chattopadhyay:2024hsf} which argue that different choices of priors can alter the mass estimates of GW230529. This is due to the event’s low signal-to-noise ratio ($\sim$11).} We have adopted the combined posterior samples, which are obtained by an equal-weight combination of the posterior samples from IMRPhenomXPHM~\cite{IMRPhenomXPHM} and SEOBNRv5PHM~\cite{SEOBNRv5PHM} with low spin prior on the secondary mass. The released LVK posterior samples of individual masses of GW230529 are derived assuming uniform priors on detector-frame masses; this does not lead to uniform priors on the binary component source-frame masses. Therefore, to obtain the posterior samples of $m_{\rm obs}$ (the primary's source-frame mass) that assumes a uniform prior, we do a prior reweighting of LVK samples of $\{ m_{\rm obs}, \chi_{\rm obs} \}$.
As the posteriors and priors on $\vec{\theta}_{\rm hc}$ are supplied as discrete samples, we use a probability density estimator fit to these samples to construct $p(\vec{\theta}_{\rm hc}|d)$ and $\pi(\vec{\theta}_{\rm hc})$. Moreover, to generate the discrete samples for the probability distribution function (PDF) $p(\vec{\theta}_{\rm p}|d)$, we use the Bayesian parameter inference library {\tt bilby}~\cite{bilby_paper} with the {\tt dynesty}~\cite{dynesty} sampler (which uses the nested sampling algorithm~\cite{Skilling}). Our choice of priors $\pi(\vec{\theta}_{\rm p}|d)$ is described in the next section.

\section{Choice of priors}\label{sec:prior}
Previous studies of the galactic NS population have reported strong evidence for bimodality in the NS mass distribution, with one peak around $\sim1.33M_{\odot}-1.39M_{\odot}$ and another peak around $\sim1.49M_{\odot}-1.81M_{\odot}$~\cite{Valentim2011MNRAS,Ozel:2012ax,Kiziltan:2013oja,Antoniadis:2016hxz,Ozel:2016oaf,Alsing:2017bbc,Farrow:2019xnc}. However, a recent study by Ref.~\cite{You:2024bmk} found that the masses of NSs follow a unimodal distribution that smoothly turns on at 1.1$M_{\odot}$, peaks at 1.27$M_{\odot}$, and then declines as a steep power law, with this turn-on power-law distribution being strongly favored over the widely adopted empirical double-Gaussian model at the 3$\sigma$ level.
The masses of the lightest and the heaviest NS known through electromagnetic observations are $1.174\pm0.004M_{\odot}$~\cite{Martinez:2015mya} (see also \cite{Tauris:2019sho}) and  $2.14^{+0.10}_{-0.09}M_{\odot}$~\cite{NANOGrav:2019jur} (see also \cite{Antoniadis:2013pzd}), respectively. The maximum possible mass of an NS is constrained to be $\lesssim2.1M_{\odot}-2.6M_{\odot}$ from electromagnetic and GW observations~\cite{Kiziltan:2013oja,Lawrence:2015oka,Fryer:2015uia,Antoniadis:2016hxz,GW170817,Margalit:2017dij,Rezzolla:2017aly,Ruiz:2017due,Alsing:2017bbc,Shao:2020bzt,Shibata:2019ctb,Nathanail:2021tay,Kashyap:2021wzs}. 

From a theoretical perspective, some equations of state can support masses up to $\sim 3 M_{\odot}$ for nonrotating NSs~\cite{RhoadesPRL1974,Mueller:1996pm,Kalogera:1996ci,Godzieba:2020tjn} and even larger masses for rotating NSs~\cite{Friedman1987ApJ,Cook1994ApJ}. Recent core-collapse supernova simulations~\cite{Suwa:2018uni, Muller:2024aod} showed that the minimum allowed mass of NS  is $1.17 M_{\odot}$.
Further, population studies of GW events containing NSs (i.e., GW170817~\cite{GW170817}, GW190425~\cite{GW190425}, GW190814~\cite{GW190814}, GW200105, and GW200115~\cite{GW200105_GW200115}) suggest that the mass distribution of NSs does not favor a pronounced single peak (in contrast to Galactic BNSs, which have a mass distribution sharply peaked around $\sim1.35M_{\odot}$)~\cite{GWTC-3-pop,Landry:2021hvl}.
Instead the mass distribution is broader and shows more support for larger masses with inferred minimum and maximum possible NS masses of $1.2_{-0.2}^{+0.1}M_{\odot}$ and $2.8_{-0.2}^{+0.2}M_{\odot}$ ($2.2_{-0.3}^{+0.7}M_{\odot}$ without GW190814), respectively~\cite{GWTC-3-pop,Landry:2021hvl}. 

In our study, we choose three different kinds of prior distributions for the NS masses as listed in Table~\ref{tab:mass-priors}: {\tt Flat}, {\tt Powerlaw}, and {\tt Truncated Gaussian}. In the {\tt Flat} distribution, the masses of NSs are drawn uniformly from the range $[1M_{\odot},\,  3M_{\odot}]$ with the mass ratio constrained between $1/3$ and unity. For the {\tt Powerlaw} prior distribution, the masses are sampled from a power-law distribution between $1.2M_{\odot}$ and $2.8M_{\odot}$ with spectral index $-2$; the mass ratio is bounded in the range $[\tfrac{1.2}{2.8},\, 1]$~\cite{GWTC-3-pop}.
In the {\tt Truncated Gaussian} case, the masses are drawn from a Gaussian distribution with mean $1.5M_{\odot}$ and standard deviation $1.1M_{\odot}$, and with a sharp cutoff at the lower end ($1.2M_{\odot}$) and the upper end ($2.7M_{\odot}$)~\cite{GWTC-3-pop}. Here, the mass ratio is constrained to the range $[\tfrac{1.2}{2.7},\, 1]$.
For each prior distribution of NS masses, the dimensionless tidal deformability parameters ($\Lambda_{\rm 1,p}$ and $\Lambda_{\rm 2,p}$) are drawn uniformly from the range $[1,\,3000]$.

\begin{figure*}[ht]
\centering
\includegraphics[scale=0.44] {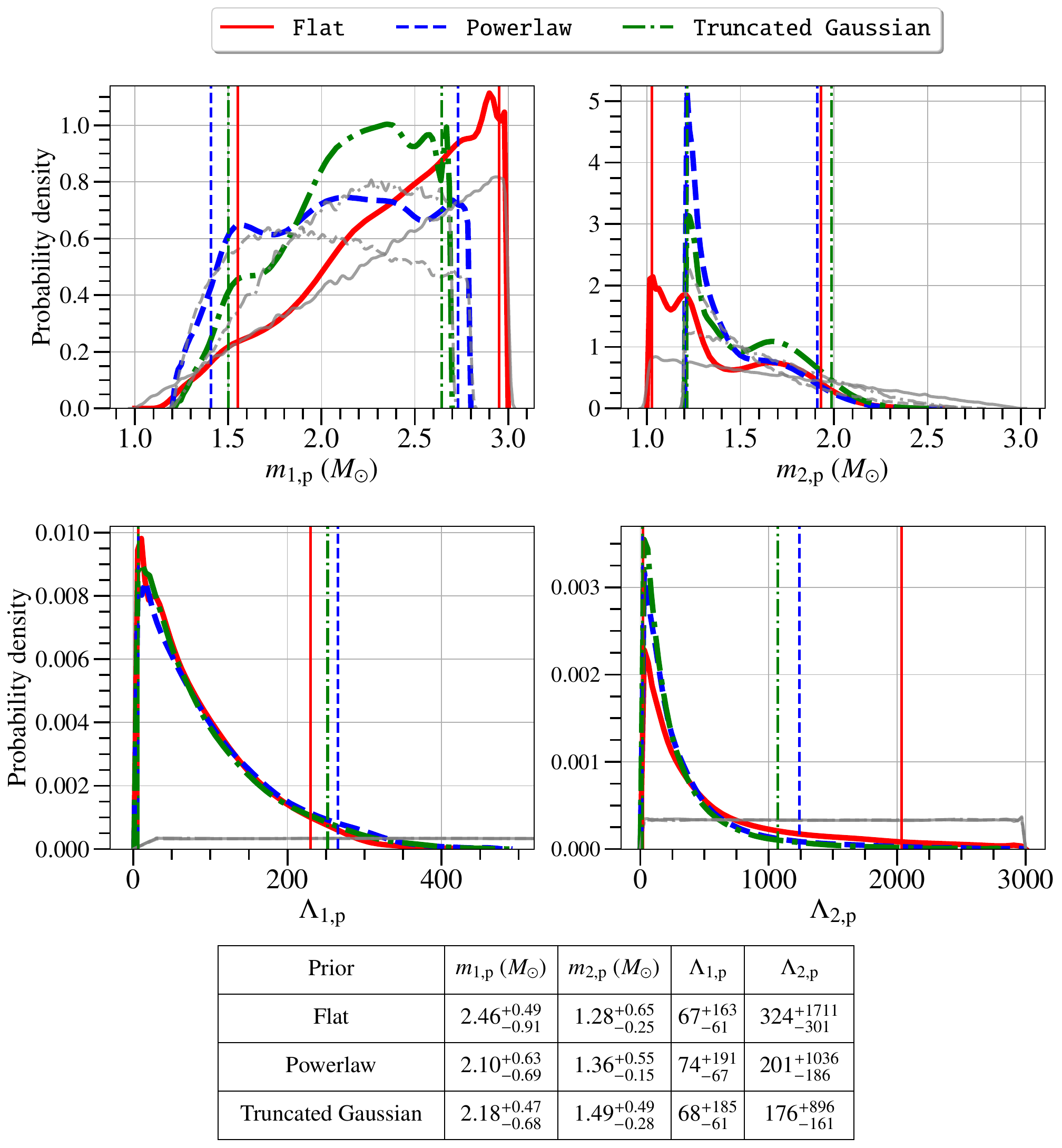}
    \caption{
    Posterior distributions of the component masses and tidal deformability parameters of the parent BNS of the primary of GW230529. The curves with different colors and line styles correspond to different prior choices for NS masses (listed in the legend). The colored vertical lines mark the 90\% credible intervals. Different prior distributions are shown in gray with varying line styles. The median values and 90\% confidence intervals for the different posterior distributions are also provided in the table.
    }\label{fig:m1-m2-lamda1-lambda2}
\end{figure*}


\begin{figure*}[ht]
\centering
\includegraphics[width=0.85\textwidth]{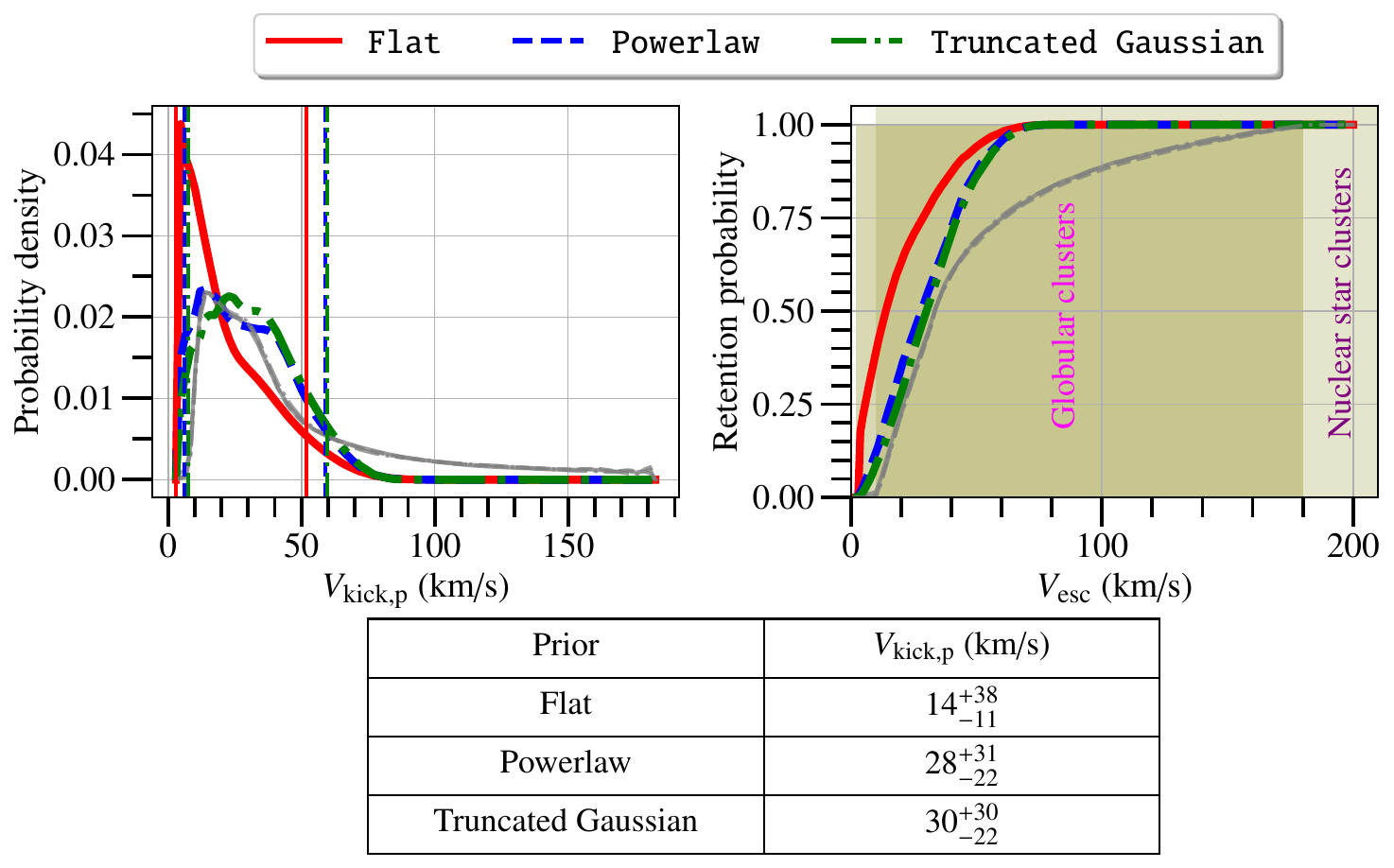}
    \caption{The left panel shows the posterior distributions of the inferred kick magnitude $V_{\rm kick,p}$ imparted to the primary component of GW230529. The curves with different colors and line styles correspond to different prior choices for NS masses (listed in the legend). The colored vertical lines mark the 90\% credible intervals. Different prior distributions are shown in gray with varying line styles. The median values and 90\% confidence intervals for the different posterior distributions are also provided in the table. The right panel shows the retention probability of the primary of GW230529 as a function of the escape speed of the host astrophysical environment. The shaded regions show the range of escape speeds for GCs and nuclear star clusters~\citep{Antonini:2016gqe}. The retention probability is computed directly from the cumulative distribution function of $V_{\rm kick,p}$ following Ref.~\cite{Mahapatra:2021hme}.
        }\label{fig:parent_kick}
\end{figure*}

\begin{figure*}[ht]
\centering
\includegraphics[scale=0.44]{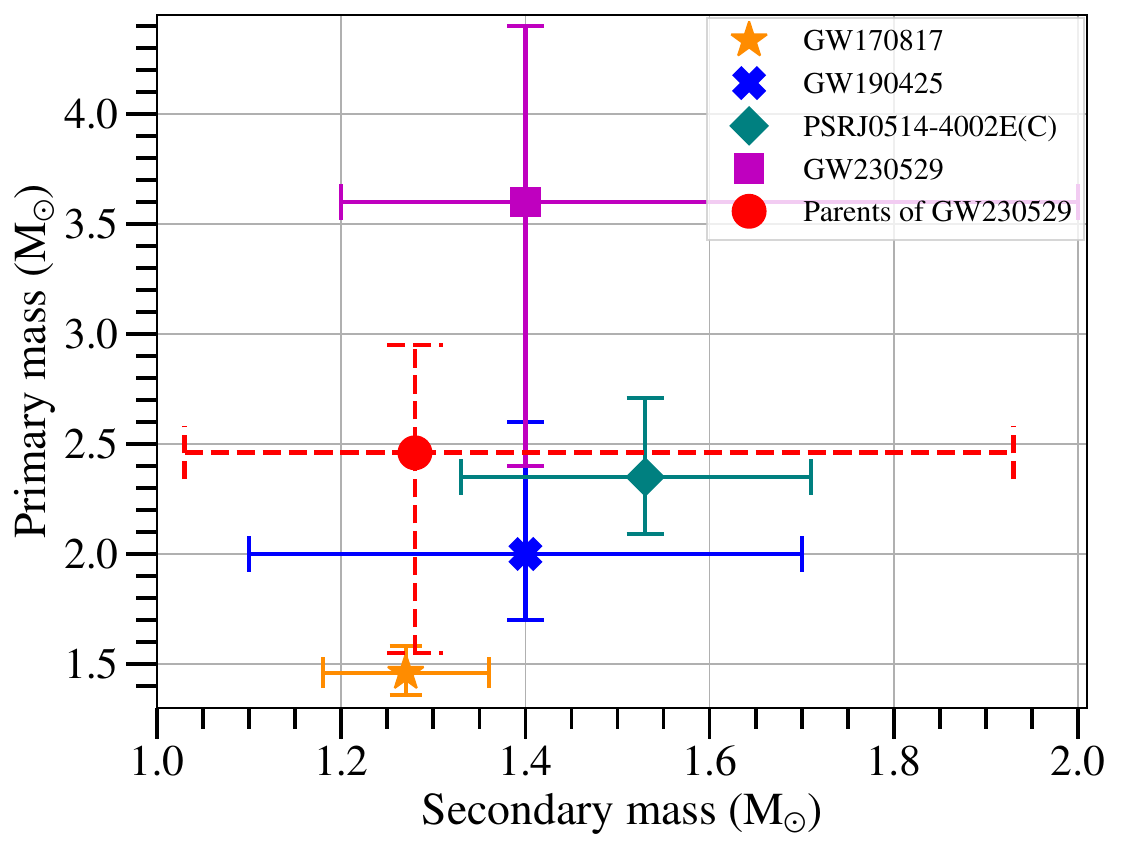}
\includegraphics[scale=0.44]{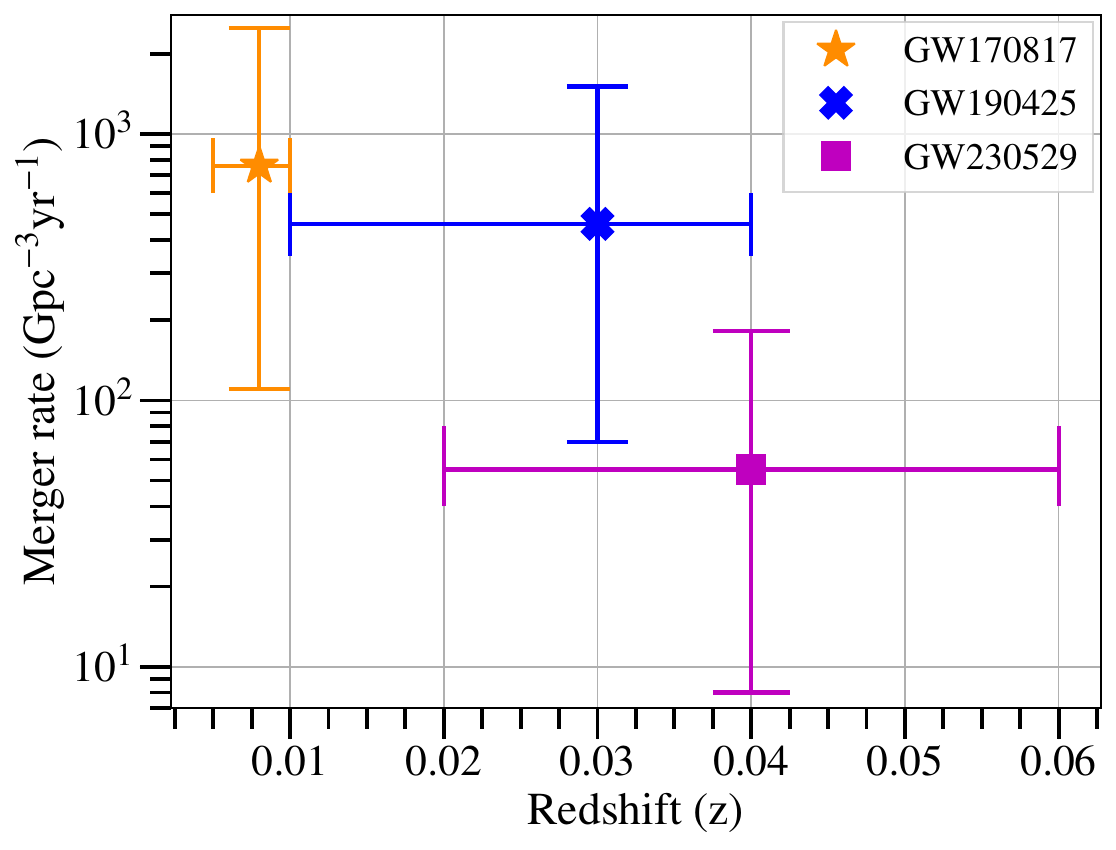}
    \caption{
    Left: comparison of the masses of the parent BNS of GW230529's primary with two BNS events in GWTC-3~\cite{GW170817,GW190425,GWTC-3} and the recently discovered binary millisecond pulsar PSR J0514--4002E~\cite{Barr:2024wwl}. For convenience, we also show the masses of GW230529. The 90\% credible intervals for the component masses of GW190425 and the masses of the parent BNS of GW230529's primary overlap. This suggests that GW190425-like binaries can serve as the potential parent of the primary of GW230529.
    Right: the 90\% credible intervals and median values for the merger rate and redshift of the two GWTC-3 BNS events and GW230529.
    }\label{fig:gw190425-proganitor}
\end{figure*}

\section{Parents of the primary of GW230529}\label{sec:results}
We now discuss what we can learn about the parents of the primary of GW230529 assuming it is a product of a previous BNS merger. There are three subsections that discuss the masses and tidal deformabilities of parent BNS, possible site of merger based on inferred GW kick, and comparison of the properties of the parent binary with other known systems from GW and radio observations.
\subsection{Masses and tidal deformability}
The parameter posterior distributions for the parents of GW230529's primary are derived using Eq.~(\ref{eq:Bayes-ancestral}) with the prior choices as described in Sec.~\ref{sec:prior}. The results are shown in Fig.~\ref{fig:m1-m2-lamda1-lambda2}. The predicted posteriors under different prior assumptions are in broad agreement with each other.
The posterior distributions on $m_{\rm 2,p}$ have the median values around $\sim1.3$--$1.5M_{\odot}$ with an extended tail beyond $2M_{\odot}$, such that the probability with which $m_{\rm 2,p}>2M_{\odot}$ is less than 3\%. The mass of the secondary in the BNS is consistent with the distribution of Galactic NS masses~\cite{Ozel:2012ax,Antoniadis:2016hxz,Ozel:2016oaf,Alsing:2017bbc,Farrow:2019xnc} as well as the NS masses from BNS systems observed through GW observations~\cite{GW170817,GW190425}. The mass of the primary of the parent BNS shows more support on the high-mass side and has a large overlap with the primary of GW190425~\cite{GW190425}; its median values span the range $2.1$--$2.5\, M_{\odot}$. (See Fig.~\ref{fig:gw190425-proganitor} for more details.) It is also consistent with the maximum possible NS mass inferred from population analyses~\cite{GWTC-3-pop}, both with and without including GW190814 (see Sec.~\ref{sec:prior}).

The dimensionless tidal deformability parameter of the primary ($\Lambda_{\rm 1,p}$) of the parent BNS is more constrained compared to the secondary ($\Lambda_{\rm 2,p}$). The posteriors on $\Lambda_{\rm 1,p}$ and $\Lambda_{\rm 2,p}$ are consistent with the measurement of tidal parameters from the BNS merger GW170817~\cite{GW170817,LIGOScientific:2018hze,LIGOScientific:2018cki,De:2018uhw,Radice2018ApJL,Radice:2018ozg,Radice:2018pdn,Coughlin:2018miv,Coughlin:2018fis}. Using GW170817, the dimensionless tidal deformability parameter of a $1.4M_{\odot}$ NS was constrained to $\Lambda(1.4 M_{\odot})<700$ at 90\% credibility~\cite{GW170817}.
The parameter $\Lambda_{\rm 1,p}$ is constrained to $\lesssim270$ with 90\% credibility ($m_{\rm 1,p}\gtrsim1.41M_{\odot}$ with 90\% credibility). The smaller magnitude of $\Lambda_{\rm 1,p}$ is due to the posteriors of $m_{\rm 1,p}$ having more support for larger mass; i.e., $m_{\rm 1,p}>2M_{\odot}$ with $>$60\% probability.  The tidal effects in the GW phasing are captured by an effective tidal parameter $\tilde{\Lambda}$~\cite{Flanagan:2007ix,Favata:2013rwa} as defined in Eq.~(5) of Ref.~\cite{Favata:2013rwa}. We also deduce constraints on the effective tidal parameter $\tilde{\Lambda}_{p}$ for the parent BNS. At 90\% confidence the constraints on $\tilde{\Lambda}_{p}$ are $150_{-113}^{+177}$, $144_{-110}^{+148}$, and $126_{-93}^{+156}$ for the {\tt Flat}, {\tt Powerlaw}, and {\tt Truncated Gaussian} priors on NS masses, respectively. The relatively smaller magnitude of $\tilde{\Lambda}_{p}$ is due to the smaller values of both $\Lambda_{\rm 1,p}$ and $\Lambda_{\rm 2,p}$. The probability distributions of mass and dimensionless tidal deformability for the parent BNS are consistent with constraints on the neutron star equation of state derived from the BNS merger GW170817~\cite{GW170817,LIGOScientific:2018cki}.

\subsection{Possible merger sites}
We have also estimated the magnitude of the recoil velocity ($V_{\rm kick,p}$) imparted to the primary component of GW230529 following the parent BNS merger using the NR fit developed in Ref.~\cite{Kulkarni:2023tex}. The kick imparted to the BNS merger remnant gets contributions from linear momentum flux due to asymmetric emission of GWs~\cite{Fitchett83, Favata:2004wz} as well as the asymmetric dynamical ejection of matter at relativistic velocities toward the end of the merger~\cite{Kulkarni:2023tex}. However, the larger recoils [$\mathcal{O}(100 \, \rm{km/s})$] in BNS mergers arise due to the dynamical ejecta. A formula for the BNS recoil as a function of the ejected mass ($M_{\rm ej}$) is provided in Eq.~(13) of Ref.~\cite{Kulkarni:2023tex}. Furthermore, we use the NR fit for $M_{\rm ej}$ as a function of the masses and dimensionless tidal deformability parameters of the BNS merger, as given in Eq.~(6) of Ref.~\cite{Nedora:2020qtd}.
Thus, the NR fit for the BNS recoil developed in Ref.~\cite{Kulkarni:2023tex} can be expressed as a function of the masses and dimensionless tidal deformability parameters to estimate $V_{\rm kick,p}$ (See the Supplemental Material for more details).

The posteriors of $V_{\rm kick,p}$ for three different choices of priors on $\vec{\theta}_{\rm p}$ (as explained in Sec.~\ref{sec:prior}) are shown in the left panel of Fig.~\ref{fig:parent_kick}. The NR fit of Ref.~\cite{Kulkarni:2023tex} is calibrated for ejected masses in the range $5 \times 10^{-6} M_{\odot} \lesssim M_{\rm ej} \lesssim 0.02 M_{\odot}$. While estimating $V_{\rm kick,p}$, we should discard posterior/prior samples of $\vec{\theta}_{\rm p}$ that result in $M_{\rm ej}$ values outside this calibration range. We found that all posterior samples of $\vec{\theta}_{\rm p}$ in all three cases resulted in $M_{\rm ej}$ values within the calibration region. However, we had to discard nearly half of the total prior samples of $\vec{\theta}_{\rm p}$ in all three cases. Therefore, the actual prior distributions of $V_{\rm kick,p}$ may be broader than those shown in Fig.~\ref{fig:parent_kick}. The mass ratio and tidal deformability parameters of the retained prior samples are quite similar across the three different prior choices, resulting in similar prior distributions. At 90\% confidence the constraints on $V_{\rm kick,p}$ are $14_{-11}^{+38}$ km/s, $28_{-22}^{+31}$ km/s, and $30_{-22}^{+30}$ km/s for the {\tt Flat}, {\tt Powerlaw}, and {\tt Truncated Gaussian} priors on NS masses, respectively. 

The right panel of Fig.~\ref{fig:parent_kick} shows the cumulative probability distribution function for $V_{\rm kick,p}$. This can be straightforwardly mapped to the retention probability of the primary of GW230529 in an astrophysical environment with an escape speed $V_{\rm esc}$ as shown in~\cite{Mahapatra:2021hme,Mahapatra:2022ngs}. If GW230529 occurred in a dense astrophysical environment, the escape speed would need to exceed 60 km/s (at 90\% credibility) for the environment to retain its primary component.
On the other hand, if GW230529 were part of a hierarchical triple (or quadruple) system, then the orbital parameters of the outer binary (or outer nested binary orbits) should be such that it could withstand a minimum recoil of magnitude $\sim 60$ km/s without getting unbound. In other words, postmerger kicks are unlikely to be a major obstacle for the hierarchical mergers of BNSs like the parent of GW230529's primary.
Even typical Milky Way GCs ($V_{\rm esc} \sim 30$ km/s) can retain $\gtrsim60\%$ of BNS remnants. Additionally, star clusters were more massive and denser at higher redshifts, leading to higher escape speeds that further increased retention probability. At birth, GCs were on average $\sim$ 4.5 times more massive than they are today, resulting in escape speeds about $\sim \sqrt{4.5}\approx2.1$ times higher~\citep{Webb2015MNRAS}. This would have allowed them to retain the majority of BNS remnants, enabling further binary formation.
This contrasts with the hierarchical mergers of BBHs, which require an astrophysical environment with $V_{\rm esc}\gtrsim 200$ km/s to retain $\gtrsim 50 \%$ of BBH remnants~\cite{Mahapatra:2021hme,Mahapatra:2022ngs,Mahapatra:2024qsy}. However, whether typical GCs allow the formation of such binaries at a rate that can explain the observed rate of GW230529 (assuming hierarchical mergers) is a more complex question that our study cannot answer.


\subsection{Comparison of the properties with other GW and radio binaries}
If the primary of GW230529 indeed comes from a previous-generation BNS merger, then it is natural to expect a greater abundance of lower-generation parent BNS mergers to happen either in dense astrophysical environments or as triples and quadruples in galactic fields.
This implies that the rate of GW230529-like mergers will be much less than the rate of mergers that led to the formation of the primary.
Therefore, it is instructive to ask whether the properties of any of the BNS events detected to date resemble the inferred parent of the primary of GW230529. 

The left panel of Fig.~\ref{fig:gw190425-proganitor} compares the inferred parent binary of GW230529's primary with two BNS events in GWTC-3~\cite{GW170817,GW190425,GWTC-3} and the recently discovered binary millisecond pulsar PSR J0514--4002E~\cite{Barr:2024wwl}.
Note that the 90\% credible intervals for the component masses of GW190425 coincide with the 90\% credible ranges for the component masses of the parent BNS of GW230529's primary. This suggests that GW190425-like BNS events could be the potential parents of the primary of GW230529. This possibility was first speculated in Ref.~\cite{GW230529} by looking at the overlap of the 90\% credible intervals for the remnant mass of GW190425 and the primary mass of GW230529. See Sec.~8 of Ref.~\cite{GW230529} for more details.
If this is the case, then the merger rate of GW190425-like binaries could be significantly higher than that of GW230529-like binaries, as explained earlier. Indeed, going by the median values, the merger rate of GW190425 ($\mathcal{R}_{190425}=460_{-390}^{+1050}$ $\rm Gpc^{-3}yr^{-1}$) is much larger than the merger rate of GW230529 ($\mathcal{R}_{230529}=55_{-47}^{+127}$ $\rm Gpc^{-3}yr^{-1}$) as can be seen in the right panel of Fig.~\ref{fig:gw190425-proganitor}. This lends further support for GW190425-like binaries as the potential parents of GW230529's primary. However, these are inferences based on a small sample of NS binaries that we have detected to date. 
These comparisons can be carried out more rigorously as more NS binaries are discovered in future observing runs of GW detectors.

We have also plotted PSRJ0514-4002E, a candidate NSBH system in the globular cluster NGC 1851, within the mass plane alongside other GW candidates. Interestingly, its parameters are comparable to those observed in GW events. However, the companion of this millisecond pulsar—the supposed BH—may not actually be a BH.
Given a mass constraint of 2.09–2.71\,M$_\odot$, it could instead be an NS formed through either from the merger of two lower-mass NSs or an NS and a heavy white dwarf, or it could simply be a massive NS~\cite{Barr:2024wwl,Ye2024}. Reference~\cite{Ye2024} suggests that while PSRJ0514-4002E may have formed in the unique environment of NGC 1851 (known for its atypical nature with hard-to-model multiple stellar populations~\cite{Carretta2011} and a highly eccentric Galactic orbit~\cite{CarballoBello2018}), the merger rate of such binaries extremely low, at approximately $\lesssim$1\,$\rm Gpc^{-3}yr^{-1}$.
This low merger rate suggests that GW230529 and PSRJ0514-4002E likely have different formation pathways even if both are products of hierarchical assembly. For example, GW230529 may have originated from a field triple system as proposed in other eccentric millisecond pulsar studies such as those on J1903+0327~\cite{TriplePortegies2011,TriplePijloo2012} or J0955-6150~\cite{TripleSerylak2022}.
Future radio observations could provide conclusive evidence about the nature of the companion of J0514-4002E if its spin is found to be relatively high (i.e., $\chi\sim 0.8$) ~\cite{Wex1999,Barr:2024wwl}. In such a case, identifying it as a BH would improve our understanding of the astrophysical scenarios leading to the formation of such compact, high-spin objects.

\subsection{Potential NSBH parents}
It is also possible that the primary of GW230529 may instead have originated from an NSBH merger. 
We have explored this alternative formation pathway and computed the properties of the potential parent NSBH system. 
To estimate the probability distributions of the parent binary parameters, we used NR fitting formulas, developed by Ref.~\cite{Zappa:2019ntl}, for the final mass and spin of NSBH mergers involving nonspinning NSs and precessing BHs, which are applied in Eq.~(\ref{eq:Bayes-ancestral}).
For the NS and BH masses, we use uniform priors in the ranges [1$M_{\odot}$, 3$M_{\odot}$] and [1$M_{\odot}$, 20$M_{\odot}$], respectively, with a mass ratio constraint between [3/20, 1]. For the tidal deformability parameter of the NS, we use a uniform prior in the range [1, 3000]. The spin magnitude of the BH follows a uniform prior in the range [0, 0.99], while the spin direction is chosen isotropically over the two-sphere. 
We find that the mass of the NS and the BH in the parent binary are constrained to $1.41_{-0.37}^{+0.64}$ and $2.46_{-0.89}^{+0.75}$, respectively, at 90\% credibility. These numbers are also similar to those presented in Fig.~\ref{fig:m1-m2-lamda1-lambda2}, where we considered the previous BNS merger scenario. The 90\% credible intervals for the component masses of GW190425 and PSRJ0514-4002E also overlap with the masses of the parent NSBH of GW230529’s primary. Therefore, the primary of GW230529 could also have originated from an NSBH, similar to systems like GW190425 or PSR J0514–4002E. The recent claim of orbital eccentricity in an NSBH merger~\cite{Morras:2025xfu} has led to speculations of this possibility. However, more observations are needed to disentangle the different formation pathways.

\section{
Fraction of GW190425-like mergers that can produce the primary of GW230529}\label{sec:branching_fraction}
We have already seen that remnants of GW190425-like mergers share similar parameters as the primary of GW230529. If there is a GW230529-like population, we can now ask what fraction of GW190425-like merger remnants will end up as primary components of a GW230529-like population. Indeed, given the complex interactions in a dense cluster, several of the GW190425-like remnants will be kicked out of the cluster, and those which are retained may not pair up with an NS in any reasonable time to produce a GW230529-like binary that merges in a Hubble time. We call this the relative branching fraction ($f_{b}$).
More precisely, the branching fraction is defined as the ratio of $\mathcal{R}_{230529}$ to $\mathcal{R}_{190425}$, where $\mathcal{R}_{230529}$ and $\mathcal{R}_{190425}$ are the merger rates of GW230529 and GW190425, respectively.

To estimate the posterior distribution on the branching fraction, $p(f_{b}|d)$, we define two variables ($\mathcal{R}_p$, $f_b$) and their inverses: 
\begin{align}
\mathcal{R}_p &\equiv \mathcal{R}_{190425} \, \mathcal{R}_{230529} \,, \;\;\;\;
f_b \equiv \frac{\mathcal{R}_{230529}}{\mathcal{R}_{190425}}\,,\\
\mathcal{R}_{190425} &= \sqrt{\frac{\mathcal{R}_p}{f_b}} \,, \;\;\;\; 
\mathcal{R}_{230529} = \sqrt{\mathcal{R}_p \, f_b } \,.\label{eq:fbdef}
\end{align}
To compute $p(f_{b}|d)$, we marginalize the joint PDF $p(\mathcal{R}_p, f_{b}|d)$ over the parameter $\mathcal{R}_p$:
\begin{equation}\label{eq:fbprobintrm}
\begin{split}
p(f_{b}|d) &= \int p(\mathcal{R}_p, f_{b}|d)\, d \mathcal{R}_p\\
&= \int p(\mathcal{R}_{190425}, \mathcal{R}_{230529}|d) \,  \frac{1}{2 \mathcal{R}_p} \,d \mathcal{R}_p\, ,
\end{split}
\end{equation}
where the factor $\tfrac{1}{2 \mathcal{R}_p}$ arises from the Jacobian of the coordinate transformation $\partial (\mathcal{R}_{190425}, \mathcal{R}_{230529})/\partial (\mathcal{R}_p, f_b)$. GW190425 and GW230529 are two independent GW event triggers. Here, we consider GW190425 and GW230529 as belonging to two different classes of compact binary mergers, and we assume the PDFs for the merger rates of GW190425- and GW230529-like binaries are independent of each other, with their joint PDF simply the product of the two.
Therefore, Eq.~(\ref{eq:fbprobintrm}) further simplifies to
\begin{equation}\label{eq:fbprob}
p(f_{b}|d) = \int p(\mathcal{R}_{190425}|d) \, p(\mathcal{R}_{230529}|d)\, \frac{1}{2 \mathcal{R}_p} \,d \mathcal{R}_p\, .  
\end{equation}
 
\begin{figure}[t]
\centering
\includegraphics[scale=0.46] {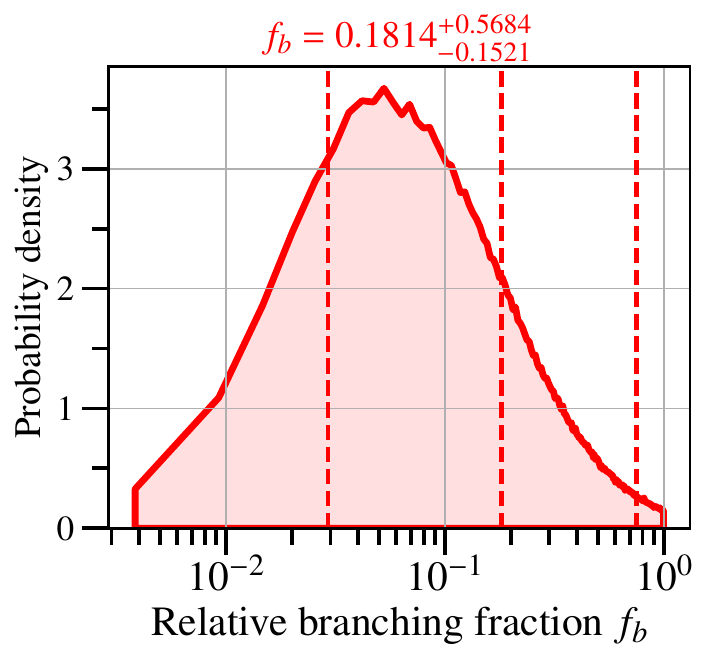}
    \caption{The posterior distribution on the relative branching fraction $f_b$ between lower-generation BNS mergers that produce BHs in the low-mass gap and higher-generation binary mergers containing low-mass gap BHs and NSs. The vertical dashed lines mark the 90\% credible intervals and median values.
    }\label{fig:branching_ratio}
\end{figure}

We next construct the PDFs $p(\mathcal{R}_{190425}|d)$ and $p(\mathcal{R}_{230529}|d)$. The PDF $p(\mathcal{R}_{190425}|d)$ is constructed using log-normal fits to the merger rate estimates $\mathcal{R}_{190425}=460_{-390}^{+1050}$ $\rm Gpc^{-3}yr^{-1}$, which has a mean of $\left\langle \log_{10}[{{\mathcal{R}}}/({\rm Gpc^{-3}yr^{-1}})] \right\rangle =2.675$ and a standard deviation of $\sigma_{\log_{10}[{{\mathcal{R}}}/({\rm Gpc^{-3}yr^{-1}})]} = 0.4$. The PDF $p(\mathcal{R}_{230529}|d)$ is constructed using log-normal fits to the data in Fig.~3 of \cite{GW230529}, which has a mean of $\left\langle \log_{10}[{{\mathcal{R}}}/({\rm Gpc^{-3}yr^{-1}})] \right\rangle =1.7404$ and a standard deviation of $\sigma_{\log_{10}[{{\mathcal{R}}}/({\rm Gpc^{-3}yr^{-1}})]} = 0.32$. These two PDFs are fed to Eq.~(\ref{eq:fbprob}) to compute the posteriors on $f_b$, $p(f_{b}|d)$. The posterior distribution on $f_b$ is shown in Fig.~\ref{fig:branching_ratio}. The branching fraction $f_b$ is constrained to $0.18_{-0.15}^{+0.57}$ at 90\% credibility. Assuming GW190425 to be representative of a typical BNS merger and GW230529 a NSBH where the BH is formed by a previous BNS merger, this number suggests that $3-75$\% of lower generation BNS mergers will lead to the formation of NSBH binaries with BHs in the low-mass gap that will merge within a Hubble time. Future detections of both BNS mergers and binaries with low-mass gap BH will put stringent constraints on $f_b$. Note that there can be NSBHs whose BH is formed by a previous BNS merger, but as long as these binaries do not merge in a Hubble time, they will not enter into $f_b$.

\section{Summary and conclusions}\label{sec:summary}
Some of the recent GW and radio observations have challenged our understanding of dynamical interactions of NSs and low-mass BHs (masses less than $5M_{\odot})$. One of the most important questions in this context is if dense star clusters can host a non-negligible fraction of BNS mergers and whether the remnants of these mergers will get a chance to further pair up with another NS or BH and merge in a Hubble time. We investigate this question in the context of GW230529---a likely NSBH merger with a BH mass less than $5 M_{\odot}$.

Assuming the massive primary component of GW230529 is a product of a previous BNS merger, we characterize the properties of its parents. We use a BNS fitting formula that maps the initial binary parameters to the parameters of the remnant. This identifies the allowed region of the initial binary configuration corresponding to the inferred properties of the remnant. We find the observed properties of the primary of GW230529 to be consistent with the merger of two NSs, one which has a (median) mass of $2.5 M_{\odot}$ and is very compact ($6\leq\Lambda\leq 230$) and another with a mass of $1.3 M_{\odot}$ whose tidal deformability we are unable to constrain. Intriguingly, the parent binary configuration of GW230529's primary closely resembles that of GW190425, the second BNS detection by the LIGO and Virgo observatories during their third observing run. Further, using the merger rates of GW230529 and GW190425, we calculate the fraction of GW190425-like merger remnants that form another binary and merge. We find this branching fraction to be $3-75\%$ at 90\% credibility, a finding that can be tested rigorously with future compact binary observations by LIGO/Virgo-like detectors~\cite{ALIGO,AVirgo} as well as next-generation observatories such as the Cosmic Explorer~\cite{CE:2019iox} and the Einstein Telescope~\cite{ETScience11,ET:2016wof}. 

If the primary of GW230529 is a product of a previous merger, it poses an important astrophysical question about the environments in which such mergers happen. Some past studies suggest the rate of BNS mergers in GCs is too small to account for the observed event rates. Given the complexities in modeling stellar dynamics in dense clusters, it is worth investigating this further, focusing on star cluster configurations that can lead to hierarchical mergers involving low-mass compact binaries~\cite{GW230529}.
Reference~\cite{Ye2024} found that GW230529-like systems could be components of dynamically assembled binaries, but the corresponding merger rate is probably $\lesssim1\,{\rm Gpc^{-3}\, yr^{-1}}$. If these configurations are indeed rare, perhaps hierarchical triples or quadruples may offer an alternative pathway to explain the inferred merger history of the primary of GW230529. Future electromagnetic (such as radio searches for NSBHs in GCs), as well as GW observations, can potentially help us resolve these conflicts.
Finally, we note that the primary of GW230529 may have originated from an NSBH merger, similar to systems like GW190425 and PSR J0514–4002E, particularly in light of recent claims of orbital eccentricity in an NSBH event~\cite{Morras:2025xfu}.

Note that the employed NR fits for BNS and NSBH remnants~\cite{Coughlin:2018fis} in our study are not as accurate as compared to the NR fits for BBH remnants~\cite{Hofmann:2016yih, Varma:2019csw}. Further, we ignore the spins of NSs in our study and use NR fits for the remnant mass and spin of nonspinning BNS components. In the future, one should revisit the results presented here when more accurate NR fits that include NS spins are available. With more accurate fitting formulas, studying the mass and spin distributions in hierarchical BNS and NSBH mergers will be a promising direction for future research.

Our method reconstructs a merger history, assuming the observed primary component of GW230529 was formed via a hierarchical merger of a BNS. We do not attempt to statistically quantify whether this hypothesis is true; rather, we present a plausible ancestral pathway based on the assumption of a BNS merger history. In the future, the detection of loud events containing low-mass gap BHs with well-constrained masses and spin magnitudes will enable us to test the hierarchical BNS merger hypothesis as an explanation for low-mass gap BHs.

\section*{Acknowledgements}
\label{sec:acknowledgements}
We thank Simon Stevenson for critically reading the manuscript and providing useful comments. P.M. thanks Thomas Dent, Juan García-Bellido, and Suvodip Mukherjee for their insightful comments on the manuscript. P.M. thanks Duncan MacLeod for his suggestions on various computing issues. D.C. thanks Vivek Venkatraman Krishnan for a useful discussion on the observations of pulsars. P.M. acknowledges the Science and Technology Facilities Council (STFC) for support through Grant No. ST/V005618/1.
P.M. and K.G.A.~acknowledge the support of the Core Research Grant No. CRG/2021/004565 of the Science and Engineering Research Board of India and a grant from the Infosys Foundation. K.G.A.~also acknowledges support from the Department of Science and Technology and the Science and Engineering Research Board (SERB) of India via the Swarnajayanti Fellowship Grant No. DST/SJF/PSA-01/2017-18. D.C. is supported by the STFC Grant No. ST/V005618/1. D.C. and B.S.S. thank the Aspen Center for Physics (ACP) summer workshop 2022 for setting up discussions that also contributed to this collaborative work. We also acknowledge National Science Foundation (NSF) support via NSF Awards No. AST-2205920 and No. PHY-2308887 to A.G., NSF CAREER Award No. PHY-1653374 to M.F., and No. AST-2307147, No. PHY-2207638, No. PHY-2308886 and No. PHY-2309064 to B.S.S. The authors are grateful for computational resources provided by the Cardiff University and support by STFC Grants No. ST/I006285/1 and No. ST/V005618/1. This manuscript has the LIGO preprint number P2500111.

This research has made use of data obtained from the Gravitational Wave Open Science Center (\url{https://www.gwosc.org/}), a service of the LIGO Laboratory, the LIGO Scientific Collaboration and the Virgo Collaboration. LIGO Laboratory and Advanced LIGO are funded by the United States National Science Foundation (NSF) as well as the Science and Technology Facilities Council (STFC) of the United Kingdom, the Max-Planck-Society (MPS), and the State of Niedersachsen/Germany for support of the construction of Advanced LIGO and construction and operation of the GEO600 detector. Additional support for Advanced LIGO was provided by the Australian Research Council. Virgo is funded, through the European Gravitational Observatory (EGO), by the French Centre National de Recherche Scientifique (CNRS), the Italian Istituto Nazionale di Fisica Nucleare (INFN) and the Dutch Nikhef, with contributions by institutions from Belgium, Germany, Greece, Hungary, Ireland, Japan, Monaco, Poland, Portugal, Spain. The construction and operation of KAGRA are funded by Ministry of Education, Culture, Sports, Science and Technology (MEXT), and Japan Society for the Promotion of Science (JSPS), National Research Foundation (NRF) and Ministry of Science and ICT (MSIT) in Korea, Academia Sinica (AS) and the Ministry of Science and Technology (MoST) in Taiwan.

\section*{DATA AVAILABILITY}
The data that support the findings of this article are not publicly available. The data are available from the authors upon reasonable request.

\bibliography{ref-list}

\begin{thebibliography}{173}%
\makeatletter
\providecommand \@ifxundefined [1]{%
 \@ifx{#1\undefined}
}%
\providecommand \@ifnum [1]{%
 \ifnum #1\expandafter \@firstoftwo
 \else \expandafter \@secondoftwo
 \fi
}%
\providecommand \@ifx [1]{%
 \ifx #1\expandafter \@firstoftwo
 \else \expandafter \@secondoftwo
 \fi
}%
\providecommand \natexlab [1]{#1}%
\providecommand \enquote  [1]{``#1''}%
\providecommand \bibnamefont  [1]{#1}%
\providecommand \bibfnamefont [1]{#1}%
\providecommand \citenamefont [1]{#1}%
\providecommand \href@noop [0]{\@secondoftwo}%
\providecommand \href [0]{\begingroup \@sanitize@url \@href}%
\providecommand \@href[1]{\@@startlink{#1}\@@href}%
\providecommand \@@href[1]{\endgroup#1\@@endlink}%
\providecommand \@sanitize@url [0]{\catcode `\\12\catcode `\$12\catcode
  `\&12\catcode `\#12\catcode `\^12\catcode `\_12\catcode `\%12\relax}%
\providecommand \@@startlink[1]{}%
\providecommand \@@endlink[0]{}%
\providecommand \url  [0]{\begingroup\@sanitize@url \@url }%
\providecommand \@url [1]{\endgroup\@href {#1}{\urlprefix }}%
\providecommand \urlprefix  [0]{URL }%
\providecommand \Eprint [0]{\href }%
\providecommand \doibase [0]{http://dx.doi.org/}%
\providecommand \selectlanguage [0]{\@gobble}%
\providecommand \bibinfo  [0]{\@secondoftwo}%
\providecommand \bibfield  [0]{\@secondoftwo}%
\providecommand \translation [1]{[#1]}%
\providecommand \BibitemOpen [0]{}%
\providecommand \bibitemStop [0]{}%
\providecommand \bibitemNoStop [0]{.\EOS\space}%
\providecommand \EOS [0]{\spacefactor3000\relax}%
\providecommand \BibitemShut  [1]{\csname bibitem#1\endcsname}%
\let\auto@bib@innerbib\@empty
\bibitem [{\citenamefont {Bailyn}\ \emph {et~al.}(1998)\citenamefont {Bailyn},
  \citenamefont {Jain}, \citenamefont {Coppi},\ and\ \citenamefont
  {Orosz}}]{Bailyn:1997xt}%
  \BibitemOpen
  \bibfield  {author} {\bibinfo {author} {\bibfnamefont {C.~D.}\ \bibnamefont
  {Bailyn}}, \bibinfo {author} {\bibfnamefont {R.~K.}\ \bibnamefont {Jain}},
  \bibinfo {author} {\bibfnamefont {P.}~\bibnamefont {Coppi}}, \ and\ \bibinfo
  {author} {\bibfnamefont {J.~A.}\ \bibnamefont {Orosz}},\ }\href {\doibase
  10.1086/305614} {\bibfield  {journal} {\bibinfo  {journal} {Astrophys. J.}\
  }\textbf {\bibinfo {volume} {499}},\ \bibinfo {pages} {367} (\bibinfo {year}
  {1998})},\ \Eprint {http://arxiv.org/abs/astro-ph/9708032}
  {arXiv:astro-ph/9708032} \BibitemShut {NoStop}%
\bibitem [{\citenamefont {{{\"O}zel}}\ \emph {et~al.}(2010)\citenamefont
  {{{\"O}zel}}, \citenamefont {{Psaltis}}, \citenamefont {{Narayan}},\ and\
  \citenamefont {{McClintock}}}]{Ozel2010ApJ}%
  \BibitemOpen
  \bibfield  {author} {\bibinfo {author} {\bibfnamefont {F.}~\bibnamefont
  {{{\"O}zel}}}, \bibinfo {author} {\bibfnamefont {D.}~\bibnamefont
  {{Psaltis}}}, \bibinfo {author} {\bibfnamefont {R.}~\bibnamefont
  {{Narayan}}}, \ and\ \bibinfo {author} {\bibfnamefont {J.~E.}\ \bibnamefont
  {{McClintock}}},\ }\href {\doibase 10.1088/0004-637X/725/2/1918} {\bibfield
  {journal} {\bibinfo  {journal} {\apj}\ }\textbf {\bibinfo {volume} {725}},\
  \bibinfo {pages} {1918} (\bibinfo {year} {2010})},\ \Eprint
  {http://arxiv.org/abs/1006.2834} {arXiv:1006.2834 [astro-ph.GA]} \BibitemShut
  {NoStop}%
\bibitem [{\citenamefont {{Farr}}\ \emph {et~al.}(2011)\citenamefont {{Farr}},
  \citenamefont {{Sravan}}, \citenamefont {{Cantrell}}, \citenamefont
  {{Kreidberg}}, \citenamefont {{Bailyn}}, \citenamefont {{Mandel}},\ and\
  \citenamefont {{Kalogera}}}]{Farr2011ApJ}%
  \BibitemOpen
  \bibfield  {author} {\bibinfo {author} {\bibfnamefont {W.~M.}\ \bibnamefont
  {{Farr}}}, \bibinfo {author} {\bibfnamefont {N.}~\bibnamefont {{Sravan}}},
  \bibinfo {author} {\bibfnamefont {A.}~\bibnamefont {{Cantrell}}}, \bibinfo
  {author} {\bibfnamefont {L.}~\bibnamefont {{Kreidberg}}}, \bibinfo {author}
  {\bibfnamefont {C.~D.}\ \bibnamefont {{Bailyn}}}, \bibinfo {author}
  {\bibfnamefont {I.}~\bibnamefont {{Mandel}}}, \ and\ \bibinfo {author}
  {\bibfnamefont {V.}~\bibnamefont {{Kalogera}}},\ }\href {\doibase
  10.1088/0004-637X/741/2/103} {\bibfield  {journal} {\bibinfo  {journal}
  {\apj}\ }\textbf {\bibinfo {volume} {741}},\ \bibinfo {eid} {103} (\bibinfo
  {year} {2011})},\ \Eprint {http://arxiv.org/abs/1011.1459} {arXiv:1011.1459
  [astro-ph.GA]} \BibitemShut {NoStop}%
\bibitem [{\citenamefont {Rhoades}\ and\ \citenamefont
  {Ruffini}(1974)}]{RhoadesPRL1974}%
  \BibitemOpen
  \bibfield  {author} {\bibinfo {author} {\bibfnamefont {C.~E.}\ \bibnamefont
  {Rhoades}}\ and\ \bibinfo {author} {\bibfnamefont {R.}~\bibnamefont
  {Ruffini}},\ }\href {\doibase 10.1103/PhysRevLett.32.324} {\bibfield
  {journal} {\bibinfo  {journal} {Phys. Rev. Lett.}\ }\textbf {\bibinfo
  {volume} {32}},\ \bibinfo {pages} {324} (\bibinfo {year} {1974})}\BibitemShut
  {NoStop}%
\bibitem [{\citenamefont {{Friedman}}\ and\ \citenamefont
  {{Ipser}}(1987)}]{Friedman1987ApJ}%
  \BibitemOpen
  \bibfield  {author} {\bibinfo {author} {\bibfnamefont {J.~L.}\ \bibnamefont
  {{Friedman}}}\ and\ \bibinfo {author} {\bibfnamefont {J.~R.}\ \bibnamefont
  {{Ipser}}},\ }\href {\doibase 10.1086/165088} {\bibfield  {journal} {\bibinfo
   {journal} {\apj}\ }\textbf {\bibinfo {volume} {314}},\ \bibinfo {pages}
  {594} (\bibinfo {year} {1987})}\BibitemShut {NoStop}%
\bibitem [{\citenamefont {{Cook}}\ \emph {et~al.}(1994)\citenamefont {{Cook}},
  \citenamefont {{Shapiro}},\ and\ \citenamefont {{Teukolsky}}}]{Cook1994ApJ}%
  \BibitemOpen
  \bibfield  {author} {\bibinfo {author} {\bibfnamefont {G.~B.}\ \bibnamefont
  {{Cook}}}, \bibinfo {author} {\bibfnamefont {S.~L.}\ \bibnamefont
  {{Shapiro}}}, \ and\ \bibinfo {author} {\bibfnamefont {S.~A.}\ \bibnamefont
  {{Teukolsky}}},\ }\href {\doibase 10.1086/173934} {\bibfield  {journal}
  {\bibinfo  {journal} {\apj}\ }\textbf {\bibinfo {volume} {424}},\ \bibinfo
  {pages} {823} (\bibinfo {year} {1994})}\BibitemShut {NoStop}%
\bibitem [{\citenamefont {Mueller}\ and\ \citenamefont
  {Serot}(1996)}]{Mueller:1996pm}%
  \BibitemOpen
  \bibfield  {author} {\bibinfo {author} {\bibfnamefont {H.}~\bibnamefont
  {Mueller}}\ and\ \bibinfo {author} {\bibfnamefont {B.~D.}\ \bibnamefont
  {Serot}},\ }\href {\doibase 10.1016/0375-9474(96)00187-X} {\bibfield
  {journal} {\bibinfo  {journal} {Nucl. Phys. A}\ }\textbf {\bibinfo {volume}
  {606}},\ \bibinfo {pages} {508} (\bibinfo {year} {1996})},\ \Eprint
  {http://arxiv.org/abs/nucl-th/9603037} {arXiv:nucl-th/9603037} \BibitemShut
  {NoStop}%
\bibitem [{\citenamefont {Kalogera}\ and\ \citenamefont
  {Baym}(1996)}]{Kalogera:1996ci}%
  \BibitemOpen
  \bibfield  {author} {\bibinfo {author} {\bibfnamefont {V.}~\bibnamefont
  {Kalogera}}\ and\ \bibinfo {author} {\bibfnamefont {G.}~\bibnamefont
  {Baym}},\ }\href {\doibase 10.1086/310296} {\bibfield  {journal} {\bibinfo
  {journal} {Astrophys. J. Lett.}\ }\textbf {\bibinfo {volume} {470}},\
  \bibinfo {pages} {L61} (\bibinfo {year} {1996})},\ \Eprint
  {http://arxiv.org/abs/astro-ph/9608059} {arXiv:astro-ph/9608059} \BibitemShut
  {NoStop}%
\bibitem [{\citenamefont {{Fryer}}\ \emph {et~al.}(2012)\citenamefont
  {{Fryer}}, \citenamefont {{Belczynski}}, \citenamefont {{Wiktorowicz}},
  \citenamefont {{Dominik}}, \citenamefont {{Kalogera}},\ and\ \citenamefont
  {{Holz}}}]{Fryer2012ApJ}%
  \BibitemOpen
  \bibfield  {author} {\bibinfo {author} {\bibfnamefont {C.~L.}\ \bibnamefont
  {{Fryer}}}, \bibinfo {author} {\bibfnamefont {K.}~\bibnamefont
  {{Belczynski}}}, \bibinfo {author} {\bibfnamefont {G.}~\bibnamefont
  {{Wiktorowicz}}}, \bibinfo {author} {\bibfnamefont {M.}~\bibnamefont
  {{Dominik}}}, \bibinfo {author} {\bibfnamefont {V.}~\bibnamefont
  {{Kalogera}}}, \ and\ \bibinfo {author} {\bibfnamefont {D.~E.}\ \bibnamefont
  {{Holz}}},\ }\href {\doibase 10.1088/0004-637X/749/1/91} {\bibfield
  {journal} {\bibinfo  {journal} {\apj}\ }\textbf {\bibinfo {volume} {749}},\
  \bibinfo {eid} {91} (\bibinfo {year} {2012})},\ \Eprint
  {http://arxiv.org/abs/1110.1726} {arXiv:1110.1726 [astro-ph.SR]} \BibitemShut
  {NoStop}%
\bibitem [{\citenamefont {{Olejak}}\ \emph {et~al.}(2022)\citenamefont
  {{Olejak}}, \citenamefont {{Fryer}}, \citenamefont {{Belczynski}},\ and\
  \citenamefont {{Baibhav}}}]{Olejak2022MNRAS}%
  \BibitemOpen
  \bibfield  {author} {\bibinfo {author} {\bibfnamefont {A.}~\bibnamefont
  {{Olejak}}}, \bibinfo {author} {\bibfnamefont {C.~L.}\ \bibnamefont
  {{Fryer}}}, \bibinfo {author} {\bibfnamefont {K.}~\bibnamefont
  {{Belczynski}}}, \ and\ \bibinfo {author} {\bibfnamefont {V.}~\bibnamefont
  {{Baibhav}}},\ }\href {\doibase 10.1093/mnras/stac2359} {\bibfield  {journal}
  {\bibinfo  {journal} {Mon. Not. Roy. Astron. Soc.}\ }\textbf {\bibinfo
  {volume} {516}},\ \bibinfo {pages} {2252} (\bibinfo {year} {2022})},\ \Eprint
  {http://arxiv.org/abs/2204.09061} {arXiv:2204.09061 [astro-ph.HE]}
  \BibitemShut {NoStop}%
\bibitem [{\citenamefont {{Muhammed}}\ \emph {et~al.}(2024)\citenamefont
  {{Muhammed}}, \citenamefont {{Duez}}, \citenamefont {{Chawhan}},
  \citenamefont {{Ghadiri}}, \citenamefont {{Buchman}}, \citenamefont
  {{Foucart}}, \citenamefont {{Chi-Kit Cheong}}, \citenamefont {{Kidder}},
  \citenamefont {{Pfeiffer}},\ and\ \citenamefont
  {{Scheel}}}]{Muhammed2024arXiv}%
  \BibitemOpen
  \bibfield  {author} {\bibinfo {author} {\bibfnamefont {N.}~\bibnamefont
  {{Muhammed}}}, \bibinfo {author} {\bibfnamefont {M.~D.}\ \bibnamefont
  {{Duez}}}, \bibinfo {author} {\bibfnamefont {P.}~\bibnamefont {{Chawhan}}},
  \bibinfo {author} {\bibfnamefont {N.}~\bibnamefont {{Ghadiri}}}, \bibinfo
  {author} {\bibfnamefont {L.~T.}\ \bibnamefont {{Buchman}}}, \bibinfo {author}
  {\bibfnamefont {F.}~\bibnamefont {{Foucart}}}, \bibinfo {author}
  {\bibfnamefont {P.}~\bibnamefont {{Chi-Kit Cheong}}}, \bibinfo {author}
  {\bibfnamefont {L.~E.}\ \bibnamefont {{Kidder}}}, \bibinfo {author}
  {\bibfnamefont {H.~P.}\ \bibnamefont {{Pfeiffer}}}, \ and\ \bibinfo {author}
  {\bibfnamefont {M.~A.}\ \bibnamefont {{Scheel}}},\ }\href {\doibase
  10.48550/arXiv.2403.05642} {\bibfield  {journal} {\bibinfo  {journal} {arXiv
  e-prints}\ ,\ \bibinfo {eid} {arXiv:2403.05642}} (\bibinfo {year} {2024})},\
  \Eprint {http://arxiv.org/abs/2403.05642} {arXiv:2403.05642 [gr-qc]}
  \BibitemShut {NoStop}%
\bibitem [{\citenamefont {{Zuraiq}}\ \emph {et~al.}(2024)\citenamefont
  {{Zuraiq}}, \citenamefont {{Mukhopadhyay}},\ and\ \citenamefont
  {{Weber}}}]{Zuraiq2024PhRvD}%
  \BibitemOpen
  \bibfield  {author} {\bibinfo {author} {\bibfnamefont {Z.}~\bibnamefont
  {{Zuraiq}}}, \bibinfo {author} {\bibfnamefont {B.}~\bibnamefont
  {{Mukhopadhyay}}}, \ and\ \bibinfo {author} {\bibfnamefont {F.}~\bibnamefont
  {{Weber}}},\ }\href {\doibase 10.1103/PhysRevD.109.023027} {\bibfield
  {journal} {\bibinfo  {journal} {\prd}\ }\textbf {\bibinfo {volume} {109}},\
  \bibinfo {eid} {023027} (\bibinfo {year} {2024})},\ \Eprint
  {http://arxiv.org/abs/2311.02169} {arXiv:2311.02169 [astro-ph.HE]}
  \BibitemShut {NoStop}%
\bibitem [{\citenamefont {{Belczynski}}\ \emph {et~al.}(2012)\citenamefont
  {{Belczynski}}, \citenamefont {{Wiktorowicz}}, \citenamefont {{Fryer}},
  \citenamefont {{Holz}},\ and\ \citenamefont
  {{Kalogera}}}]{Belczynski2012ApJ}%
  \BibitemOpen
  \bibfield  {author} {\bibinfo {author} {\bibfnamefont {K.}~\bibnamefont
  {{Belczynski}}}, \bibinfo {author} {\bibfnamefont {G.}~\bibnamefont
  {{Wiktorowicz}}}, \bibinfo {author} {\bibfnamefont {C.~L.}\ \bibnamefont
  {{Fryer}}}, \bibinfo {author} {\bibfnamefont {D.~E.}\ \bibnamefont {{Holz}}},
  \ and\ \bibinfo {author} {\bibfnamefont {V.}~\bibnamefont {{Kalogera}}},\
  }\href {\doibase 10.1088/0004-637X/757/1/91} {\bibfield  {journal} {\bibinfo
  {journal} {\apj}\ }\textbf {\bibinfo {volume} {757}},\ \bibinfo {eid} {91}
  (\bibinfo {year} {2012})},\ \Eprint {http://arxiv.org/abs/1110.1635}
  {arXiv:1110.1635 [astro-ph.GA]} \BibitemShut {NoStop}%
\bibitem [{\citenamefont {{Thompson}}\ \emph {et~al.}(2019)\citenamefont
  {{Thompson}} \emph {et~al.}}]{Thompson2019Sci}%
  \BibitemOpen
  \bibfield  {author} {\bibinfo {author} {\bibfnamefont {T.~A.}\ \bibnamefont
  {{Thompson}}} \emph {et~al.},\ }\href {\doibase 10.1126/science.aau4005}
  {\bibfield  {journal} {\bibinfo  {journal} {Science}\ }\textbf {\bibinfo
  {volume} {366}},\ \bibinfo {pages} {637} (\bibinfo {year} {2019})},\ \Eprint
  {http://arxiv.org/abs/1806.02751} {arXiv:1806.02751 [astro-ph.HE]}
  \BibitemShut {NoStop}%
\bibitem [{\citenamefont {van~den Heuvel}\ and\ \citenamefont
  {Tauris}(2020)}]{vandenHeuvel:2020chh}%
  \BibitemOpen
  \bibfield  {author} {\bibinfo {author} {\bibfnamefont {P.~J.}\ \bibnamefont
  {van~den Heuvel}}\ and\ \bibinfo {author} {\bibfnamefont {T.~M.}\
  \bibnamefont {Tauris}},\ }\href {\doibase 10.1126/science.aba3282} {\
  (\bibinfo {year} {2020}),\ 10.1126/science.aba3282},\ \Eprint
  {http://arxiv.org/abs/2005.04896} {arXiv:2005.04896 [astro-ph.SR]}
  \BibitemShut {NoStop}%
\bibitem [{\citenamefont {Thompson}\ \emph {et~al.}(2020)\citenamefont
  {Thompson} \emph {et~al.}}]{Thompson:2020nbd}%
  \BibitemOpen
  \bibfield  {author} {\bibinfo {author} {\bibfnamefont {T.~A.}\ \bibnamefont
  {Thompson}} \emph {et~al.},\ }\href {\doibase 10.1126/science.aba4356} {\
  (\bibinfo {year} {2020}),\ 10.1126/science.aba4356},\ \Eprint
  {http://arxiv.org/abs/2005.07653} {arXiv:2005.07653 [astro-ph.HE]}
  \BibitemShut {NoStop}%
\bibitem [{\citenamefont {Jayasinghe}\ \emph {et~al.}(2021)\citenamefont
  {Jayasinghe} \emph {et~al.}}]{Jayasinghe:2021uqb}%
  \BibitemOpen
  \bibfield  {author} {\bibinfo {author} {\bibfnamefont {T.}~\bibnamefont
  {Jayasinghe}} \emph {et~al.},\ }\href {\doibase 10.1093/mnras/stab907}
  {\bibfield  {journal} {\bibinfo  {journal} {Mon. Not. Roy. Astron. Soc.}\
  }\textbf {\bibinfo {volume} {504}},\ \bibinfo {pages} {2577} (\bibinfo {year}
  {2021})},\ \Eprint {http://arxiv.org/abs/2101.02212} {arXiv:2101.02212
  [astro-ph.SR]} \BibitemShut {NoStop}%
\bibitem [{\citenamefont {{El-Badry}}\ \emph {et~al.}(2022)\citenamefont
  {{El-Badry}}, \citenamefont {{Seeburger}}, \citenamefont {{Jayasinghe}},
  \citenamefont {{Rix}}, \citenamefont {{Almada}}, \citenamefont {{Conroy}},
  \citenamefont {{Price-Whelan}},\ and\ \citenamefont
  {{Burdge}}}]{Kareem2022MNRAS}%
  \BibitemOpen
  \bibfield  {author} {\bibinfo {author} {\bibfnamefont {K.}~\bibnamefont
  {{El-Badry}}}, \bibinfo {author} {\bibfnamefont {R.}~\bibnamefont
  {{Seeburger}}}, \bibinfo {author} {\bibfnamefont {T.}~\bibnamefont
  {{Jayasinghe}}}, \bibinfo {author} {\bibfnamefont {H.-W.}\ \bibnamefont
  {{Rix}}}, \bibinfo {author} {\bibfnamefont {S.}~\bibnamefont {{Almada}}},
  \bibinfo {author} {\bibfnamefont {C.}~\bibnamefont {{Conroy}}}, \bibinfo
  {author} {\bibfnamefont {A.~M.}\ \bibnamefont {{Price-Whelan}}}, \ and\
  \bibinfo {author} {\bibfnamefont {K.}~\bibnamefont {{Burdge}}},\ }\href
  {\doibase 10.1093/mnras/stac815} {\bibfield  {journal} {\bibinfo  {journal}
  {Mon. Not. Roy. Astron. Soc.}\ }\textbf {\bibinfo {volume} {512}},\ \bibinfo
  {pages} {5620} (\bibinfo {year} {2022})},\ \Eprint
  {http://arxiv.org/abs/2203.06348} {arXiv:2203.06348 [astro-ph.SR]}
  \BibitemShut {NoStop}%
\bibitem [{\citenamefont {Barr}\ \emph {et~al.}(2024)\citenamefont {Barr} \emph
  {et~al.}}]{Barr:2024wwl}%
  \BibitemOpen
  \bibfield  {author} {\bibinfo {author} {\bibfnamefont {E.~D.}\ \bibnamefont
  {Barr}} \emph {et~al.},\ }\href {\doibase 10.1126/science.adg3005} {\bibfield
   {journal} {\bibinfo  {journal} {Science}\ }\textbf {\bibinfo {volume}
  {383}},\ \bibinfo {pages} {275} (\bibinfo {year} {2024})},\ \Eprint
  {http://arxiv.org/abs/2401.09872} {arXiv:2401.09872 [astro-ph.HE]}
  \BibitemShut {NoStop}%
\bibitem [{\citenamefont {Song}\ \emph {et~al.}(2024)\citenamefont {Song} \emph
  {et~al.}}]{Song:2024tqr}%
  \BibitemOpen
  \bibfield  {author} {\bibinfo {author} {\bibfnamefont {W.}~\bibnamefont
  {Song}} \emph {et~al.},\ }\href {\doibase 10.1038/s41550-024-02359-9}
  {\bibfield  {journal} {\bibinfo  {journal} {Nature Astron.}\ }\textbf
  {\bibinfo {volume} {8}},\ \bibinfo {pages} {1583} (\bibinfo {year} {2024})},\
  \Eprint {http://arxiv.org/abs/2409.06352} {arXiv:2409.06352 [astro-ph.SR]}
  \BibitemShut {NoStop}%
\bibitem [{\citenamefont {Abbott}\ \emph
  {et~al.}(2020{\natexlab{a}})\citenamefont {Abbott} \emph
  {et~al.}}]{GW190814}%
  \BibitemOpen
  \bibfield  {author} {\bibinfo {author} {\bibfnamefont {R.}~\bibnamefont
  {Abbott}} \emph {et~al.} (\bibinfo {collaboration} {LIGO Scientific,
  Virgo}),\ }\href {\doibase 10.3847/2041-8213/ab960f} {\bibfield  {journal}
  {\bibinfo  {journal} {Astrophys. J. Lett.}\ }\textbf {\bibinfo {volume}
  {896}},\ \bibinfo {pages} {L44} (\bibinfo {year} {2020}{\natexlab{a}})},\
  \Eprint {http://arxiv.org/abs/2006.12611} {arXiv:2006.12611 [astro-ph.HE]}
  \BibitemShut {NoStop}%
\bibitem [{\citenamefont {Abbott}\ \emph
  {et~al.}(2021{\natexlab{a}})\citenamefont {Abbott} \emph {et~al.}}]{GWTC-3}%
  \BibitemOpen
  \bibfield  {author} {\bibinfo {author} {\bibfnamefont {R.}~\bibnamefont
  {Abbott}} \emph {et~al.},\ }\href {\doibase 10.48550/arXiv.2111.03606}
  {\bibfield  {journal} {\bibinfo  {journal} {arXiv e-prints}\ ,\ \bibinfo
  {eid} {arXiv:2111.03606}} (\bibinfo {year} {2021}{\natexlab{a}})},\ \Eprint
  {http://arxiv.org/abs/2111.03606} {arXiv:2111.03606 [gr-qc]} \BibitemShut
  {NoStop}%
\bibitem [{\citenamefont {{El-Badry}}\ \emph {et~al.}(2024)\citenamefont
  {{El-Badry}}, \citenamefont {{Rix}}, \citenamefont {{Latham}}, \citenamefont
  {{Shahaf}}, \citenamefont {{Mazeh}}, \citenamefont {{Bieryla}}, \citenamefont
  {{Buchhave}}, \citenamefont {{Andrae}}, \citenamefont {{Yamaguchi}},
  \citenamefont {{Isaacson}}, \citenamefont {{Howard}}, \citenamefont
  {{Savino}},\ and\ \citenamefont {{Ilyin}}}]{Kareem2024OJAp}%
  \BibitemOpen
  \bibfield  {author} {\bibinfo {author} {\bibfnamefont {K.}~\bibnamefont
  {{El-Badry}}}, \bibinfo {author} {\bibfnamefont {H.-W.}\ \bibnamefont
  {{Rix}}}, \bibinfo {author} {\bibfnamefont {D.~W.}\ \bibnamefont {{Latham}}},
  \bibinfo {author} {\bibfnamefont {S.}~\bibnamefont {{Shahaf}}}, \bibinfo
  {author} {\bibfnamefont {T.}~\bibnamefont {{Mazeh}}}, \bibinfo {author}
  {\bibfnamefont {A.}~\bibnamefont {{Bieryla}}}, \bibinfo {author}
  {\bibfnamefont {L.~A.}\ \bibnamefont {{Buchhave}}}, \bibinfo {author}
  {\bibfnamefont {R.}~\bibnamefont {{Andrae}}}, \bibinfo {author}
  {\bibfnamefont {N.}~\bibnamefont {{Yamaguchi}}}, \bibinfo {author}
  {\bibfnamefont {H.}~\bibnamefont {{Isaacson}}}, \bibinfo {author}
  {\bibfnamefont {A.~W.}\ \bibnamefont {{Howard}}}, \bibinfo {author}
  {\bibfnamefont {A.}~\bibnamefont {{Savino}}}, \ and\ \bibinfo {author}
  {\bibfnamefont {I.~V.}\ \bibnamefont {{Ilyin}}},\ }\href {\doibase
  10.33232/001c.121261} {\bibfield  {journal} {\bibinfo  {journal} {The Open
  Journal of Astrophysics}\ }\textbf {\bibinfo {volume} {7}},\ \bibinfo {eid}
  {58} (\bibinfo {year} {2024})},\ \Eprint {http://arxiv.org/abs/2405.00089}
  {arXiv:2405.00089 [astro-ph.SR]} \BibitemShut {NoStop}%
\bibitem [{\citenamefont {{Wyrzykowski}}\ and\ \citenamefont
  {{Mandel}}(2020)}]{Wyrzykowski2020A&A}%
  \BibitemOpen
  \bibfield  {author} {\bibinfo {author} {\bibfnamefont {{\L}.}~\bibnamefont
  {{Wyrzykowski}}}\ and\ \bibinfo {author} {\bibfnamefont {I.}~\bibnamefont
  {{Mandel}}},\ }\href {\doibase 10.1051/0004-6361/201935842} {\bibfield
  {journal} {\bibinfo  {journal} {Astronomy \& Astrophysics}\ }\textbf
  {\bibinfo {volume} {636}},\ \bibinfo {eid} {A20} (\bibinfo {year} {2020})},\
  \Eprint {http://arxiv.org/abs/1904.07789} {arXiv:1904.07789 [astro-ph.SR]}
  \BibitemShut {NoStop}%
\bibitem [{\citenamefont {Wyrzykowski}\ \emph {et~al.}(2016)\citenamefont
  {Wyrzykowski} \emph {et~al.}}]{Wyrzykowski:2015ppa}%
  \BibitemOpen
  \bibfield  {author} {\bibinfo {author} {\bibfnamefont {L.}~\bibnamefont
  {Wyrzykowski}} \emph {et~al.},\ }\href {\doibase 10.1093/mnras/stw426}
  {\bibfield  {journal} {\bibinfo  {journal} {Mon. Not. Roy. Astron. Soc.}\
  }\textbf {\bibinfo {volume} {458}},\ \bibinfo {pages} {3012} (\bibinfo {year}
  {2016})},\ \Eprint {http://arxiv.org/abs/1509.04899} {arXiv:1509.04899
  [astro-ph.SR]} \BibitemShut {NoStop}%
\bibitem [{\citenamefont {Abac}\ \emph {et~al.}(2024)\citenamefont {Abac} \emph
  {et~al.}}]{GW230529}%
  \BibitemOpen
  \bibfield  {author} {\bibinfo {author} {\bibfnamefont {A.~G.}\ \bibnamefont
  {Abac}} \emph {et~al.} (\bibinfo {collaboration} {LIGO Scientific, Virgo,,
  KAGRA, VIRGO}),\ }\href {\doibase 10.3847/2041-8213/ad5beb} {\bibfield
  {journal} {\bibinfo  {journal} {Astrophys. J. Lett.}\ }\textbf {\bibinfo
  {volume} {970}},\ \bibinfo {pages} {L34} (\bibinfo {year} {2024})},\ \Eprint
  {http://arxiv.org/abs/2404.04248} {arXiv:2404.04248 [astro-ph.HE]}
  \BibitemShut {NoStop}%
\bibitem [{\citenamefont {Koehn}\ \emph {et~al.}(2024)\citenamefont {Koehn},
  \citenamefont {Wouters}, \citenamefont {Rose}, \citenamefont {Pang},
  \citenamefont {Somasundaram}, \citenamefont {Tews},\ and\ \citenamefont
  {Dietrich}}]{Koehn:2024ape}%
  \BibitemOpen
  \bibfield  {author} {\bibinfo {author} {\bibfnamefont {H.}~\bibnamefont
  {Koehn}}, \bibinfo {author} {\bibfnamefont {T.}~\bibnamefont {Wouters}},
  \bibinfo {author} {\bibfnamefont {H.}~\bibnamefont {Rose}}, \bibinfo {author}
  {\bibfnamefont {P.~T.~H.}\ \bibnamefont {Pang}}, \bibinfo {author}
  {\bibfnamefont {R.}~\bibnamefont {Somasundaram}}, \bibinfo {author}
  {\bibfnamefont {I.}~\bibnamefont {Tews}}, \ and\ \bibinfo {author}
  {\bibfnamefont {T.}~\bibnamefont {Dietrich}},\ }\href {\doibase
  10.1103/PhysRevD.110.103015} {\bibfield  {journal} {\bibinfo  {journal}
  {Phys. Rev. D}\ }\textbf {\bibinfo {volume} {110}},\ \bibinfo {pages}
  {103015} (\bibinfo {year} {2024})},\ \Eprint
  {http://arxiv.org/abs/2407.07837} {arXiv:2407.07837 [astro-ph.HE]}
  \BibitemShut {NoStop}%
\bibitem [{\citenamefont {Mahapatra}\ \emph {et~al.}(2024)\citenamefont
  {Mahapatra}, \citenamefont {Chattopadhyay}, \citenamefont {Gupta},
  \citenamefont {Antonini}, \citenamefont {Favata}, \citenamefont
  {Sathyaprakash},\ and\ \citenamefont {Arun}}]{Mahapatra:2024qsy}%
  \BibitemOpen
  \bibfield  {author} {\bibinfo {author} {\bibfnamefont {P.}~\bibnamefont
  {Mahapatra}}, \bibinfo {author} {\bibfnamefont {D.}~\bibnamefont
  {Chattopadhyay}}, \bibinfo {author} {\bibfnamefont {A.}~\bibnamefont
  {Gupta}}, \bibinfo {author} {\bibfnamefont {F.}~\bibnamefont {Antonini}},
  \bibinfo {author} {\bibfnamefont {M.}~\bibnamefont {Favata}}, \bibinfo
  {author} {\bibfnamefont {B.~S.}\ \bibnamefont {Sathyaprakash}}, \ and\
  \bibinfo {author} {\bibfnamefont {K.~G.}\ \bibnamefont {Arun}},\ }\href
  {\doibase 10.3847/1538-4357/ad781b} {\bibfield  {journal} {\bibinfo
  {journal} {Astrophys. J.}\ }\textbf {\bibinfo {volume} {975}},\ \bibinfo
  {pages} {117} (\bibinfo {year} {2024})},\ \Eprint
  {http://arxiv.org/abs/2406.06390} {arXiv:2406.06390 [astro-ph.HE]}
  \BibitemShut {NoStop}%
\bibitem [{\citenamefont {Fryer}\ and\ \citenamefont
  {Kalogera}(2001)}]{Fryer:1999ht}%
  \BibitemOpen
  \bibfield  {author} {\bibinfo {author} {\bibfnamefont {C.~L.}\ \bibnamefont
  {Fryer}}\ and\ \bibinfo {author} {\bibfnamefont {V.}~\bibnamefont
  {Kalogera}},\ }\href {\doibase 10.1086/321359} {\bibfield  {journal}
  {\bibinfo  {journal} {Astrophys. J.}\ }\textbf {\bibinfo {volume} {554}},\
  \bibinfo {pages} {548} (\bibinfo {year} {2001})},\ \Eprint
  {http://arxiv.org/abs/astro-ph/9911312} {arXiv:astro-ph/9911312} \BibitemShut
  {NoStop}%
\bibitem [{\citenamefont {Kushnir}(2015)}]{Kushnir:2015mca}%
  \BibitemOpen
  \bibfield  {author} {\bibinfo {author} {\bibfnamefont {D.}~\bibnamefont
  {Kushnir}},\ }\href@noop {} {\  (\bibinfo {year} {2015})},\ \Eprint
  {http://arxiv.org/abs/1502.03111} {arXiv:1502.03111 [astro-ph.HE]}
  \BibitemShut {NoStop}%
\bibitem [{\citenamefont {Chattopadhyay}\ \emph {et~al.}(2022)\citenamefont
  {Chattopadhyay}, \citenamefont {Stevenson}, \citenamefont {Broekgaarden},
  \citenamefont {Antonini},\ and\ \citenamefont
  {Belczynski}}]{Chattopadhyay:2022cnp}%
  \BibitemOpen
  \bibfield  {author} {\bibinfo {author} {\bibfnamefont {D.}~\bibnamefont
  {Chattopadhyay}}, \bibinfo {author} {\bibfnamefont {S.}~\bibnamefont
  {Stevenson}}, \bibinfo {author} {\bibfnamefont {F.}~\bibnamefont
  {Broekgaarden}}, \bibinfo {author} {\bibfnamefont {F.}~\bibnamefont
  {Antonini}}, \ and\ \bibinfo {author} {\bibfnamefont {K.}~\bibnamefont
  {Belczynski}},\ }\href {\doibase 10.1093/mnras/stac1283} {\bibfield
  {journal} {\bibinfo  {journal} {Mon. Not. Roy. Astron. Soc.}\ }\textbf
  {\bibinfo {volume} {513}},\ \bibinfo {pages} {5780} (\bibinfo {year}
  {2022})},\ \Eprint {http://arxiv.org/abs/2203.05850} {arXiv:2203.05850
  [astro-ph.HE]} \BibitemShut {NoStop}%
\bibitem [{\citenamefont {Zhu}\ \emph {et~al.}(2024{\natexlab{a}})\citenamefont
  {Zhu}, \citenamefont {Hu}, \citenamefont {Kang}, \citenamefont {Zhang},
  \citenamefont {Tong}, \citenamefont {Shao},\ and\ \citenamefont
  {Qin}}]{Zhu:2024cvt}%
  \BibitemOpen
  \bibfield  {author} {\bibinfo {author} {\bibfnamefont {J.-P.}\ \bibnamefont
  {Zhu}}, \bibinfo {author} {\bibfnamefont {R.-C.}\ \bibnamefont {Hu}},
  \bibinfo {author} {\bibfnamefont {Y.}~\bibnamefont {Kang}}, \bibinfo {author}
  {\bibfnamefont {B.}~\bibnamefont {Zhang}}, \bibinfo {author} {\bibfnamefont
  {H.}~\bibnamefont {Tong}}, \bibinfo {author} {\bibfnamefont {L.}~\bibnamefont
  {Shao}}, \ and\ \bibinfo {author} {\bibfnamefont {Y.}~\bibnamefont {Qin}},\
  }\href {\doibase 10.3847/1538-4357/ad72f0} {\bibfield  {journal} {\bibinfo
  {journal} {Astrophys. J.}\ }\textbf {\bibinfo {volume} {974}},\ \bibinfo
  {pages} {211} (\bibinfo {year} {2024}{\natexlab{a}})},\ \Eprint
  {http://arxiv.org/abs/2404.10596} {arXiv:2404.10596 [astro-ph.HE]}
  \BibitemShut {NoStop}%
\bibitem [{\citenamefont {Xing}\ \emph {et~al.}(2024)\citenamefont {Xing} \emph
  {et~al.}}]{Xing:2024ydg}%
  \BibitemOpen
  \bibfield  {author} {\bibinfo {author} {\bibfnamefont {Z.}~\bibnamefont
  {Xing}} \emph {et~al.},\ }\href@noop {} {\  (\bibinfo {year} {2024})},\
  \Eprint {http://arxiv.org/abs/2410.20415} {arXiv:2410.20415 [astro-ph.HE]}
  \BibitemShut {NoStop}%
\bibitem [{\citenamefont {Qin}\ \emph {et~al.}(2024)\citenamefont {Qin} \emph
  {et~al.}}]{Qin:2024ojw}%
  \BibitemOpen
  \bibfield  {author} {\bibinfo {author} {\bibfnamefont {Y.}~\bibnamefont
  {Qin}} \emph {et~al.},\ }\href {\doibase 10.1051/0004-6361/202452335}
  {\bibfield  {journal} {\bibinfo  {journal} {Astron. Astrophys.}\ }\textbf
  {\bibinfo {volume} {691}},\ \bibinfo {pages} {L19} (\bibinfo {year}
  {2024})},\ \Eprint {http://arxiv.org/abs/2409.14476} {arXiv:2409.14476
  [astro-ph.HE]} \BibitemShut {NoStop}%
\bibitem [{\citenamefont {Chattopadhyay}\ \emph {et~al.}(2024)\citenamefont
  {Chattopadhyay}, \citenamefont {Al-Shammari}, \citenamefont {Antonini},
  \citenamefont {Fairhurst}, \citenamefont {Miles},\ and\ \citenamefont
  {Raymond}}]{Chattopadhyay:2024hsf}%
  \BibitemOpen
  \bibfield  {author} {\bibinfo {author} {\bibfnamefont {D.}~\bibnamefont
  {Chattopadhyay}}, \bibinfo {author} {\bibfnamefont {S.}~\bibnamefont
  {Al-Shammari}}, \bibinfo {author} {\bibfnamefont {F.}~\bibnamefont
  {Antonini}}, \bibinfo {author} {\bibfnamefont {S.}~\bibnamefont {Fairhurst}},
  \bibinfo {author} {\bibfnamefont {B.}~\bibnamefont {Miles}}, \ and\ \bibinfo
  {author} {\bibfnamefont {V.}~\bibnamefont {Raymond}},\ }\href {\doibase
  10.1093/mnrasl/slae099} {\bibfield  {journal} {\bibinfo  {journal} {Monthly
  Notices of the Royal Astronomical Society: Letters}\ }\textbf {\bibinfo
  {volume} {536}},\ \bibinfo {pages} {L19} (\bibinfo {year} {2024})},\ \Eprint
  {http://arxiv.org/abs/https://academic.oup.com/mnrasl/article-pdf/536/1/L19/60671977/slae099.pdf}
  {https://academic.oup.com/mnrasl/article-pdf/536/1/L19/60671977/slae099.pdf}
  \BibitemShut {NoStop}%
\bibitem [{\citenamefont {Chandra}\ \emph {et~al.}(2024)\citenamefont
  {Chandra}, \citenamefont {Gupta}, \citenamefont {Gamba}, \citenamefont
  {Kashyap}, \citenamefont {Chattopadhyay}, \citenamefont {Gonzalez},
  \citenamefont {Bernuzzi},\ and\ \citenamefont
  {Sathyaprakash}}]{Chandra:2024ila}%
  \BibitemOpen
  \bibfield  {author} {\bibinfo {author} {\bibfnamefont {K.}~\bibnamefont
  {Chandra}}, \bibinfo {author} {\bibfnamefont {I.}~\bibnamefont {Gupta}},
  \bibinfo {author} {\bibfnamefont {R.}~\bibnamefont {Gamba}}, \bibinfo
  {author} {\bibfnamefont {R.}~\bibnamefont {Kashyap}}, \bibinfo {author}
  {\bibfnamefont {D.}~\bibnamefont {Chattopadhyay}}, \bibinfo {author}
  {\bibfnamefont {A.}~\bibnamefont {Gonzalez}}, \bibinfo {author}
  {\bibfnamefont {S.}~\bibnamefont {Bernuzzi}}, \ and\ \bibinfo {author}
  {\bibfnamefont {B.~S.}\ \bibnamefont {Sathyaprakash}},\ }\href {\doibase
  10.3847/1538-4357/ad90bd} {\bibfield  {journal} {\bibinfo  {journal}
  {Astrophys. J.}\ }\textbf {\bibinfo {volume} {977}},\ \bibinfo {pages} {167}
  (\bibinfo {year} {2024})},\ \Eprint {http://arxiv.org/abs/2405.03841}
  {arXiv:2405.03841 [astro-ph.HE]} \BibitemShut {NoStop}%
\bibitem [{\citenamefont {O'Connor}\ and\ \citenamefont
  {Ott}(2011)}]{OConnor:2010moj}%
  \BibitemOpen
  \bibfield  {author} {\bibinfo {author} {\bibfnamefont {E.}~\bibnamefont
  {O'Connor}}\ and\ \bibinfo {author} {\bibfnamefont {C.~D.}\ \bibnamefont
  {Ott}},\ }\href {\doibase 10.1088/0004-637X/730/2/70} {\bibfield  {journal}
  {\bibinfo  {journal} {Astrophys. J.}\ }\textbf {\bibinfo {volume} {730}},\
  \bibinfo {pages} {70} (\bibinfo {year} {2011})},\ \Eprint
  {http://arxiv.org/abs/1010.5550} {arXiv:1010.5550 [astro-ph.HE]} \BibitemShut
  {NoStop}%
\bibitem [{\citenamefont {Janka}(2012)}]{Janka:2012wk}%
  \BibitemOpen
  \bibfield  {author} {\bibinfo {author} {\bibfnamefont {H.-T.}\ \bibnamefont
  {Janka}},\ }\href {\doibase 10.1146/annurev-nucl-102711-094901} {\bibfield
  {journal} {\bibinfo  {journal} {Ann. Rev. Nucl. Part. Sci.}\ }\textbf
  {\bibinfo {volume} {62}},\ \bibinfo {pages} {407} (\bibinfo {year} {2012})},\
  \Eprint {http://arxiv.org/abs/1206.2503} {arXiv:1206.2503 [astro-ph.SR]}
  \BibitemShut {NoStop}%
\bibitem [{\citenamefont {Sukhbold}\ \emph {et~al.}(2016)\citenamefont
  {Sukhbold}, \citenamefont {Ertl}, \citenamefont {Woosley}, \citenamefont
  {Brown},\ and\ \citenamefont {Janka}}]{Sukhbold:2015wba}%
  \BibitemOpen
  \bibfield  {author} {\bibinfo {author} {\bibfnamefont {T.}~\bibnamefont
  {Sukhbold}}, \bibinfo {author} {\bibfnamefont {T.}~\bibnamefont {Ertl}},
  \bibinfo {author} {\bibfnamefont {S.~E.}\ \bibnamefont {Woosley}}, \bibinfo
  {author} {\bibfnamefont {J.~M.}\ \bibnamefont {Brown}}, \ and\ \bibinfo
  {author} {\bibfnamefont {H.~T.}\ \bibnamefont {Janka}},\ }\href {\doibase
  10.3847/0004-637X/821/1/38} {\bibfield  {journal} {\bibinfo  {journal}
  {Astrophys. J.}\ }\textbf {\bibinfo {volume} {821}},\ \bibinfo {pages} {38}
  (\bibinfo {year} {2016})},\ \Eprint {http://arxiv.org/abs/1510.04643}
  {arXiv:1510.04643 [astro-ph.HE]} \BibitemShut {NoStop}%
\bibitem [{\citenamefont {M\"uller}\ \emph {et~al.}(2016)\citenamefont
  {M\"uller}, \citenamefont {Heger}, \citenamefont {Liptai},\ and\
  \citenamefont {Cameron}}]{Muller:2016ujh}%
  \BibitemOpen
  \bibfield  {author} {\bibinfo {author} {\bibfnamefont {B.}~\bibnamefont
  {M\"uller}}, \bibinfo {author} {\bibfnamefont {A.}~\bibnamefont {Heger}},
  \bibinfo {author} {\bibfnamefont {D.}~\bibnamefont {Liptai}}, \ and\ \bibinfo
  {author} {\bibfnamefont {J.~B.}\ \bibnamefont {Cameron}},\ }\href {\doibase
  10.1093/mnras/stw1083} {\bibfield  {journal} {\bibinfo  {journal} {Mon. Not.
  Roy. Astron. Soc.}\ }\textbf {\bibinfo {volume} {460}},\ \bibinfo {pages}
  {742} (\bibinfo {year} {2016})},\ \Eprint {http://arxiv.org/abs/1602.05956}
  {arXiv:1602.05956 [astro-ph.SR]} \BibitemShut {NoStop}%
\bibitem [{\citenamefont {Couch}\ \emph {et~al.}(2019)\citenamefont {Couch},
  \citenamefont {Warren},\ and\ \citenamefont {O'Connor}}]{Couch:2019mrd}%
  \BibitemOpen
  \bibfield  {author} {\bibinfo {author} {\bibfnamefont {S.~M.}\ \bibnamefont
  {Couch}}, \bibinfo {author} {\bibfnamefont {M.~L.}\ \bibnamefont {Warren}}, \
  and\ \bibinfo {author} {\bibfnamefont {E.~P.}\ \bibnamefont {O'Connor}},\
  }\href {\doibase 10.3847/1538-4357/ab609e} {\  (\bibinfo {year} {2019}),\
  10.3847/1538-4357/ab609e},\ \Eprint {http://arxiv.org/abs/1902.01340}
  {arXiv:1902.01340 [astro-ph.HE]} \BibitemShut {NoStop}%
\bibitem [{\citenamefont {Ertl}\ \emph {et~al.}(2019)\citenamefont {Ertl},
  \citenamefont {Woosley}, \citenamefont {Sukhbold},\ and\ \citenamefont
  {Janka}}]{Ertl:2019zks}%
  \BibitemOpen
  \bibfield  {author} {\bibinfo {author} {\bibfnamefont {T.}~\bibnamefont
  {Ertl}}, \bibinfo {author} {\bibfnamefont {S.~E.}\ \bibnamefont {Woosley}},
  \bibinfo {author} {\bibfnamefont {T.}~\bibnamefont {Sukhbold}}, \ and\
  \bibinfo {author} {\bibfnamefont {H.~T.}\ \bibnamefont {Janka}},\ }\href
  {\doibase 10.3847/1538-4357/ab6458} {\  (\bibinfo {year} {2019}),\
  10.3847/1538-4357/ab6458},\ \Eprint {http://arxiv.org/abs/1910.01641}
  {arXiv:1910.01641 [astro-ph.HE]} \BibitemShut {NoStop}%
\bibitem [{\citenamefont {Siegel}\ \emph {et~al.}(2023)\citenamefont {Siegel}
  \emph {et~al.}}]{Siegel:2022gwc}%
  \BibitemOpen
  \bibfield  {author} {\bibinfo {author} {\bibfnamefont {J.~C.}\ \bibnamefont
  {Siegel}} \emph {et~al.},\ }\href {\doibase 10.3847/1538-4357/ace9d9}
  {\bibfield  {journal} {\bibinfo  {journal} {Astrophys. J.}\ }\textbf
  {\bibinfo {volume} {954}},\ \bibinfo {pages} {212} (\bibinfo {year}
  {2023})},\ \Eprint {http://arxiv.org/abs/2209.06844} {arXiv:2209.06844
  [astro-ph.HE]} \BibitemShut {NoStop}%
\bibitem [{\citenamefont {Zhu}\ \emph {et~al.}(2024{\natexlab{b}})\citenamefont
  {Zhu}, \citenamefont {Qin}, \citenamefont {Wang}, \citenamefont {Hu},
  \citenamefont {Zhang},\ and\ \citenamefont {Wu}}]{Zhu:2023nhy}%
  \BibitemOpen
  \bibfield  {author} {\bibinfo {author} {\bibfnamefont {J.-P.}\ \bibnamefont
  {Zhu}}, \bibinfo {author} {\bibfnamefont {Y.}~\bibnamefont {Qin}}, \bibinfo
  {author} {\bibfnamefont {Z.-H.-T.}\ \bibnamefont {Wang}}, \bibinfo {author}
  {\bibfnamefont {R.-C.}\ \bibnamefont {Hu}}, \bibinfo {author} {\bibfnamefont
  {B.}~\bibnamefont {Zhang}}, \ and\ \bibinfo {author} {\bibfnamefont
  {S.}~\bibnamefont {Wu}},\ }\href {\doibase 10.1093/mnras/stae815} {\bibfield
  {journal} {\bibinfo  {journal} {Mon. Not. Roy. Astron. Soc.}\ }\textbf
  {\bibinfo {volume} {529}},\ \bibinfo {pages} {4554} (\bibinfo {year}
  {2024}{\natexlab{b}})},\ \Eprint {http://arxiv.org/abs/2310.14256}
  {arXiv:2310.14256 [astro-ph.HE]} \BibitemShut {NoStop}%
\bibitem [{\citenamefont {Gupta}\ \emph {et~al.}(2020)\citenamefont {Gupta},
  \citenamefont {Gerosa}, \citenamefont {Arun}, \citenamefont {Berti},
  \citenamefont {Farr},\ and\ \citenamefont {Sathyaprakash}}]{Gupta:2019nwj}%
  \BibitemOpen
  \bibfield  {author} {\bibinfo {author} {\bibfnamefont {A.}~\bibnamefont
  {Gupta}}, \bibinfo {author} {\bibfnamefont {D.}~\bibnamefont {Gerosa}},
  \bibinfo {author} {\bibfnamefont {K.~G.}\ \bibnamefont {Arun}}, \bibinfo
  {author} {\bibfnamefont {E.}~\bibnamefont {Berti}}, \bibinfo {author}
  {\bibfnamefont {W.~M.}\ \bibnamefont {Farr}}, \ and\ \bibinfo {author}
  {\bibfnamefont {B.~S.}\ \bibnamefont {Sathyaprakash}},\ }\href {\doibase
  10.1103/PhysRevD.101.103036} {\bibfield  {journal} {\bibinfo  {journal}
  {Phys. Rev. D}\ }\textbf {\bibinfo {volume} {101}},\ \bibinfo {pages}
  {103036} (\bibinfo {year} {2020})},\ \Eprint
  {http://arxiv.org/abs/1909.05804} {arXiv:1909.05804 [gr-qc]} \BibitemShut
  {NoStop}%
\bibitem [{\citenamefont {Ye}\ \emph {et~al.}(2020)\citenamefont {Ye},
  \citenamefont {Fong}, \citenamefont {Kremer}, \citenamefont {Rodriguez},
  \citenamefont {Chatterjee}, \citenamefont {Fragione},\ and\ \citenamefont
  {Rasio}}]{Ye:2019xvf}%
  \BibitemOpen
  \bibfield  {author} {\bibinfo {author} {\bibfnamefont {C.~S.}\ \bibnamefont
  {Ye}}, \bibinfo {author} {\bibfnamefont {W.-f.}\ \bibnamefont {Fong}},
  \bibinfo {author} {\bibfnamefont {K.}~\bibnamefont {Kremer}}, \bibinfo
  {author} {\bibfnamefont {C.~L.}\ \bibnamefont {Rodriguez}}, \bibinfo {author}
  {\bibfnamefont {S.}~\bibnamefont {Chatterjee}}, \bibinfo {author}
  {\bibfnamefont {G.}~\bibnamefont {Fragione}}, \ and\ \bibinfo {author}
  {\bibfnamefont {F.~A.}\ \bibnamefont {Rasio}},\ }\href {\doibase
  10.3847/2041-8213/ab5dc5} {\bibfield  {journal} {\bibinfo  {journal}
  {Astrophys. J. Lett.}\ }\textbf {\bibinfo {volume} {888}},\ \bibinfo {pages}
  {L10} (\bibinfo {year} {2020})},\ \Eprint {http://arxiv.org/abs/1910.10740}
  {arXiv:1910.10740 [astro-ph.HE]} \BibitemShut {NoStop}%
\bibitem [{\citenamefont {{Phinney}}(1991)}]{Phinney1991ApJL}%
  \BibitemOpen
  \bibfield  {author} {\bibinfo {author} {\bibfnamefont {E.~S.}\ \bibnamefont
  {{Phinney}}},\ }\href {\doibase 10.1086/186163} {\bibfield  {journal}
  {\bibinfo  {journal} {"Astrophys. J. Lett."}\ }\textbf {\bibinfo {volume}
  {380}},\ \bibinfo {pages} {L17} (\bibinfo {year} {1991})}\BibitemShut
  {NoStop}%
\bibitem [{\citenamefont {Bae}\ \emph {et~al.}(2014)\citenamefont {Bae},
  \citenamefont {Kim},\ and\ \citenamefont {Lee}}]{Bae:2013fna}%
  \BibitemOpen
  \bibfield  {author} {\bibinfo {author} {\bibfnamefont {Y.-B.}\ \bibnamefont
  {Bae}}, \bibinfo {author} {\bibfnamefont {C.}~\bibnamefont {Kim}}, \ and\
  \bibinfo {author} {\bibfnamefont {H.~M.}\ \bibnamefont {Lee}},\ }\href
  {\doibase 10.1093/mnras/stu381} {\bibfield  {journal} {\bibinfo  {journal}
  {Mon. Not. Roy. Astron. Soc.}\ }\textbf {\bibinfo {volume} {440}},\ \bibinfo
  {pages} {2714} (\bibinfo {year} {2014})},\ \Eprint
  {http://arxiv.org/abs/1308.1641} {arXiv:1308.1641 [astro-ph.HE]} \BibitemShut
  {NoStop}%
\bibitem [{\citenamefont {{Clausen}}\ \emph {et~al.}(2013)\citenamefont
  {{Clausen}}, \citenamefont {{Sigurdsson}},\ and\ \citenamefont
  {{Chernoff}}}]{Clausen2013MNRAS}%
  \BibitemOpen
  \bibfield  {author} {\bibinfo {author} {\bibfnamefont {D.}~\bibnamefont
  {{Clausen}}}, \bibinfo {author} {\bibfnamefont {S.}~\bibnamefont
  {{Sigurdsson}}}, \ and\ \bibinfo {author} {\bibfnamefont {D.~F.}\
  \bibnamefont {{Chernoff}}},\ }\href {\doibase 10.1093/mnras/sts295}
  {\bibfield  {journal} {\bibinfo  {journal} {"Mon. Not. Roy. Astron. Soc."}\
  }\textbf {\bibinfo {volume} {428}},\ \bibinfo {pages} {3618} (\bibinfo {year}
  {2013})},\ \Eprint {http://arxiv.org/abs/1210.8153} {arXiv:1210.8153
  [astro-ph.HE]} \BibitemShut {NoStop}%
\bibitem [{\citenamefont {Belczynski}\ \emph {et~al.}(2018)\citenamefont
  {Belczynski} \emph {et~al.}}]{Belczynski:2017mqx}%
  \BibitemOpen
  \bibfield  {author} {\bibinfo {author} {\bibfnamefont {K.}~\bibnamefont
  {Belczynski}} \emph {et~al.},\ }\href {\doibase 10.1051/0004-6361/201732428}
  {\bibfield  {journal} {\bibinfo  {journal} {Astron. Astrophys.}\ }\textbf
  {\bibinfo {volume} {615}},\ \bibinfo {pages} {A91} (\bibinfo {year}
  {2018})},\ \Eprint {http://arxiv.org/abs/1712.00632} {arXiv:1712.00632
  [astro-ph.HE]} \BibitemShut {NoStop}%
\bibitem [{\citenamefont {Sigurdsson}\ and\ \citenamefont
  {Phinney}(1995)}]{Sigurdsson:1994ju}%
  \BibitemOpen
  \bibfield  {author} {\bibinfo {author} {\bibfnamefont {S.}~\bibnamefont
  {Sigurdsson}}\ and\ \bibinfo {author} {\bibfnamefont {E.~S.}\ \bibnamefont
  {Phinney}},\ }\href {\doibase 10.1086/192199} {\bibfield  {journal} {\bibinfo
   {journal} {Astrophys. J. Suppl.}\ }\textbf {\bibinfo {volume} {99}},\
  \bibinfo {pages} {609} (\bibinfo {year} {1995})},\ \Eprint
  {http://arxiv.org/abs/astro-ph/9412078} {arXiv:astro-ph/9412078} \BibitemShut
  {NoStop}%
\bibitem [{\citenamefont {Fragione}\ \emph {et~al.}(2018)\citenamefont
  {Fragione}, \citenamefont {Pavl\'\i{}k},\ and\ \citenamefont
  {Banerjee}}]{Fragione:2018jxd}%
  \BibitemOpen
  \bibfield  {author} {\bibinfo {author} {\bibfnamefont {G.}~\bibnamefont
  {Fragione}}, \bibinfo {author} {\bibfnamefont {V.}~\bibnamefont
  {Pavl\'\i{}k}}, \ and\ \bibinfo {author} {\bibfnamefont {S.}~\bibnamefont
  {Banerjee}},\ }\href {\doibase 10.1093/mnras/sty2234} {\bibfield  {journal}
  {\bibinfo  {journal} {Mon. Not. Roy. Astron. Soc.}\ }\textbf {\bibinfo
  {volume} {480}},\ \bibinfo {pages} {4955} (\bibinfo {year} {2018})},\ \Eprint
  {http://arxiv.org/abs/1804.04856} {arXiv:1804.04856 [astro-ph.GA]}
  \BibitemShut {NoStop}%
\bibitem [{\citenamefont {Ye}\ \emph {et~al.}(2019)\citenamefont {Ye},
  \citenamefont {Kremer}, \citenamefont {Chatterjee}, \citenamefont
  {Rodriguez},\ and\ \citenamefont {Rasio}}]{Ye:2019luh}%
  \BibitemOpen
  \bibfield  {author} {\bibinfo {author} {\bibfnamefont {C.~S.}\ \bibnamefont
  {Ye}}, \bibinfo {author} {\bibfnamefont {K.}~\bibnamefont {Kremer}}, \bibinfo
  {author} {\bibfnamefont {S.}~\bibnamefont {Chatterjee}}, \bibinfo {author}
  {\bibfnamefont {C.~L.}\ \bibnamefont {Rodriguez}}, \ and\ \bibinfo {author}
  {\bibfnamefont {F.~A.}\ \bibnamefont {Rasio}},\ }\href {\doibase
  10.3847/1538-4357/ab1b21} {\bibfield  {journal} {\bibinfo  {journal}
  {Astrophys. J.}\ }\textbf {\bibinfo {volume} {877}},\ \bibinfo {pages} {122}
  (\bibinfo {year} {2019})},\ \Eprint {http://arxiv.org/abs/1902.05963}
  {arXiv:1902.05963 [astro-ph.HE]} \BibitemShut {NoStop}%
\bibitem [{\citenamefont {Grindlay}\ \emph {et~al.}(2006)\citenamefont
  {Grindlay}, \citenamefont {Portegies~Zwart},\ and\ \citenamefont
  {McMillan}}]{Grindlay:2005ym}%
  \BibitemOpen
  \bibfield  {author} {\bibinfo {author} {\bibfnamefont {J.}~\bibnamefont
  {Grindlay}}, \bibinfo {author} {\bibfnamefont {S.}~\bibnamefont
  {Portegies~Zwart}}, \ and\ \bibinfo {author} {\bibfnamefont {S.}~\bibnamefont
  {McMillan}},\ }\href {\doibase 10.1038/nphys214} {\bibfield  {journal}
  {\bibinfo  {journal} {Nature Phys.}\ }\textbf {\bibinfo {volume} {2}},\
  \bibinfo {pages} {116} (\bibinfo {year} {2006})},\ \Eprint
  {http://arxiv.org/abs/astro-ph/0512654} {arXiv:astro-ph/0512654} \BibitemShut
  {NoStop}%
\bibitem [{\citenamefont {{Lee}}\ \emph {et~al.}(2010)\citenamefont {{Lee}},
  \citenamefont {{Ramirez-Ruiz}},\ and\ \citenamefont {{van de
  Ven}}}]{Lee2010ApJ}%
  \BibitemOpen
  \bibfield  {author} {\bibinfo {author} {\bibfnamefont {W.~H.}\ \bibnamefont
  {{Lee}}}, \bibinfo {author} {\bibfnamefont {E.}~\bibnamefont
  {{Ramirez-Ruiz}}}, \ and\ \bibinfo {author} {\bibfnamefont {G.}~\bibnamefont
  {{van de Ven}}},\ }\href {\doibase 10.1088/0004-637X/720/1/953} {\bibfield
  {journal} {\bibinfo  {journal} {"Astrophys. J."}\ }\textbf {\bibinfo {volume}
  {720}},\ \bibinfo {pages} {953} (\bibinfo {year} {2010})},\ \Eprint
  {http://arxiv.org/abs/0909.2884} {arXiv:0909.2884 [astro-ph.HE]} \BibitemShut
  {NoStop}%
\bibitem [{\citenamefont {{Guetta}}\ and\ \citenamefont
  {{Stella}}(2009)}]{Guetta2009A&A}%
  \BibitemOpen
  \bibfield  {author} {\bibinfo {author} {\bibfnamefont {D.}~\bibnamefont
  {{Guetta}}}\ and\ \bibinfo {author} {\bibfnamefont {L.}~\bibnamefont
  {{Stella}}},\ }\href {\doibase 10.1051/0004-6361:200810493} {\bibfield
  {journal} {\bibinfo  {journal} {Astronomy and Astrophysics}\ }\textbf
  {\bibinfo {volume} {498}},\ \bibinfo {pages} {329} (\bibinfo {year}
  {2009})},\ \Eprint {http://arxiv.org/abs/0811.0684} {arXiv:0811.0684
  [astro-ph]} \BibitemShut {NoStop}%
\bibitem [{\citenamefont {Andrews}\ and\ \citenamefont
  {Mandel}(2019)}]{Andrews:2019vou}%
  \BibitemOpen
  \bibfield  {author} {\bibinfo {author} {\bibfnamefont {J.~J.}\ \bibnamefont
  {Andrews}}\ and\ \bibinfo {author} {\bibfnamefont {I.}~\bibnamefont
  {Mandel}},\ }\href {\doibase 10.3847/2041-8213/ab2ed1} {\bibfield  {journal}
  {\bibinfo  {journal} {Astrophys. J. Lett.}\ }\textbf {\bibinfo {volume}
  {880}},\ \bibinfo {pages} {L8} (\bibinfo {year} {2019})},\ \Eprint
  {http://arxiv.org/abs/1904.12745} {arXiv:1904.12745 [astro-ph.HE]}
  \BibitemShut {NoStop}%
\bibitem [{\citenamefont {Bagchi}\ \emph {et~al.}(2013)\citenamefont {Bagchi},
  \citenamefont {Lorimer},\ and\ \citenamefont {Wolfe}}]{Bagchi:2013wga}%
  \BibitemOpen
  \bibfield  {author} {\bibinfo {author} {\bibfnamefont {M.}~\bibnamefont
  {Bagchi}}, \bibinfo {author} {\bibfnamefont {D.~R.}\ \bibnamefont {Lorimer}},
  \ and\ \bibinfo {author} {\bibfnamefont {S.}~\bibnamefont {Wolfe}},\ }\href
  {\doibase 10.1093/mnras/stt559} {\bibfield  {journal} {\bibinfo  {journal}
  {Mon. Not. Roy. Astron. Soc.}\ }\textbf {\bibinfo {volume} {432}},\ \bibinfo
  {pages} {1303} (\bibinfo {year} {2013})},\ \Eprint
  {http://arxiv.org/abs/1302.4914} {arXiv:1302.4914 [astro-ph.SR]} \BibitemShut
  {NoStop}%
\bibitem [{\citenamefont {Chattopadhyay}\ \emph {et~al.}(2021)\citenamefont
  {Chattopadhyay}, \citenamefont {Stevenson}, \citenamefont {Hurley},
  \citenamefont {Bailes},\ and\ \citenamefont
  {Broekgaarden}}]{Chattopadhyay:2020lff}%
  \BibitemOpen
  \bibfield  {author} {\bibinfo {author} {\bibfnamefont {D.}~\bibnamefont
  {Chattopadhyay}}, \bibinfo {author} {\bibfnamefont {S.}~\bibnamefont
  {Stevenson}}, \bibinfo {author} {\bibfnamefont {J.~R.}\ \bibnamefont
  {Hurley}}, \bibinfo {author} {\bibfnamefont {M.}~\bibnamefont {Bailes}}, \
  and\ \bibinfo {author} {\bibfnamefont {F.}~\bibnamefont {Broekgaarden}},\
  }\href {\doibase 10.1093/mnras/stab973} {\bibfield  {journal} {\bibinfo
  {journal} {Mon. Not. Roy. Astron. Soc.}\ }\textbf {\bibinfo {volume} {504}},\
  \bibinfo {pages} {3682} (\bibinfo {year} {2021})},\ \Eprint
  {http://arxiv.org/abs/2011.13503} {arXiv:2011.13503 [astro-ph.HE]}
  \BibitemShut {NoStop}%
\bibitem [{\citenamefont {Fortin}\ \emph {et~al.}(2024)\citenamefont {Fortin},
  \citenamefont {Kalsi}, \citenamefont {Garc\'\i{}a}, \citenamefont
  {Simaz-Bunzel},\ and\ \citenamefont {Chaty}}]{Fortin:2024siz}%
  \BibitemOpen
  \bibfield  {author} {\bibinfo {author} {\bibfnamefont {F.}~\bibnamefont
  {Fortin}}, \bibinfo {author} {\bibfnamefont {A.}~\bibnamefont {Kalsi}},
  \bibinfo {author} {\bibfnamefont {F.}~\bibnamefont {Garc\'\i{}a}}, \bibinfo
  {author} {\bibfnamefont {A.}~\bibnamefont {Simaz-Bunzel}}, \ and\ \bibinfo
  {author} {\bibfnamefont {S.}~\bibnamefont {Chaty}},\ }\href {\doibase
  10.1051/0004-6361/202347908} {\bibfield  {journal} {\bibinfo  {journal}
  {Astron. Astrophys.}\ }\textbf {\bibinfo {volume} {684}},\ \bibinfo {pages}
  {A124} (\bibinfo {year} {2024})},\ \Eprint {http://arxiv.org/abs/2401.11931}
  {arXiv:2401.11931 [astro-ph.HE]} \BibitemShut {NoStop}%
\bibitem [{\citenamefont {{Ye}}\ \emph {et~al.}(2024)\citenamefont {{Ye}},
  \citenamefont {{Kremer}}, \citenamefont {{Ransom}},\ and\ \citenamefont
  {{Rasio}}}]{Ye2024}%
  \BibitemOpen
  \bibfield  {author} {\bibinfo {author} {\bibfnamefont {C.~S.}\ \bibnamefont
  {{Ye}}}, \bibinfo {author} {\bibfnamefont {K.}~\bibnamefont {{Kremer}}},
  \bibinfo {author} {\bibfnamefont {S.~M.}\ \bibnamefont {{Ransom}}}, \ and\
  \bibinfo {author} {\bibfnamefont {F.~A.}\ \bibnamefont {{Rasio}}},\ }\href
  {\doibase 10.3847/1538-4357/ad76a0} {\bibfield  {journal} {\bibinfo
  {journal} {\apj}\ }\textbf {\bibinfo {volume} {975}},\ \bibinfo {eid} {77}
  (\bibinfo {year} {2024})},\ \Eprint {http://arxiv.org/abs/2408.00076}
  {arXiv:2408.00076 [astro-ph.HE]} \BibitemShut {NoStop}%
\bibitem [{\citenamefont {{Hobbs}}\ \emph {et~al.}(2005)\citenamefont
  {{Hobbs}}, \citenamefont {{Lorimer}}, \citenamefont {{Lyne}},\ and\
  \citenamefont {{Kramer}}}]{Hobbs2005MNRAS}%
  \BibitemOpen
  \bibfield  {author} {\bibinfo {author} {\bibfnamefont {G.}~\bibnamefont
  {{Hobbs}}}, \bibinfo {author} {\bibfnamefont {D.~R.}\ \bibnamefont
  {{Lorimer}}}, \bibinfo {author} {\bibfnamefont {A.~G.}\ \bibnamefont
  {{Lyne}}}, \ and\ \bibinfo {author} {\bibfnamefont {M.}~\bibnamefont
  {{Kramer}}},\ }\href {\doibase 10.1111/j.1365-2966.2005.09087.x} {\bibfield
  {journal} {\bibinfo  {journal} {Mon. Not. Roy. Astron. Soc.}\ }\textbf
  {\bibinfo {volume} {360}},\ \bibinfo {pages} {974} (\bibinfo {year}
  {2005})},\ \Eprint {http://arxiv.org/abs/astro-ph/0504584}
  {arXiv:astro-ph/0504584 [astro-ph]} \BibitemShut {NoStop}%
\bibitem [{\citenamefont {{Igoshev}}\ \emph {et~al.}(2021)\citenamefont
  {{Igoshev}}, \citenamefont {{Chruslinska}}, \citenamefont {{Dorozsmai}},\
  and\ \citenamefont {{Toonen}}}]{Igoshev2021}%
  \BibitemOpen
  \bibfield  {author} {\bibinfo {author} {\bibfnamefont {A.~P.}\ \bibnamefont
  {{Igoshev}}}, \bibinfo {author} {\bibfnamefont {M.}~\bibnamefont
  {{Chruslinska}}}, \bibinfo {author} {\bibfnamefont {A.}~\bibnamefont
  {{Dorozsmai}}}, \ and\ \bibinfo {author} {\bibfnamefont {S.}~\bibnamefont
  {{Toonen}}},\ }\href {\doibase 10.1093/mnras/stab2734} {\bibfield  {journal}
  {\bibinfo  {journal} {Mon. Not. Roy. Astron. Soc.}\ }\textbf {\bibinfo
  {volume} {508}},\ \bibinfo {pages} {3345} (\bibinfo {year} {2021})},\ \Eprint
  {http://arxiv.org/abs/2109.10362} {arXiv:2109.10362 [astro-ph.HE]}
  \BibitemShut {NoStop}%
\bibitem [{\citenamefont {Fortin}\ \emph {et~al.}(2022)\citenamefont {Fortin},
  \citenamefont {Garcia}, \citenamefont {Chaty}, \citenamefont
  {Chassande-Mottin},\ and\ \citenamefont {Bunzel}}]{Fortin:2022ukx}%
  \BibitemOpen
  \bibfield  {author} {\bibinfo {author} {\bibfnamefont {F.}~\bibnamefont
  {Fortin}}, \bibinfo {author} {\bibfnamefont {F.}~\bibnamefont {Garcia}},
  \bibinfo {author} {\bibfnamefont {S.}~\bibnamefont {Chaty}}, \bibinfo
  {author} {\bibfnamefont {E.}~\bibnamefont {Chassande-Mottin}}, \ and\
  \bibinfo {author} {\bibfnamefont {A.~S.}\ \bibnamefont {Bunzel}},\ }\href
  {\doibase 10.1051/0004-6361/202140853} {\bibfield  {journal} {\bibinfo
  {journal} {Astron. Astrophys.}\ }\textbf {\bibinfo {volume} {665}},\ \bibinfo
  {pages} {A31} (\bibinfo {year} {2022})},\ \Eprint
  {http://arxiv.org/abs/2206.03904} {arXiv:2206.03904 [astro-ph.HE]}
  \BibitemShut {NoStop}%
\bibitem [{\citenamefont {Disberg}\ \emph {et~al.}(2024)\citenamefont
  {Disberg}, \citenamefont {Gaspari},\ and\ \citenamefont
  {Levan}}]{Disberg:2024yrd}%
  \BibitemOpen
  \bibfield  {author} {\bibinfo {author} {\bibfnamefont {P.}~\bibnamefont
  {Disberg}}, \bibinfo {author} {\bibfnamefont {N.}~\bibnamefont {Gaspari}}, \
  and\ \bibinfo {author} {\bibfnamefont {A.~J.}\ \bibnamefont {Levan}},\ }\href
  {\doibase 10.1051/0004-6361/202450790} {\bibfield  {journal} {\bibinfo
  {journal} {Astron. Astrophys.}\ }\textbf {\bibinfo {volume} {689}},\ \bibinfo
  {pages} {A348} (\bibinfo {year} {2024})},\ \Eprint
  {http://arxiv.org/abs/2407.13667} {arXiv:2407.13667 [astro-ph.HE]}
  \BibitemShut {NoStop}%
\bibitem [{\citenamefont {Manchester}\ \emph {et~al.}(2005)\citenamefont
  {Manchester}, \citenamefont {Hobbs}, \citenamefont {Teoh},\ and\
  \citenamefont {Hobbs}}]{Manchester:2004bp}%
  \BibitemOpen
  \bibfield  {author} {\bibinfo {author} {\bibfnamefont {R.~N.}\ \bibnamefont
  {Manchester}}, \bibinfo {author} {\bibfnamefont {G.~B.}\ \bibnamefont
  {Hobbs}}, \bibinfo {author} {\bibfnamefont {A.}~\bibnamefont {Teoh}}, \ and\
  \bibinfo {author} {\bibfnamefont {M.}~\bibnamefont {Hobbs}},\ }\href
  {\doibase 10.1086/428488} {\bibfield  {journal} {\bibinfo  {journal} {Astron.
  J.}\ }\textbf {\bibinfo {volume} {129}},\ \bibinfo {pages} {1993} (\bibinfo
  {year} {2005})},\ \Eprint {http://arxiv.org/abs/astro-ph/0412641}
  {arXiv:astro-ph/0412641} \BibitemShut {NoStop}%
\bibitem [{\citenamefont {{Harris}}(1996)}]{Harris1996}%
  \BibitemOpen
  \bibfield  {author} {\bibinfo {author} {\bibfnamefont {W.~E.}\ \bibnamefont
  {{Harris}}},\ }\href {\doibase 10.1086/118116} {\bibfield  {journal}
  {\bibinfo  {journal} {The Astronomical Journal}\ }\textbf {\bibinfo {volume}
  {112}},\ \bibinfo {pages} {1487} (\bibinfo {year} {1996})}\BibitemShut
  {NoStop}%
\bibitem [{\citenamefont {https://www3.mpifr
  bonn.mpg.de/staff/pfreire/GCpsr.html}()}]{pulsars_gc_link}%
  \BibitemOpen
  \bibfield  {author} {\bibinfo {author} {\bibnamefont {https://www3.mpifr
  bonn.mpg.de/staff/pfreire/GCpsr.html}},\ }\href@noop {} {\ }\BibitemShut
  {NoStop}%
\bibitem [{\citenamefont {{Belczynski}}\ \emph {et~al.}(2010)\citenamefont
  {{Belczynski}}, \citenamefont {{Bulik}}, \citenamefont {{Fryer}},
  \citenamefont {{Ruiter}}, \citenamefont {{Valsecchi}}, \citenamefont
  {{Vink}},\ and\ \citenamefont {{Hurley}}}]{Belczynski2010}%
  \BibitemOpen
  \bibfield  {author} {\bibinfo {author} {\bibfnamefont {K.}~\bibnamefont
  {{Belczynski}}}, \bibinfo {author} {\bibfnamefont {T.}~\bibnamefont
  {{Bulik}}}, \bibinfo {author} {\bibfnamefont {C.~L.}\ \bibnamefont
  {{Fryer}}}, \bibinfo {author} {\bibfnamefont {A.}~\bibnamefont {{Ruiter}}},
  \bibinfo {author} {\bibfnamefont {F.}~\bibnamefont {{Valsecchi}}}, \bibinfo
  {author} {\bibfnamefont {J.~S.}\ \bibnamefont {{Vink}}}, \ and\ \bibinfo
  {author} {\bibfnamefont {J.~R.}\ \bibnamefont {{Hurley}}},\ }\href {\doibase
  10.1088/0004-637X/714/2/1217} {\bibfield  {journal} {\bibinfo  {journal}
  {\apj}\ }\textbf {\bibinfo {volume} {714}},\ \bibinfo {pages} {1217}
  (\bibinfo {year} {2010})},\ \Eprint {http://arxiv.org/abs/0904.2784}
  {arXiv:0904.2784 [astro-ph.SR]} \BibitemShut {NoStop}%
\bibitem [{\citenamefont {Ziosi}\ \emph {et~al.}(2014)\citenamefont {Ziosi},
  \citenamefont {Mapelli}, \citenamefont {Branchesi},\ and\ \citenamefont
  {Tormen}}]{Ziosi:2014sra}%
  \BibitemOpen
  \bibfield  {author} {\bibinfo {author} {\bibfnamefont {B.~M.}\ \bibnamefont
  {Ziosi}}, \bibinfo {author} {\bibfnamefont {M.}~\bibnamefont {Mapelli}},
  \bibinfo {author} {\bibfnamefont {M.}~\bibnamefont {Branchesi}}, \ and\
  \bibinfo {author} {\bibfnamefont {G.}~\bibnamefont {Tormen}},\ }\href
  {\doibase 10.1093/mnras/stu824} {\bibfield  {journal} {\bibinfo  {journal}
  {Mon. Not. Roy. Astron. Soc.}\ }\textbf {\bibinfo {volume} {441}},\ \bibinfo
  {pages} {3703} (\bibinfo {year} {2014})},\ \Eprint
  {http://arxiv.org/abs/1404.7147} {arXiv:1404.7147 [astro-ph.GA]} \BibitemShut
  {NoStop}%
\bibitem [{\citenamefont {Rastello}\ \emph {et~al.}(2020)\citenamefont
  {Rastello}, \citenamefont {Mapelli}, \citenamefont {Di~Carlo}, \citenamefont
  {Giacobbo}, \citenamefont {Santoliquido}, \citenamefont {Spera},
  \citenamefont {Ballone},\ and\ \citenamefont {Iorio}}]{Rastello:2020sru}%
  \BibitemOpen
  \bibfield  {author} {\bibinfo {author} {\bibfnamefont {S.}~\bibnamefont
  {Rastello}}, \bibinfo {author} {\bibfnamefont {M.}~\bibnamefont {Mapelli}},
  \bibinfo {author} {\bibfnamefont {U.~N.}\ \bibnamefont {Di~Carlo}}, \bibinfo
  {author} {\bibfnamefont {N.}~\bibnamefont {Giacobbo}}, \bibinfo {author}
  {\bibfnamefont {F.}~\bibnamefont {Santoliquido}}, \bibinfo {author}
  {\bibfnamefont {M.}~\bibnamefont {Spera}}, \bibinfo {author} {\bibfnamefont
  {A.}~\bibnamefont {Ballone}}, \ and\ \bibinfo {author} {\bibfnamefont
  {G.}~\bibnamefont {Iorio}},\ }\href {\doibase 10.1093/mnras/staa2018}
  {\bibfield  {journal} {\bibinfo  {journal} {Mon. Not. Roy. Astron. Soc.}\
  }\textbf {\bibinfo {volume} {497}},\ \bibinfo {pages} {1563} (\bibinfo {year}
  {2020})},\ \Eprint {http://arxiv.org/abs/2003.02277} {arXiv:2003.02277
  [astro-ph.HE]} \BibitemShut {NoStop}%
\bibitem [{\citenamefont {Arca~Sedda}(2021)}]{ArcaSedda:2021zmm}%
  \BibitemOpen
  \bibfield  {author} {\bibinfo {author} {\bibfnamefont {M.}~\bibnamefont
  {Arca~Sedda}},\ }\href {\doibase 10.3847/2041-8213/abdfcd} {\bibfield
  {journal} {\bibinfo  {journal} {Astrophys. J. Lett.}\ }\textbf {\bibinfo
  {volume} {908}},\ \bibinfo {pages} {L38} (\bibinfo {year} {2021})},\ \Eprint
  {http://arxiv.org/abs/2102.03364} {arXiv:2102.03364 [astro-ph.HE]}
  \BibitemShut {NoStop}%
\bibitem [{\citenamefont {Petrovich}\ and\ \citenamefont
  {Antonini}(2017)}]{Petrovich:2017otm}%
  \BibitemOpen
  \bibfield  {author} {\bibinfo {author} {\bibfnamefont {C.}~\bibnamefont
  {Petrovich}}\ and\ \bibinfo {author} {\bibfnamefont {F.}~\bibnamefont
  {Antonini}},\ }\href {\doibase 10.3847/1538-4357/aa8628} {\bibfield
  {journal} {\bibinfo  {journal} {Astrophys. J.}\ }\textbf {\bibinfo {volume}
  {846}},\ \bibinfo {pages} {146} (\bibinfo {year} {2017})},\ \Eprint
  {http://arxiv.org/abs/1705.05848} {arXiv:1705.05848 [astro-ph.HE]}
  \BibitemShut {NoStop}%
\bibitem [{\citenamefont {McKernan}\ \emph {et~al.}(2020)\citenamefont
  {McKernan}, \citenamefont {Ford},\ and\ \citenamefont
  {O'Shaughnessy}}]{McKernan:2020lgr}%
  \BibitemOpen
  \bibfield  {author} {\bibinfo {author} {\bibfnamefont {B.}~\bibnamefont
  {McKernan}}, \bibinfo {author} {\bibfnamefont {K.~E.~S.}\ \bibnamefont
  {Ford}}, \ and\ \bibinfo {author} {\bibfnamefont {R.}~\bibnamefont
  {O'Shaughnessy}},\ }\href {\doibase 10.1093/mnras/staa2681} {\bibfield
  {journal} {\bibinfo  {journal} {Mon. Not. Roy. Astron. Soc.}\ }\textbf
  {\bibinfo {volume} {498}},\ \bibinfo {pages} {4088} (\bibinfo {year}
  {2020})},\ \Eprint {http://arxiv.org/abs/2002.00046} {arXiv:2002.00046
  [astro-ph.HE]} \BibitemShut {NoStop}%
\bibitem [{\citenamefont {Yang}\ \emph {et~al.}(2020)\citenamefont {Yang},
  \citenamefont {Gayathri}, \citenamefont {Bartos}, \citenamefont {Haiman},
  \citenamefont {Safarzadeh},\ and\ \citenamefont {Tagawa}}]{Yang:2020xyi}%
  \BibitemOpen
  \bibfield  {author} {\bibinfo {author} {\bibfnamefont {Y.}~\bibnamefont
  {Yang}}, \bibinfo {author} {\bibfnamefont {V.}~\bibnamefont {Gayathri}},
  \bibinfo {author} {\bibfnamefont {I.}~\bibnamefont {Bartos}}, \bibinfo
  {author} {\bibfnamefont {Z.}~\bibnamefont {Haiman}}, \bibinfo {author}
  {\bibfnamefont {M.}~\bibnamefont {Safarzadeh}}, \ and\ \bibinfo {author}
  {\bibfnamefont {H.}~\bibnamefont {Tagawa}},\ }\href {\doibase
  10.3847/2041-8213/abb940} {\bibfield  {journal} {\bibinfo  {journal}
  {Astrophys. J. Lett.}\ }\textbf {\bibinfo {volume} {901}},\ \bibinfo {pages}
  {L34} (\bibinfo {year} {2020})},\ \Eprint {http://arxiv.org/abs/2007.04781}
  {arXiv:2007.04781 [astro-ph.HE]} \BibitemShut {NoStop}%
\bibitem [{\citenamefont {Tagawa}\ \emph {et~al.}(2021)\citenamefont {Tagawa},
  \citenamefont {Kocsis}, \citenamefont {Haiman}, \citenamefont {Bartos},
  \citenamefont {Omukai},\ and\ \citenamefont {Samsing}}]{Tagawa:2020qll}%
  \BibitemOpen
  \bibfield  {author} {\bibinfo {author} {\bibfnamefont {H.}~\bibnamefont
  {Tagawa}}, \bibinfo {author} {\bibfnamefont {B.}~\bibnamefont {Kocsis}},
  \bibinfo {author} {\bibfnamefont {Z.}~\bibnamefont {Haiman}}, \bibinfo
  {author} {\bibfnamefont {I.}~\bibnamefont {Bartos}}, \bibinfo {author}
  {\bibfnamefont {K.}~\bibnamefont {Omukai}}, \ and\ \bibinfo {author}
  {\bibfnamefont {J.}~\bibnamefont {Samsing}},\ }\href {\doibase
  10.3847/1538-4357/abd555} {\bibfield  {journal} {\bibinfo  {journal}
  {Astrophys. J.}\ }\textbf {\bibinfo {volume} {908}},\ \bibinfo {pages} {194}
  (\bibinfo {year} {2021})},\ \Eprint {http://arxiv.org/abs/2012.00011}
  {arXiv:2012.00011 [astro-ph.HE]} \BibitemShut {NoStop}%
\bibitem [{\citenamefont {Lu}\ \emph {et~al.}(2020)\citenamefont {Lu},
  \citenamefont {Beniamini},\ and\ \citenamefont {Bonnerot}}]{Lu:2020gfh}%
  \BibitemOpen
  \bibfield  {author} {\bibinfo {author} {\bibfnamefont {W.}~\bibnamefont
  {Lu}}, \bibinfo {author} {\bibfnamefont {P.}~\bibnamefont {Beniamini}}, \
  and\ \bibinfo {author} {\bibfnamefont {C.}~\bibnamefont {Bonnerot}},\ }\href
  {\doibase 10.1093/mnras/staa3372} {\bibfield  {journal} {\bibinfo  {journal}
  {Mon. Not. Roy. Astron. Soc.}\ }\textbf {\bibinfo {volume} {500}},\ \bibinfo
  {pages} {1817} (\bibinfo {year} {2020})},\ \Eprint
  {http://arxiv.org/abs/2009.10082} {arXiv:2009.10082 [astro-ph.HE]}
  \BibitemShut {NoStop}%
\bibitem [{\citenamefont {Liu}\ and\ \citenamefont {Lai}(2021)}]{Liu:2020gif}%
  \BibitemOpen
  \bibfield  {author} {\bibinfo {author} {\bibfnamefont {B.}~\bibnamefont
  {Liu}}\ and\ \bibinfo {author} {\bibfnamefont {D.}~\bibnamefont {Lai}},\
  }\href {\doibase 10.1093/mnras/stab178} {\bibfield  {journal} {\bibinfo
  {journal} {Mon. Not. Roy. Astron. Soc.}\ }\textbf {\bibinfo {volume} {502}},\
  \bibinfo {pages} {2049} (\bibinfo {year} {2021})},\ \Eprint
  {http://arxiv.org/abs/2009.10068} {arXiv:2009.10068 [astro-ph.HE]}
  \BibitemShut {NoStop}%
\bibitem [{\citenamefont {Bartos}\ \emph {et~al.}(2023)\citenamefont {Bartos},
  \citenamefont {Rosswog}, \citenamefont {Gayathri}, \citenamefont {Miller},
  \citenamefont {Veske},\ and\ \citenamefont {Marka}}]{Bartos:2023lfu}%
  \BibitemOpen
  \bibfield  {author} {\bibinfo {author} {\bibfnamefont {I.}~\bibnamefont
  {Bartos}}, \bibinfo {author} {\bibfnamefont {S.}~\bibnamefont {Rosswog}},
  \bibinfo {author} {\bibfnamefont {V.}~\bibnamefont {Gayathri}}, \bibinfo
  {author} {\bibfnamefont {M.~C.}\ \bibnamefont {Miller}}, \bibinfo {author}
  {\bibfnamefont {D.}~\bibnamefont {Veske}}, \ and\ \bibinfo {author}
  {\bibfnamefont {S.}~\bibnamefont {Marka}},\ }\href@noop {} {\  (\bibinfo
  {year} {2023})},\ \Eprint {http://arxiv.org/abs/2302.10350} {arXiv:2302.10350
  [astro-ph.HE]} \BibitemShut {NoStop}%
\bibitem [{\citenamefont {Gayathri}\ \emph {et~al.}(2023)\citenamefont
  {Gayathri}, \citenamefont {Bartos}, \citenamefont {Rosswog}, \citenamefont
  {Miller}, \citenamefont {Veske}, \citenamefont {Lu},\ and\ \citenamefont
  {Marka}}]{Gayathri:2023met}%
  \BibitemOpen
  \bibfield  {author} {\bibinfo {author} {\bibfnamefont {V.}~\bibnamefont
  {Gayathri}}, \bibinfo {author} {\bibfnamefont {I.}~\bibnamefont {Bartos}},
  \bibinfo {author} {\bibfnamefont {S.}~\bibnamefont {Rosswog}}, \bibinfo
  {author} {\bibfnamefont {M.~C.}\ \bibnamefont {Miller}}, \bibinfo {author}
  {\bibfnamefont {D.}~\bibnamefont {Veske}}, \bibinfo {author} {\bibfnamefont
  {W.}~\bibnamefont {Lu}}, \ and\ \bibinfo {author} {\bibfnamefont
  {S.}~\bibnamefont {Marka}},\ }\href@noop {} {\  (\bibinfo {year} {2023})},\
  \Eprint {http://arxiv.org/abs/2307.09097} {arXiv:2307.09097 [astro-ph.HE]}
  \BibitemShut {NoStop}%
\bibitem [{\citenamefont {Safarzadeh}\ \emph {et~al.}(2020)\citenamefont
  {Safarzadeh}, \citenamefont {Hamers}, \citenamefont {Loeb},\ and\
  \citenamefont {Berger}}]{Safarzadeh:2019qkk}%
  \BibitemOpen
  \bibfield  {author} {\bibinfo {author} {\bibfnamefont {M.}~\bibnamefont
  {Safarzadeh}}, \bibinfo {author} {\bibfnamefont {A.~S.}\ \bibnamefont
  {Hamers}}, \bibinfo {author} {\bibfnamefont {A.}~\bibnamefont {Loeb}}, \ and\
  \bibinfo {author} {\bibfnamefont {E.}~\bibnamefont {Berger}},\ }\href
  {\doibase 10.3847/2041-8213/ab5dc8} {\bibfield  {journal} {\bibinfo
  {journal} {Astrophys. J. Lett.}\ }\textbf {\bibinfo {volume} {888}},\
  \bibinfo {pages} {L3} (\bibinfo {year} {2020})},\ \Eprint
  {http://arxiv.org/abs/1911.04495} {arXiv:1911.04495 [astro-ph.HE]}
  \BibitemShut {NoStop}%
\bibitem [{\citenamefont {Fragione}\ \emph {et~al.}(2020)\citenamefont
  {Fragione}, \citenamefont {Loeb},\ and\ \citenamefont
  {Rasio}}]{Fragione:2020aki}%
  \BibitemOpen
  \bibfield  {author} {\bibinfo {author} {\bibfnamefont {G.}~\bibnamefont
  {Fragione}}, \bibinfo {author} {\bibfnamefont {A.}~\bibnamefont {Loeb}}, \
  and\ \bibinfo {author} {\bibfnamefont {F.~A.}\ \bibnamefont {Rasio}},\ }\href
  {\doibase 10.3847/2041-8213/ab9093} {\bibfield  {journal} {\bibinfo
  {journal} {Astrophys. J. Lett.}\ }\textbf {\bibinfo {volume} {895}},\
  \bibinfo {pages} {L15} (\bibinfo {year} {2020})},\ \Eprint
  {http://arxiv.org/abs/2002.11278} {arXiv:2002.11278 [astro-ph.GA]}
  \BibitemShut {NoStop}%
\bibitem [{\citenamefont {Vigna-G\'omez}\ \emph {et~al.}(2021)\citenamefont
  {Vigna-G\'omez}, \citenamefont {Toonen}, \citenamefont {Ramirez-Ruiz},
  \citenamefont {Leigh}, \citenamefont {Riley},\ and\ \citenamefont
  {Haster}}]{Vigna-Gomez:2020fvw}%
  \BibitemOpen
  \bibfield  {author} {\bibinfo {author} {\bibfnamefont {A.}~\bibnamefont
  {Vigna-G\'omez}}, \bibinfo {author} {\bibfnamefont {S.}~\bibnamefont
  {Toonen}}, \bibinfo {author} {\bibfnamefont {E.}~\bibnamefont
  {Ramirez-Ruiz}}, \bibinfo {author} {\bibfnamefont {N.~W.~C.}\ \bibnamefont
  {Leigh}}, \bibinfo {author} {\bibfnamefont {J.}~\bibnamefont {Riley}}, \ and\
  \bibinfo {author} {\bibfnamefont {C.-J.}\ \bibnamefont {Haster}},\ }\href
  {\doibase 10.3847/2041-8213/abd5b7} {\bibfield  {journal} {\bibinfo
  {journal} {Astrophys. J. Lett.}\ }\textbf {\bibinfo {volume} {907}},\
  \bibinfo {pages} {L19} (\bibinfo {year} {2021})},\ \Eprint
  {http://arxiv.org/abs/2010.13669} {arXiv:2010.13669 [astro-ph.HE]}
  \BibitemShut {NoStop}%
\bibitem [{\citenamefont {Hamers}\ \emph {et~al.}(2021)\citenamefont {Hamers},
  \citenamefont {Fragione}, \citenamefont {Neunteufel},\ and\ \citenamefont
  {Kocsis}}]{Hamers:2021olp}%
  \BibitemOpen
  \bibfield  {author} {\bibinfo {author} {\bibfnamefont {A.~S.}\ \bibnamefont
  {Hamers}}, \bibinfo {author} {\bibfnamefont {G.}~\bibnamefont {Fragione}},
  \bibinfo {author} {\bibfnamefont {P.}~\bibnamefont {Neunteufel}}, \ and\
  \bibinfo {author} {\bibfnamefont {B.}~\bibnamefont {Kocsis}},\ }\href
  {\doibase 10.1093/mnras/stab2136} {\bibfield  {journal} {\bibinfo  {journal}
  {Mon. Not. Roy. Astron. Soc.}\ }\textbf {\bibinfo {volume} {506}},\ \bibinfo
  {pages} {5345} (\bibinfo {year} {2021})},\ \Eprint
  {http://arxiv.org/abs/2103.03782} {arXiv:2103.03782 [astro-ph.HE]}
  \BibitemShut {NoStop}%
\bibitem [{\citenamefont {Vynatheya}\ and\ \citenamefont
  {Hamers}(2022)}]{Vynatheya:2021mgl}%
  \BibitemOpen
  \bibfield  {author} {\bibinfo {author} {\bibfnamefont {P.}~\bibnamefont
  {Vynatheya}}\ and\ \bibinfo {author} {\bibfnamefont {A.~S.}\ \bibnamefont
  {Hamers}},\ }\href {\doibase 10.3847/1538-4357/ac4892} {\bibfield  {journal}
  {\bibinfo  {journal} {Astrophys. J.}\ }\textbf {\bibinfo {volume} {926}},\
  \bibinfo {pages} {195} (\bibinfo {year} {2022})},\ \Eprint
  {http://arxiv.org/abs/2110.14680} {arXiv:2110.14680 [astro-ph.HE]}
  \BibitemShut {NoStop}%
\bibitem [{\citenamefont {Abbott}\ \emph
  {et~al.}(2021{\natexlab{b}})\citenamefont {Abbott} \emph {et~al.}}]{GWTC2}%
  \BibitemOpen
  \bibfield  {author} {\bibinfo {author} {\bibfnamefont {R.}~\bibnamefont
  {Abbott}} \emph {et~al.} (\bibinfo {collaboration} {LIGO Scientific,
  Virgo}),\ }\href {\doibase 10.1103/PhysRevX.11.021053} {\bibfield  {journal}
  {\bibinfo  {journal} {Phys. Rev. X}\ }\textbf {\bibinfo {volume} {11}},\
  \bibinfo {pages} {021053} (\bibinfo {year} {2021}{\natexlab{b}})},\ \Eprint
  {http://arxiv.org/abs/2010.14527} {arXiv:2010.14527 [gr-qc]} \BibitemShut
  {NoStop}%
\bibitem [{\citenamefont {Afroz}\ and\ \citenamefont
  {Mukherjee}(2024)}]{Afroz:2024fzp}%
  \BibitemOpen
  \bibfield  {author} {\bibinfo {author} {\bibfnamefont {S.}~\bibnamefont
  {Afroz}}\ and\ \bibinfo {author} {\bibfnamefont {S.}~\bibnamefont
  {Mukherjee}},\ }\href@noop {} {\  (\bibinfo {year} {2024})},\ \Eprint
  {http://arxiv.org/abs/2411.07304} {arXiv:2411.07304 [astro-ph.HE]}
  \BibitemShut {NoStop}%
\bibitem [{\citenamefont {Abbott}\ \emph
  {et~al.}(2020{\natexlab{b}})\citenamefont {Abbott} \emph
  {et~al.}}]{GW190425}%
  \BibitemOpen
  \bibfield  {author} {\bibinfo {author} {\bibfnamefont {B.~P.}\ \bibnamefont
  {Abbott}} \emph {et~al.} (\bibinfo {collaboration} {LIGO Scientific,
  Virgo}),\ }\href {\doibase 10.3847/2041-8213/ab75f5} {\bibfield  {journal}
  {\bibinfo  {journal} {Astrophys. J. Lett.}\ }\textbf {\bibinfo {volume}
  {892}},\ \bibinfo {pages} {L3} (\bibinfo {year} {2020}{\natexlab{b}})},\
  \Eprint {http://arxiv.org/abs/2001.01761} {arXiv:2001.01761 [astro-ph.HE]}
  \BibitemShut {NoStop}%
\bibitem [{\citenamefont {Clesse}\ and\ \citenamefont
  {Garcia-Bellido}(2022)}]{Clesse:2020ghq}%
  \BibitemOpen
  \bibfield  {author} {\bibinfo {author} {\bibfnamefont {S.}~\bibnamefont
  {Clesse}}\ and\ \bibinfo {author} {\bibfnamefont {J.}~\bibnamefont
  {Garcia-Bellido}},\ }\href {\doibase 10.1016/j.dark.2022.101111} {\bibfield
  {journal} {\bibinfo  {journal} {Phys. Dark Univ.}\ }\textbf {\bibinfo
  {volume} {38}},\ \bibinfo {pages} {101111} (\bibinfo {year} {2022})},\
  \Eprint {http://arxiv.org/abs/2007.06481} {arXiv:2007.06481 [astro-ph.CO]}
  \BibitemShut {NoStop}%
\bibitem [{\citenamefont {Mr\'oz}\ \emph {et~al.}(2024)\citenamefont {Mr\'oz}
  \emph {et~al.}}]{Mroz:2024mse}%
  \BibitemOpen
  \bibfield  {author} {\bibinfo {author} {\bibfnamefont {P.}~\bibnamefont
  {Mr\'oz}} \emph {et~al.},\ }\href {\doibase 10.1038/s41586-024-07704-6}
  {\bibfield  {journal} {\bibinfo  {journal} {Nature}\ }\textbf {\bibinfo
  {volume} {632}},\ \bibinfo {pages} {749} (\bibinfo {year} {2024})},\ \Eprint
  {http://arxiv.org/abs/2403.02386} {arXiv:2403.02386 [astro-ph.GA]}
  \BibitemShut {NoStop}%
\bibitem [{\citenamefont {Garcia-Bellido}\ and\ \citenamefont
  {Hawkins}(2024)}]{Garcia-Bellido:2024yaz}%
  \BibitemOpen
  \bibfield  {author} {\bibinfo {author} {\bibfnamefont {J.}~\bibnamefont
  {Garcia-Bellido}}\ and\ \bibinfo {author} {\bibfnamefont {M.}~\bibnamefont
  {Hawkins}},\ }\href {\doibase 10.3390/universe10120449} {\bibfield  {journal}
  {\bibinfo  {journal} {Universe}\ }\textbf {\bibinfo {volume} {10}},\ \bibinfo
  {pages} {449} (\bibinfo {year} {2024})},\ \Eprint
  {http://arxiv.org/abs/2402.00212} {arXiv:2402.00212 [astro-ph.GA]}
  \BibitemShut {NoStop}%
\bibitem [{\citenamefont {Coughlin}\ \emph {et~al.}(2019)\citenamefont
  {Coughlin}, \citenamefont {Dietrich}, \citenamefont {Margalit},\ and\
  \citenamefont {Metzger}}]{Coughlin:2018fis}%
  \BibitemOpen
  \bibfield  {author} {\bibinfo {author} {\bibfnamefont {M.~W.}\ \bibnamefont
  {Coughlin}}, \bibinfo {author} {\bibfnamefont {T.}~\bibnamefont {Dietrich}},
  \bibinfo {author} {\bibfnamefont {B.}~\bibnamefont {Margalit}}, \ and\
  \bibinfo {author} {\bibfnamefont {B.~D.}\ \bibnamefont {Metzger}},\ }\href
  {\doibase 10.1093/mnrasl/slz133} {\bibfield  {journal} {\bibinfo  {journal}
  {Mon. Not. Roy. Astron. Soc.}\ }\textbf {\bibinfo {volume} {489}},\ \bibinfo
  {pages} {L91} (\bibinfo {year} {2019})},\ \Eprint
  {http://arxiv.org/abs/1812.04803} {arXiv:1812.04803 [astro-ph.HE]}
  \BibitemShut {NoStop}%
\bibitem [{\citenamefont {Zappa}\ \emph {et~al.}(2019)\citenamefont {Zappa},
  \citenamefont {Bernuzzi}, \citenamefont {Pannarale}, \citenamefont
  {Mapelli},\ and\ \citenamefont {Giacobbo}}]{Zappa:2019ntl}%
  \BibitemOpen
  \bibfield  {author} {\bibinfo {author} {\bibfnamefont {F.}~\bibnamefont
  {Zappa}}, \bibinfo {author} {\bibfnamefont {S.}~\bibnamefont {Bernuzzi}},
  \bibinfo {author} {\bibfnamefont {F.}~\bibnamefont {Pannarale}}, \bibinfo
  {author} {\bibfnamefont {M.}~\bibnamefont {Mapelli}}, \ and\ \bibinfo
  {author} {\bibfnamefont {N.}~\bibnamefont {Giacobbo}},\ }\href {\doibase
  10.1103/PhysRevLett.123.041102} {\bibfield  {journal} {\bibinfo  {journal}
  {Phys. Rev. Lett.}\ }\textbf {\bibinfo {volume} {123}},\ \bibinfo {pages}
  {041102} (\bibinfo {year} {2019})},\ \Eprint
  {http://arxiv.org/abs/1903.11622} {arXiv:1903.11622 [gr-qc]} \BibitemShut
  {NoStop}%
\bibitem [{\citenamefont {Abbott}\ \emph
  {et~al.}(2017{\natexlab{a}})\citenamefont {Abbott} \emph
  {et~al.}}]{GW170817}%
  \BibitemOpen
  \bibfield  {author} {\bibinfo {author} {\bibfnamefont {B.}~\bibnamefont
  {Abbott}} \emph {et~al.} (\bibinfo {collaboration} {Virgo, LIGO
  Scientific}),\ }\href {\doibase 10.1103/PhysRevLett.119.161101} {\bibfield
  {journal} {\bibinfo  {journal} {Phys. Rev. Lett.}\ }\textbf {\bibinfo
  {volume} {119}},\ \bibinfo {pages} {161101} (\bibinfo {year}
  {2017}{\natexlab{a}})},\ \Eprint {http://arxiv.org/abs/1710.05832}
  {arXiv:1710.05832 [gr-qc]} \BibitemShut {NoStop}%
\bibitem [{\citenamefont {Pretorius}(2005)}]{Pretorius:2005gq}%
  \BibitemOpen
  \bibfield  {author} {\bibinfo {author} {\bibfnamefont {F.}~\bibnamefont
  {Pretorius}},\ }\href {\doibase 10.1103/PhysRevLett.95.121101} {\bibfield
  {journal} {\bibinfo  {journal} {Phys. Rev. Lett.}\ }\textbf {\bibinfo
  {volume} {95}},\ \bibinfo {pages} {121101} (\bibinfo {year} {2005})},\
  \Eprint {http://arxiv.org/abs/gr-qc/0507014} {arXiv:gr-qc/0507014}
  \BibitemShut {NoStop}%
\bibitem [{\citenamefont {Campanelli}\ \emph {et~al.}(2006)\citenamefont
  {Campanelli}, \citenamefont {Lousto}, \citenamefont {Marronetti},\ and\
  \citenamefont {Zlochower}}]{Campanelli:2005dd}%
  \BibitemOpen
  \bibfield  {author} {\bibinfo {author} {\bibfnamefont {M.}~\bibnamefont
  {Campanelli}}, \bibinfo {author} {\bibfnamefont {C.~O.}\ \bibnamefont
  {Lousto}}, \bibinfo {author} {\bibfnamefont {P.}~\bibnamefont {Marronetti}},
  \ and\ \bibinfo {author} {\bibfnamefont {Y.}~\bibnamefont {Zlochower}},\
  }\href {\doibase 10.1103/PhysRevLett.96.111101} {\bibfield  {journal}
  {\bibinfo  {journal} {Phys. Rev. Lett.}\ }\textbf {\bibinfo {volume} {96}},\
  \bibinfo {pages} {111101} (\bibinfo {year} {2006})},\ \Eprint
  {http://arxiv.org/abs/gr-qc/0511048} {arXiv:gr-qc/0511048} \BibitemShut
  {NoStop}%
\bibitem [{\citenamefont {Baker}\ \emph {et~al.}(2006)\citenamefont {Baker},
  \citenamefont {Centrella}, \citenamefont {Choi}, \citenamefont {Koppitz},\
  and\ \citenamefont {van Meter}}]{Baker:2005vv}%
  \BibitemOpen
  \bibfield  {author} {\bibinfo {author} {\bibfnamefont {J.~G.}\ \bibnamefont
  {Baker}}, \bibinfo {author} {\bibfnamefont {J.}~\bibnamefont {Centrella}},
  \bibinfo {author} {\bibfnamefont {D.-I.}\ \bibnamefont {Choi}}, \bibinfo
  {author} {\bibfnamefont {M.}~\bibnamefont {Koppitz}}, \ and\ \bibinfo
  {author} {\bibfnamefont {J.}~\bibnamefont {van Meter}},\ }\href {\doibase
  10.1103/PhysRevLett.96.111102} {\bibfield  {journal} {\bibinfo  {journal}
  {Phys. Rev. Lett.}\ }\textbf {\bibinfo {volume} {96}},\ \bibinfo {pages}
  {111102} (\bibinfo {year} {2006})},\ \Eprint
  {http://arxiv.org/abs/gr-qc/0511103} {arXiv:gr-qc/0511103} \BibitemShut
  {NoStop}%
\bibitem [{\citenamefont {Lousto}\ \emph {et~al.}(2010)\citenamefont {Lousto},
  \citenamefont {Campanelli}, \citenamefont {Zlochower},\ and\ \citenamefont
  {Nakano}}]{Lousto:2009mf}%
  \BibitemOpen
  \bibfield  {author} {\bibinfo {author} {\bibfnamefont {C.~O.}\ \bibnamefont
  {Lousto}}, \bibinfo {author} {\bibfnamefont {M.}~\bibnamefont {Campanelli}},
  \bibinfo {author} {\bibfnamefont {Y.}~\bibnamefont {Zlochower}}, \ and\
  \bibinfo {author} {\bibfnamefont {H.}~\bibnamefont {Nakano}},\ }\href
  {\doibase 10.1088/0264-9381/27/11/114006} {\bibfield  {journal} {\bibinfo
  {journal} {Class. Quant. Grav.}\ }\textbf {\bibinfo {volume} {27}},\ \bibinfo
  {pages} {114006} (\bibinfo {year} {2010})},\ \Eprint
  {http://arxiv.org/abs/0904.3541} {arXiv:0904.3541 [gr-qc]} \BibitemShut
  {NoStop}%
\bibitem [{\citenamefont {Barausse}\ \emph {et~al.}(2012)\citenamefont
  {Barausse}, \citenamefont {Morozova},\ and\ \citenamefont
  {Rezzolla}}]{Barausse:2012qz}%
  \BibitemOpen
  \bibfield  {author} {\bibinfo {author} {\bibfnamefont {E.}~\bibnamefont
  {Barausse}}, \bibinfo {author} {\bibfnamefont {V.}~\bibnamefont {Morozova}},
  \ and\ \bibinfo {author} {\bibfnamefont {L.}~\bibnamefont {Rezzolla}},\
  }\href {\doibase 10.1088/0004-637X/758/1/63} {\bibfield  {journal} {\bibinfo
  {journal} {Astrophys. J.}\ }\textbf {\bibinfo {volume} {758}},\ \bibinfo
  {pages} {63} (\bibinfo {year} {2012})},\ \bibinfo {note} {[Erratum:
  Astrophys.J. 786, 76 (2014)]},\ \Eprint {http://arxiv.org/abs/1206.3803}
  {arXiv:1206.3803 [gr-qc]} \BibitemShut {NoStop}%
\bibitem [{\citenamefont {Hofmann}\ \emph {et~al.}(2016)\citenamefont
  {Hofmann}, \citenamefont {Barausse},\ and\ \citenamefont
  {Rezzolla}}]{Hofmann:2016yih}%
  \BibitemOpen
  \bibfield  {author} {\bibinfo {author} {\bibfnamefont {F.}~\bibnamefont
  {Hofmann}}, \bibinfo {author} {\bibfnamefont {E.}~\bibnamefont {Barausse}}, \
  and\ \bibinfo {author} {\bibfnamefont {L.}~\bibnamefont {Rezzolla}},\ }\href
  {\doibase 10.3847/2041-8205/825/2/L19} {\bibfield  {journal} {\bibinfo
  {journal} {Astrophys. J. Lett.}\ }\textbf {\bibinfo {volume} {825}},\
  \bibinfo {pages} {L19} (\bibinfo {year} {2016})},\ \Eprint
  {http://arxiv.org/abs/1605.01938} {arXiv:1605.01938 [gr-qc]} \BibitemShut
  {NoStop}%
\bibitem [{\citenamefont {Varma}\ \emph {et~al.}(2019)\citenamefont {Varma},
  \citenamefont {Field}, \citenamefont {Scheel}, \citenamefont {Blackman},
  \citenamefont {Gerosa}, \citenamefont {Stein}, \citenamefont {Kidder},\ and\
  \citenamefont {Pfeiffer}}]{Varma:2019csw}%
  \BibitemOpen
  \bibfield  {author} {\bibinfo {author} {\bibfnamefont {V.}~\bibnamefont
  {Varma}}, \bibinfo {author} {\bibfnamefont {S.~E.}\ \bibnamefont {Field}},
  \bibinfo {author} {\bibfnamefont {M.~A.}\ \bibnamefont {Scheel}}, \bibinfo
  {author} {\bibfnamefont {J.}~\bibnamefont {Blackman}}, \bibinfo {author}
  {\bibfnamefont {D.}~\bibnamefont {Gerosa}}, \bibinfo {author} {\bibfnamefont
  {L.~C.}\ \bibnamefont {Stein}}, \bibinfo {author} {\bibfnamefont {L.~E.}\
  \bibnamefont {Kidder}}, \ and\ \bibinfo {author} {\bibfnamefont {H.~P.}\
  \bibnamefont {Pfeiffer}},\ }\href {\doibase 10.1103/PhysRevResearch.1.033015}
  {\bibfield  {journal} {\bibinfo  {journal} {Phys. Rev. Research.}\ }\textbf
  {\bibinfo {volume} {1}},\ \bibinfo {pages} {033015} (\bibinfo {year}
  {2019})},\ \Eprint {http://arxiv.org/abs/1905.09300} {arXiv:1905.09300
  [gr-qc]} \BibitemShut {NoStop}%
\bibitem [{\citenamefont {Gonzalez}\ \emph
  {et~al.}(2023{\natexlab{a}})\citenamefont {Gonzalez}, \citenamefont {Gamba},
  \citenamefont {Breschi}, \citenamefont {Zappa}, \citenamefont {Carullo},
  \citenamefont {Bernuzzi},\ and\ \citenamefont {Nagar}}]{Gonzalez:2022prs}%
  \BibitemOpen
  \bibfield  {author} {\bibinfo {author} {\bibfnamefont {A.}~\bibnamefont
  {Gonzalez}}, \bibinfo {author} {\bibfnamefont {R.}~\bibnamefont {Gamba}},
  \bibinfo {author} {\bibfnamefont {M.}~\bibnamefont {Breschi}}, \bibinfo
  {author} {\bibfnamefont {F.}~\bibnamefont {Zappa}}, \bibinfo {author}
  {\bibfnamefont {G.}~\bibnamefont {Carullo}}, \bibinfo {author} {\bibfnamefont
  {S.}~\bibnamefont {Bernuzzi}}, \ and\ \bibinfo {author} {\bibfnamefont
  {A.}~\bibnamefont {Nagar}},\ }\href {\doibase 10.1103/PhysRevD.107.084026}
  {\bibfield  {journal} {\bibinfo  {journal} {Phys. Rev. D}\ }\textbf {\bibinfo
  {volume} {107}},\ \bibinfo {pages} {084026} (\bibinfo {year}
  {2023}{\natexlab{a}})},\ \Eprint {http://arxiv.org/abs/2212.03909}
  {arXiv:2212.03909 [gr-qc]} \BibitemShut {NoStop}%
\bibitem [{\citenamefont {Baibhav}\ \emph {et~al.}(2021)\citenamefont
  {Baibhav}, \citenamefont {Berti}, \citenamefont {Gerosa}, \citenamefont
  {Mould},\ and\ \citenamefont {Wong}}]{Baibhav:2021qzw}%
  \BibitemOpen
  \bibfield  {author} {\bibinfo {author} {\bibfnamefont {V.}~\bibnamefont
  {Baibhav}}, \bibinfo {author} {\bibfnamefont {E.}~\bibnamefont {Berti}},
  \bibinfo {author} {\bibfnamefont {D.}~\bibnamefont {Gerosa}}, \bibinfo
  {author} {\bibfnamefont {M.}~\bibnamefont {Mould}}, \ and\ \bibinfo {author}
  {\bibfnamefont {K.~W.~K.}\ \bibnamefont {Wong}},\ }\href {\doibase
  10.1103/PhysRevD.104.084002} {\bibfield  {journal} {\bibinfo  {journal}
  {Phys. Rev. D}\ }\textbf {\bibinfo {volume} {104}},\ \bibinfo {pages}
  {084002} (\bibinfo {year} {2021})},\ \Eprint
  {http://arxiv.org/abs/2105.12140} {arXiv:2105.12140 [gr-qc]} \BibitemShut
  {NoStop}%
\bibitem [{\citenamefont {Barrera}\ and\ \citenamefont
  {Bartos}(2022)}]{Barrera:2022yfj}%
  \BibitemOpen
  \bibfield  {author} {\bibinfo {author} {\bibfnamefont {O.}~\bibnamefont
  {Barrera}}\ and\ \bibinfo {author} {\bibfnamefont {I.}~\bibnamefont
  {Bartos}},\ }\href {\doibase 10.3847/2041-8213/ac5f47} {\bibfield  {journal}
  {\bibinfo  {journal} {Astrophys. J. Lett.}\ }\textbf {\bibinfo {volume}
  {929}},\ \bibinfo {pages} {L1} (\bibinfo {year} {2022})},\ \Eprint
  {http://arxiv.org/abs/2201.09943} {arXiv:2201.09943 [astro-ph.HE]}
  \BibitemShut {NoStop}%
\bibitem [{\citenamefont {Barrera}\ and\ \citenamefont
  {Bartos}(2024)}]{Barrera:2023qde}%
  \BibitemOpen
  \bibfield  {author} {\bibinfo {author} {\bibfnamefont {O.}~\bibnamefont
  {Barrera}}\ and\ \bibinfo {author} {\bibfnamefont {I.}~\bibnamefont
  {Bartos}},\ }\href {\doibase 10.1016/j.astropartphys.2023.102919} {\bibfield
  {journal} {\bibinfo  {journal} {Astropart. Phys.}\ }\textbf {\bibinfo
  {volume} {156}},\ \bibinfo {pages} {102919} (\bibinfo {year} {2024})},\
  \Eprint {http://arxiv.org/abs/2307.11856} {arXiv:2307.11856 [astro-ph.HE]}
  \BibitemShut {NoStop}%
\bibitem [{\citenamefont {Ara\'ujo-\'Alvarez}\ \emph
  {et~al.}(2024)\citenamefont {Ara\'ujo-\'Alvarez}, \citenamefont {Wong},
  \citenamefont {Liu},\ and\ \citenamefont {Bustillo}}]{Alvarez:2024dpd}%
  \BibitemOpen
  \bibfield  {author} {\bibinfo {author} {\bibfnamefont {C.}~\bibnamefont
  {Ara\'ujo-\'Alvarez}}, \bibinfo {author} {\bibfnamefont {H.~W.~Y.}\
  \bibnamefont {Wong}}, \bibinfo {author} {\bibfnamefont {A.}~\bibnamefont
  {Liu}}, \ and\ \bibinfo {author} {\bibfnamefont {J.~C.}\ \bibnamefont
  {Bustillo}},\ }\href {\doibase 10.3847/1538-4357/ad90a9} {\bibfield
  {journal} {\bibinfo  {journal} {Astrophys. J.}\ }\textbf {\bibinfo {volume}
  {977}},\ \bibinfo {pages} {220} (\bibinfo {year} {2024})},\ \Eprint
  {http://arxiv.org/abs/2404.00720} {arXiv:2404.00720 [astro-ph.HE]}
  \BibitemShut {NoStop}%
\bibitem [{\citenamefont {Lorimer}(2008)}]{Lorimer:2008se}%
  \BibitemOpen
  \bibfield  {author} {\bibinfo {author} {\bibfnamefont {D.~R.}\ \bibnamefont
  {Lorimer}},\ }\href {\doibase 10.12942/lrr-2008-8} {\bibfield  {journal}
  {\bibinfo  {journal} {Living Rev. Rel.}\ }\textbf {\bibinfo {volume} {11}},\
  \bibinfo {pages} {8} (\bibinfo {year} {2008})},\ \Eprint
  {http://arxiv.org/abs/0811.0762} {arXiv:0811.0762 [astro-ph]} \BibitemShut
  {NoStop}%
\bibitem [{\citenamefont {Chakrabarty}(2008)}]{Chakrabarty:2008gz}%
  \BibitemOpen
  \bibfield  {author} {\bibinfo {author} {\bibfnamefont {D.}~\bibnamefont
  {Chakrabarty}},\ }\href {\doibase 10.1063/1.3031208} {\bibfield  {journal}
  {\bibinfo  {journal} {AIP Conf. Proc.}\ }\textbf {\bibinfo {volume} {1068}},\
  \bibinfo {pages} {67} (\bibinfo {year} {2008})},\ \Eprint
  {http://arxiv.org/abs/0809.4031} {arXiv:0809.4031 [astro-ph]} \BibitemShut
  {NoStop}%
\bibitem [{\citenamefont {Burgay}\ \emph {et~al.}(2003)\citenamefont {Burgay}
  \emph {et~al.}}]{Burgay:2003jj}%
  \BibitemOpen
  \bibfield  {author} {\bibinfo {author} {\bibfnamefont {M.}~\bibnamefont
  {Burgay}} \emph {et~al.},\ }\href {\doibase 10.1038/nature02124} {\bibfield
  {journal} {\bibinfo  {journal} {Nature}\ }\textbf {\bibinfo {volume} {426}},\
  \bibinfo {pages} {531} (\bibinfo {year} {2003})},\ \Eprint
  {http://arxiv.org/abs/astro-ph/0312071} {arXiv:astro-ph/0312071} \BibitemShut
  {NoStop}%
\bibitem [{\citenamefont {Pratten}\ \emph {et~al.}(2021)\citenamefont {Pratten}
  \emph {et~al.}}]{IMRPhenomXPHM}%
  \BibitemOpen
  \bibfield  {author} {\bibinfo {author} {\bibfnamefont {G.}~\bibnamefont
  {Pratten}} \emph {et~al.},\ }\href {\doibase 10.1103/PhysRevD.103.104056}
  {\bibfield  {journal} {\bibinfo  {journal} {Phys. Rev. D}\ }\textbf {\bibinfo
  {volume} {103}},\ \bibinfo {pages} {104056} (\bibinfo {year} {2021})},\
  \Eprint {http://arxiv.org/abs/2004.06503} {arXiv:2004.06503 [gr-qc]}
  \BibitemShut {NoStop}%
\bibitem [{\citenamefont {Ramos-Buades}\ \emph {et~al.}(2023)\citenamefont
  {Ramos-Buades}, \citenamefont {Buonanno}, \citenamefont {Estell\'es},
  \citenamefont {Khalil}, \citenamefont {Mihaylov}, \citenamefont {Ossokine},
  \citenamefont {Pompili},\ and\ \citenamefont {Shiferaw}}]{SEOBNRv5PHM}%
  \BibitemOpen
  \bibfield  {author} {\bibinfo {author} {\bibfnamefont {A.}~\bibnamefont
  {Ramos-Buades}}, \bibinfo {author} {\bibfnamefont {A.}~\bibnamefont
  {Buonanno}}, \bibinfo {author} {\bibfnamefont {H.}~\bibnamefont
  {Estell\'es}}, \bibinfo {author} {\bibfnamefont {M.}~\bibnamefont {Khalil}},
  \bibinfo {author} {\bibfnamefont {D.~P.}\ \bibnamefont {Mihaylov}}, \bibinfo
  {author} {\bibfnamefont {S.}~\bibnamefont {Ossokine}}, \bibinfo {author}
  {\bibfnamefont {L.}~\bibnamefont {Pompili}}, \ and\ \bibinfo {author}
  {\bibfnamefont {M.}~\bibnamefont {Shiferaw}},\ }\href {\doibase
  10.1103/PhysRevD.108.124037} {\bibfield  {journal} {\bibinfo  {journal}
  {Phys. Rev. D}\ }\textbf {\bibinfo {volume} {108}},\ \bibinfo {pages}
  {124037} (\bibinfo {year} {2023})},\ \Eprint
  {http://arxiv.org/abs/2303.18046} {arXiv:2303.18046 [gr-qc]} \BibitemShut
  {NoStop}%
\bibitem [{\citenamefont {Ashton}\ \emph {et~al.}(2019)\citenamefont {Ashton}
  \emph {et~al.}}]{bilby_paper}%
  \BibitemOpen
  \bibfield  {author} {\bibinfo {author} {\bibfnamefont {G.}~\bibnamefont
  {Ashton}} \emph {et~al.},\ }\href {\doibase 10.3847/1538-4365/ab06fc}
  {\bibfield  {journal} {\bibinfo  {journal} {Astrophys. J. Suppl.}\ }\textbf
  {\bibinfo {volume} {241}},\ \bibinfo {pages} {27} (\bibinfo {year} {2019})},\
  \Eprint {http://arxiv.org/abs/1811.02042} {arXiv:1811.02042 [astro-ph.IM]}
  \BibitemShut {NoStop}%
\bibitem [{\citenamefont {{Speagle}}(2020)}]{dynesty}%
  \BibitemOpen
  \bibfield  {author} {\bibinfo {author} {\bibfnamefont {J.~S.}\ \bibnamefont
  {{Speagle}}},\ }\href {\doibase 10.1093/mnras/staa278} {\bibfield  {journal}
  {\bibinfo  {journal} {Mon. Not. Roy. Astron. Soc.}\ }\textbf {\bibinfo
  {volume} {493}},\ \bibinfo {pages} {3132} (\bibinfo {year} {2020})},\ \Eprint
  {http://arxiv.org/abs/1904.02180} {arXiv:1904.02180 [astro-ph.IM]}
  \BibitemShut {NoStop}%
\bibitem [{\citenamefont {{Skilling}}(2004)}]{Skilling}%
  \BibitemOpen
  \bibfield  {author} {\bibinfo {author} {\bibfnamefont {J.}~\bibnamefont
  {{Skilling}}},\ }in\ \href {\doibase 10.1063/1.1835238} {\emph {\bibinfo
  {booktitle} {Bayesian Inference and Maximum Entropy Methods in Science and
  Engineering: 24th International Workshop on Bayesian Inference and Maximum
  Entropy Methods in Science and Engineering}}},\ \bibinfo {series} {American
  Institute of Physics Conference Series}, Vol.\ \bibinfo {volume} {735},\
  \bibinfo {editor} {edited by\ \bibinfo {editor} {\bibfnamefont
  {R.}~\bibnamefont {{Fischer}}}, \bibinfo {editor} {\bibfnamefont
  {R.}~\bibnamefont {{Preuss}}}, \ and\ \bibinfo {editor} {\bibfnamefont
  {U.~V.}\ \bibnamefont {{Toussaint}}}}\ (\bibinfo {year} {2004})\ pp.\
  \bibinfo {pages} {395--405}\BibitemShut {NoStop}%
\bibitem [{\citenamefont {{Valentim}}\ \emph {et~al.}(2011)\citenamefont
  {{Valentim}}, \citenamefont {{Rangel}},\ and\ \citenamefont
  {{Horvath}}}]{Valentim2011MNRAS}%
  \BibitemOpen
  \bibfield  {author} {\bibinfo {author} {\bibfnamefont {R.}~\bibnamefont
  {{Valentim}}}, \bibinfo {author} {\bibfnamefont {E.}~\bibnamefont
  {{Rangel}}}, \ and\ \bibinfo {author} {\bibfnamefont {J.~E.}\ \bibnamefont
  {{Horvath}}},\ }\href {\doibase 10.1111/j.1365-2966.2011.18477.x} {\bibfield
  {journal} {\bibinfo  {journal} {"Mon. Not. Roy. Astron. Soc."}\ }\textbf
  {\bibinfo {volume} {414}},\ \bibinfo {pages} {1427} (\bibinfo {year}
  {2011})},\ \Eprint {http://arxiv.org/abs/1101.4872} {arXiv:1101.4872
  [astro-ph.SR]} \BibitemShut {NoStop}%
\bibitem [{\citenamefont {Ozel}\ \emph {et~al.}(2012)\citenamefont {Ozel},
  \citenamefont {Psaltis}, \citenamefont {Narayan},\ and\ \citenamefont
  {Villarreal}}]{Ozel:2012ax}%
  \BibitemOpen
  \bibfield  {author} {\bibinfo {author} {\bibfnamefont {F.}~\bibnamefont
  {Ozel}}, \bibinfo {author} {\bibfnamefont {D.}~\bibnamefont {Psaltis}},
  \bibinfo {author} {\bibfnamefont {R.}~\bibnamefont {Narayan}}, \ and\
  \bibinfo {author} {\bibfnamefont {A.~S.}\ \bibnamefont {Villarreal}},\ }\href
  {\doibase 10.1088/0004-637X/757/1/55} {\bibfield  {journal} {\bibinfo
  {journal} {Astrophys. J.}\ }\textbf {\bibinfo {volume} {757}},\ \bibinfo
  {pages} {55} (\bibinfo {year} {2012})},\ \Eprint
  {http://arxiv.org/abs/1201.1006} {arXiv:1201.1006 [astro-ph.HE]} \BibitemShut
  {NoStop}%
\bibitem [{\citenamefont {Kiziltan}\ \emph {et~al.}(2013)\citenamefont
  {Kiziltan}, \citenamefont {Kottas}, \citenamefont {De~Yoreo},\ and\
  \citenamefont {Thorsett}}]{Kiziltan:2013oja}%
  \BibitemOpen
  \bibfield  {author} {\bibinfo {author} {\bibfnamefont {B.}~\bibnamefont
  {Kiziltan}}, \bibinfo {author} {\bibfnamefont {A.}~\bibnamefont {Kottas}},
  \bibinfo {author} {\bibfnamefont {M.}~\bibnamefont {De~Yoreo}}, \ and\
  \bibinfo {author} {\bibfnamefont {S.~E.}\ \bibnamefont {Thorsett}},\ }\href
  {\doibase 10.1088/0004-637X/778/1/66} {\bibfield  {journal} {\bibinfo
  {journal} {Astrophys. J.}\ }\textbf {\bibinfo {volume} {778}},\ \bibinfo
  {pages} {66} (\bibinfo {year} {2013})},\ \Eprint
  {http://arxiv.org/abs/1309.6635} {arXiv:1309.6635 [astro-ph.SR]} \BibitemShut
  {NoStop}%
\bibitem [{\citenamefont {Antoniadis}\ \emph {et~al.}(2016)\citenamefont
  {Antoniadis}, \citenamefont {Tauris}, \citenamefont {Ozel}, \citenamefont
  {Barr}, \citenamefont {Champion},\ and\ \citenamefont
  {Freire}}]{Antoniadis:2016hxz}%
  \BibitemOpen
  \bibfield  {author} {\bibinfo {author} {\bibfnamefont {J.}~\bibnamefont
  {Antoniadis}}, \bibinfo {author} {\bibfnamefont {T.~M.}\ \bibnamefont
  {Tauris}}, \bibinfo {author} {\bibfnamefont {F.}~\bibnamefont {Ozel}},
  \bibinfo {author} {\bibfnamefont {E.}~\bibnamefont {Barr}}, \bibinfo {author}
  {\bibfnamefont {D.~J.}\ \bibnamefont {Champion}}, \ and\ \bibinfo {author}
  {\bibfnamefont {P.~C.~C.}\ \bibnamefont {Freire}},\ }\href@noop {} {\
  (\bibinfo {year} {2016})},\ \Eprint {http://arxiv.org/abs/1605.01665}
  {arXiv:1605.01665 [astro-ph.HE]} \BibitemShut {NoStop}%
\bibitem [{\citenamefont {\"Ozel}\ and\ \citenamefont
  {Freire}(2016)}]{Ozel:2016oaf}%
  \BibitemOpen
  \bibfield  {author} {\bibinfo {author} {\bibfnamefont {F.}~\bibnamefont
  {\"Ozel}}\ and\ \bibinfo {author} {\bibfnamefont {P.}~\bibnamefont
  {Freire}},\ }\href {\doibase 10.1146/annurev-astro-081915-023322} {\bibfield
  {journal} {\bibinfo  {journal} {Ann. Rev. Astron. Astrophys.}\ }\textbf
  {\bibinfo {volume} {54}},\ \bibinfo {pages} {401} (\bibinfo {year} {2016})},\
  \Eprint {http://arxiv.org/abs/1603.02698} {arXiv:1603.02698 [astro-ph.HE]}
  \BibitemShut {NoStop}%
\bibitem [{\citenamefont {Alsing}\ \emph {et~al.}(2018)\citenamefont {Alsing},
  \citenamefont {Silva},\ and\ \citenamefont {Berti}}]{Alsing:2017bbc}%
  \BibitemOpen
  \bibfield  {author} {\bibinfo {author} {\bibfnamefont {J.}~\bibnamefont
  {Alsing}}, \bibinfo {author} {\bibfnamefont {H.~O.}\ \bibnamefont {Silva}}, \
  and\ \bibinfo {author} {\bibfnamefont {E.}~\bibnamefont {Berti}},\ }\href
  {\doibase 10.1093/mnras/sty1065} {\bibfield  {journal} {\bibinfo  {journal}
  {Mon. Not. Roy. Astron. Soc.}\ }\textbf {\bibinfo {volume} {478}},\ \bibinfo
  {pages} {1377} (\bibinfo {year} {2018})},\ \Eprint
  {http://arxiv.org/abs/1709.07889} {arXiv:1709.07889 [astro-ph.HE]}
  \BibitemShut {NoStop}%
\bibitem [{\citenamefont {Farrow}\ \emph {et~al.}(2019)\citenamefont {Farrow},
  \citenamefont {Zhu},\ and\ \citenamefont {Thrane}}]{Farrow:2019xnc}%
  \BibitemOpen
  \bibfield  {author} {\bibinfo {author} {\bibfnamefont {N.}~\bibnamefont
  {Farrow}}, \bibinfo {author} {\bibfnamefont {X.-J.}\ \bibnamefont {Zhu}}, \
  and\ \bibinfo {author} {\bibfnamefont {E.}~\bibnamefont {Thrane}},\ }\href
  {\doibase 10.3847/1538-4357/ab12e3} {\bibfield  {journal} {\bibinfo
  {journal} {Astrophys. J.}\ }\textbf {\bibinfo {volume} {876}},\ \bibinfo
  {pages} {18} (\bibinfo {year} {2019})},\ \Eprint
  {http://arxiv.org/abs/1902.03300} {arXiv:1902.03300 [astro-ph.HE]}
  \BibitemShut {NoStop}%
\bibitem [{\citenamefont {You}\ \emph {et~al.}(2024)\citenamefont {You} \emph
  {et~al.}}]{You:2024bmk}%
  \BibitemOpen
  \bibfield  {author} {\bibinfo {author} {\bibfnamefont {Z.-Q.}\ \bibnamefont
  {You}} \emph {et~al.},\ }\href {\doibase 10.1038/s41550-025-02487-w} {\
  (\bibinfo {year} {2024}),\ 10.1038/s41550-025-02487-w},\ \Eprint
  {http://arxiv.org/abs/2412.05524} {arXiv:2412.05524 [astro-ph.HE]}
  \BibitemShut {NoStop}%
\bibitem [{\citenamefont {Martinez}\ \emph {et~al.}(2015)\citenamefont
  {Martinez}, \citenamefont {Stovall}, \citenamefont {Freire}, \citenamefont
  {Deneva}, \citenamefont {Jenet}, \citenamefont {McLaughlin}, \citenamefont
  {Bagchi}, \citenamefont {Bates},\ and\ \citenamefont
  {Ridolfi}}]{Martinez:2015mya}%
  \BibitemOpen
  \bibfield  {author} {\bibinfo {author} {\bibfnamefont {J.~G.}\ \bibnamefont
  {Martinez}}, \bibinfo {author} {\bibfnamefont {K.}~\bibnamefont {Stovall}},
  \bibinfo {author} {\bibfnamefont {P.~C.~C.}\ \bibnamefont {Freire}}, \bibinfo
  {author} {\bibfnamefont {J.~S.}\ \bibnamefont {Deneva}}, \bibinfo {author}
  {\bibfnamefont {F.~A.}\ \bibnamefont {Jenet}}, \bibinfo {author}
  {\bibfnamefont {M.~A.}\ \bibnamefont {McLaughlin}}, \bibinfo {author}
  {\bibfnamefont {M.}~\bibnamefont {Bagchi}}, \bibinfo {author} {\bibfnamefont
  {S.~D.}\ \bibnamefont {Bates}}, \ and\ \bibinfo {author} {\bibfnamefont
  {A.}~\bibnamefont {Ridolfi}},\ }\href {\doibase 10.1088/0004-637X/812/2/143}
  {\bibfield  {journal} {\bibinfo  {journal} {Astrophys. J.}\ }\textbf
  {\bibinfo {volume} {812}},\ \bibinfo {pages} {143} (\bibinfo {year}
  {2015})},\ \Eprint {http://arxiv.org/abs/1509.08805} {arXiv:1509.08805
  [astro-ph.HE]} \BibitemShut {NoStop}%
\bibitem [{\citenamefont {Tauris}\ and\ \citenamefont
  {Janka}(2019)}]{Tauris:2019sho}%
  \BibitemOpen
  \bibfield  {author} {\bibinfo {author} {\bibfnamefont {T.~M.}\ \bibnamefont
  {Tauris}}\ and\ \bibinfo {author} {\bibfnamefont {H.-T.}\ \bibnamefont
  {Janka}},\ }\href {\doibase 10.3847/2041-8213/ab5642} {\bibfield  {journal}
  {\bibinfo  {journal} {Astrophys. J. Lett.}\ }\textbf {\bibinfo {volume}
  {886}},\ \bibinfo {pages} {L20} (\bibinfo {year} {2019})},\ \Eprint
  {http://arxiv.org/abs/1909.12318} {arXiv:1909.12318 [astro-ph.SR]}
  \BibitemShut {NoStop}%
\bibitem [{\citenamefont {Cromartie}\ \emph {et~al.}(2019)\citenamefont
  {Cromartie} \emph {et~al.}}]{NANOGrav:2019jur}%
  \BibitemOpen
  \bibfield  {author} {\bibinfo {author} {\bibfnamefont {H.~T.}\ \bibnamefont
  {Cromartie}} \emph {et~al.} (\bibinfo {collaboration} {NANOGrav}),\ }\href
  {\doibase 10.1038/s41550-019-0880-2} {\bibfield  {journal} {\bibinfo
  {journal} {Nature Astron.}\ }\textbf {\bibinfo {volume} {4}},\ \bibinfo
  {pages} {72} (\bibinfo {year} {2019})},\ \Eprint
  {http://arxiv.org/abs/1904.06759} {arXiv:1904.06759 [astro-ph.HE]}
  \BibitemShut {NoStop}%
\bibitem [{\citenamefont {Antoniadis}\ \emph {et~al.}(2013)\citenamefont
  {Antoniadis} \emph {et~al.}}]{Antoniadis:2013pzd}%
  \BibitemOpen
  \bibfield  {author} {\bibinfo {author} {\bibfnamefont {J.}~\bibnamefont
  {Antoniadis}} \emph {et~al.},\ }\href {\doibase 10.1126/science.1233232}
  {\bibfield  {journal} {\bibinfo  {journal} {Science}\ }\textbf {\bibinfo
  {volume} {340}},\ \bibinfo {pages} {6131} (\bibinfo {year} {2013})},\ \Eprint
  {http://arxiv.org/abs/1304.6875} {arXiv:1304.6875 [astro-ph.HE]} \BibitemShut
  {NoStop}%
\bibitem [{\citenamefont {Lawrence}\ \emph {et~al.}(2015)\citenamefont
  {Lawrence}, \citenamefont {Tervala}, \citenamefont {Bedaque},\ and\
  \citenamefont {Miller}}]{Lawrence:2015oka}%
  \BibitemOpen
  \bibfield  {author} {\bibinfo {author} {\bibfnamefont {S.}~\bibnamefont
  {Lawrence}}, \bibinfo {author} {\bibfnamefont {J.~G.}\ \bibnamefont
  {Tervala}}, \bibinfo {author} {\bibfnamefont {P.~F.}\ \bibnamefont
  {Bedaque}}, \ and\ \bibinfo {author} {\bibfnamefont {M.~C.}\ \bibnamefont
  {Miller}},\ }\href {\doibase 10.1088/0004-637X/808/2/186} {\bibfield
  {journal} {\bibinfo  {journal} {Astrophys. J.}\ }\textbf {\bibinfo {volume}
  {808}},\ \bibinfo {pages} {186} (\bibinfo {year} {2015})},\ \Eprint
  {http://arxiv.org/abs/1505.00231} {arXiv:1505.00231 [astro-ph.HE]}
  \BibitemShut {NoStop}%
\bibitem [{\citenamefont {Fryer}\ \emph {et~al.}(2015)\citenamefont {Fryer},
  \citenamefont {Belczynski}, \citenamefont {Ramirez-Ruiz}, \citenamefont
  {Rosswog}, \citenamefont {Shen},\ and\ \citenamefont
  {Steiner}}]{Fryer:2015uia}%
  \BibitemOpen
  \bibfield  {author} {\bibinfo {author} {\bibfnamefont {C.~L.}\ \bibnamefont
  {Fryer}}, \bibinfo {author} {\bibfnamefont {K.}~\bibnamefont {Belczynski}},
  \bibinfo {author} {\bibfnamefont {E.}~\bibnamefont {Ramirez-Ruiz}}, \bibinfo
  {author} {\bibfnamefont {S.}~\bibnamefont {Rosswog}}, \bibinfo {author}
  {\bibfnamefont {G.}~\bibnamefont {Shen}}, \ and\ \bibinfo {author}
  {\bibfnamefont {A.~W.}\ \bibnamefont {Steiner}},\ }\href {\doibase
  10.1088/0004-637X/812/1/24} {\bibfield  {journal} {\bibinfo  {journal}
  {Astrophys. J.}\ }\textbf {\bibinfo {volume} {812}},\ \bibinfo {pages} {24}
  (\bibinfo {year} {2015})},\ \Eprint {http://arxiv.org/abs/1504.07605}
  {arXiv:1504.07605 [astro-ph.HE]} \BibitemShut {NoStop}%
\bibitem [{\citenamefont {Margalit}\ and\ \citenamefont
  {Metzger}(2017)}]{Margalit:2017dij}%
  \BibitemOpen
  \bibfield  {author} {\bibinfo {author} {\bibfnamefont {B.}~\bibnamefont
  {Margalit}}\ and\ \bibinfo {author} {\bibfnamefont {B.~D.}\ \bibnamefont
  {Metzger}},\ }\href {\doibase 10.3847/2041-8213/aa991c} {\bibfield  {journal}
  {\bibinfo  {journal} {Astrophys. J. Lett.}\ }\textbf {\bibinfo {volume}
  {850}},\ \bibinfo {pages} {L19} (\bibinfo {year} {2017})},\ \Eprint
  {http://arxiv.org/abs/1710.05938} {arXiv:1710.05938 [astro-ph.HE]}
  \BibitemShut {NoStop}%
\bibitem [{\citenamefont {Rezzolla}\ \emph {et~al.}(2018)\citenamefont
  {Rezzolla}, \citenamefont {Most},\ and\ \citenamefont
  {Weih}}]{Rezzolla:2017aly}%
  \BibitemOpen
  \bibfield  {author} {\bibinfo {author} {\bibfnamefont {L.}~\bibnamefont
  {Rezzolla}}, \bibinfo {author} {\bibfnamefont {E.~R.}\ \bibnamefont {Most}},
  \ and\ \bibinfo {author} {\bibfnamefont {L.~R.}\ \bibnamefont {Weih}},\
  }\href {\doibase 10.3847/2041-8213/aaa401} {\bibfield  {journal} {\bibinfo
  {journal} {Astrophys. J. Lett.}\ }\textbf {\bibinfo {volume} {852}},\
  \bibinfo {pages} {L25} (\bibinfo {year} {2018})},\ \Eprint
  {http://arxiv.org/abs/1711.00314} {arXiv:1711.00314 [astro-ph.HE]}
  \BibitemShut {NoStop}%
\bibitem [{\citenamefont {Ruiz}\ \emph {et~al.}(2018)\citenamefont {Ruiz},
  \citenamefont {Shapiro},\ and\ \citenamefont {Tsokaros}}]{Ruiz:2017due}%
  \BibitemOpen
  \bibfield  {author} {\bibinfo {author} {\bibfnamefont {M.}~\bibnamefont
  {Ruiz}}, \bibinfo {author} {\bibfnamefont {S.~L.}\ \bibnamefont {Shapiro}}, \
  and\ \bibinfo {author} {\bibfnamefont {A.}~\bibnamefont {Tsokaros}},\ }\href
  {\doibase 10.1103/PhysRevD.97.021501} {\bibfield  {journal} {\bibinfo
  {journal} {Phys. Rev. D}\ }\textbf {\bibinfo {volume} {97}},\ \bibinfo
  {pages} {021501} (\bibinfo {year} {2018})},\ \Eprint
  {http://arxiv.org/abs/1711.00473} {arXiv:1711.00473 [astro-ph.HE]}
  \BibitemShut {NoStop}%
\bibitem [{\citenamefont {Shao}\ \emph {et~al.}(2020)\citenamefont {Shao},
  \citenamefont {Tang}, \citenamefont {Jiang},\ and\ \citenamefont
  {Fan}}]{Shao:2020bzt}%
  \BibitemOpen
  \bibfield  {author} {\bibinfo {author} {\bibfnamefont {D.-S.}\ \bibnamefont
  {Shao}}, \bibinfo {author} {\bibfnamefont {S.-P.}\ \bibnamefont {Tang}},
  \bibinfo {author} {\bibfnamefont {J.-L.}\ \bibnamefont {Jiang}}, \ and\
  \bibinfo {author} {\bibfnamefont {Y.-Z.}\ \bibnamefont {Fan}},\ }\href
  {\doibase 10.1103/PhysRevD.102.063006} {\bibfield  {journal} {\bibinfo
  {journal} {Phys. Rev. D}\ }\textbf {\bibinfo {volume} {102}},\ \bibinfo
  {pages} {063006} (\bibinfo {year} {2020})},\ \Eprint
  {http://arxiv.org/abs/2009.04275} {arXiv:2009.04275 [astro-ph.HE]}
  \BibitemShut {NoStop}%
\bibitem [{\citenamefont {Shibata}\ \emph {et~al.}(2019)\citenamefont
  {Shibata}, \citenamefont {Zhou}, \citenamefont {Kiuchi},\ and\ \citenamefont
  {Fujibayashi}}]{Shibata:2019ctb}%
  \BibitemOpen
  \bibfield  {author} {\bibinfo {author} {\bibfnamefont {M.}~\bibnamefont
  {Shibata}}, \bibinfo {author} {\bibfnamefont {E.}~\bibnamefont {Zhou}},
  \bibinfo {author} {\bibfnamefont {K.}~\bibnamefont {Kiuchi}}, \ and\ \bibinfo
  {author} {\bibfnamefont {S.}~\bibnamefont {Fujibayashi}},\ }\href {\doibase
  10.1103/PhysRevD.100.023015} {\bibfield  {journal} {\bibinfo  {journal}
  {Phys. Rev. D}\ }\textbf {\bibinfo {volume} {100}},\ \bibinfo {pages}
  {023015} (\bibinfo {year} {2019})},\ \Eprint
  {http://arxiv.org/abs/1905.03656} {arXiv:1905.03656 [astro-ph.HE]}
  \BibitemShut {NoStop}%
\bibitem [{\citenamefont {Nathanail}\ \emph {et~al.}(2021)\citenamefont
  {Nathanail}, \citenamefont {Most},\ and\ \citenamefont
  {Rezzolla}}]{Nathanail:2021tay}%
  \BibitemOpen
  \bibfield  {author} {\bibinfo {author} {\bibfnamefont {A.}~\bibnamefont
  {Nathanail}}, \bibinfo {author} {\bibfnamefont {E.~R.}\ \bibnamefont {Most}},
  \ and\ \bibinfo {author} {\bibfnamefont {L.}~\bibnamefont {Rezzolla}},\
  }\href {\doibase 10.3847/2041-8213/abdfc6} {\bibfield  {journal} {\bibinfo
  {journal} {Astrophys. J. Lett.}\ }\textbf {\bibinfo {volume} {908}},\
  \bibinfo {pages} {L28} (\bibinfo {year} {2021})},\ \Eprint
  {http://arxiv.org/abs/2101.01735} {arXiv:2101.01735 [astro-ph.HE]}
  \BibitemShut {NoStop}%
\bibitem [{\citenamefont {Kashyap}\ \emph {et~al.}(2022)\citenamefont {Kashyap}
  \emph {et~al.}}]{Kashyap:2021wzs}%
  \BibitemOpen
  \bibfield  {author} {\bibinfo {author} {\bibfnamefont {R.}~\bibnamefont
  {Kashyap}} \emph {et~al.},\ }\href {\doibase 10.1103/PhysRevD.105.103022}
  {\bibfield  {journal} {\bibinfo  {journal} {Phys. Rev. D}\ }\textbf {\bibinfo
  {volume} {105}},\ \bibinfo {pages} {103022} (\bibinfo {year} {2022})},\
  \Eprint {http://arxiv.org/abs/2111.05183} {arXiv:2111.05183 [astro-ph.HE]}
  \BibitemShut {NoStop}%
\bibitem [{\citenamefont {Godzieba}\ \emph {et~al.}(2021)\citenamefont
  {Godzieba}, \citenamefont {Radice},\ and\ \citenamefont
  {Bernuzzi}}]{Godzieba:2020tjn}%
  \BibitemOpen
  \bibfield  {author} {\bibinfo {author} {\bibfnamefont {D.~A.}\ \bibnamefont
  {Godzieba}}, \bibinfo {author} {\bibfnamefont {D.}~\bibnamefont {Radice}}, \
  and\ \bibinfo {author} {\bibfnamefont {S.}~\bibnamefont {Bernuzzi}},\ }\href
  {\doibase 10.3847/1538-4357/abd4dd} {\bibfield  {journal} {\bibinfo
  {journal} {Astrophys. J.}\ }\textbf {\bibinfo {volume} {908}},\ \bibinfo
  {pages} {122} (\bibinfo {year} {2021})},\ \Eprint
  {http://arxiv.org/abs/2007.10999} {arXiv:2007.10999 [astro-ph.HE]}
  \BibitemShut {NoStop}%
\bibitem [{\citenamefont {Suwa}\ \emph {et~al.}(2018)\citenamefont {Suwa},
  \citenamefont {Yoshida}, \citenamefont {Shibata}, \citenamefont {Umeda},\
  and\ \citenamefont {Takahashi}}]{Suwa:2018uni}%
  \BibitemOpen
  \bibfield  {author} {\bibinfo {author} {\bibfnamefont {Y.}~\bibnamefont
  {Suwa}}, \bibinfo {author} {\bibfnamefont {T.}~\bibnamefont {Yoshida}},
  \bibinfo {author} {\bibfnamefont {M.}~\bibnamefont {Shibata}}, \bibinfo
  {author} {\bibfnamefont {H.}~\bibnamefont {Umeda}}, \ and\ \bibinfo {author}
  {\bibfnamefont {K.}~\bibnamefont {Takahashi}},\ }\href {\doibase
  10.1093/mnras/sty2460} {\bibfield  {journal} {\bibinfo  {journal} {Mon. Not.
  Roy. Astron. Soc.}\ }\textbf {\bibinfo {volume} {481}},\ \bibinfo {pages}
  {3305} (\bibinfo {year} {2018})},\ \Eprint {http://arxiv.org/abs/1808.02328}
  {arXiv:1808.02328 [astro-ph.HE]} \BibitemShut {NoStop}%
\bibitem [{\citenamefont {M\"uller}\ \emph {et~al.}(2025)\citenamefont
  {M\"uller}, \citenamefont {Heger},\ and\ \citenamefont
  {Powell}}]{Muller:2024aod}%
  \BibitemOpen
  \bibfield  {author} {\bibinfo {author} {\bibfnamefont {B.}~\bibnamefont
  {M\"uller}}, \bibinfo {author} {\bibfnamefont {A.}~\bibnamefont {Heger}}, \
  and\ \bibinfo {author} {\bibfnamefont {J.}~\bibnamefont {Powell}},\ }\href
  {\doibase 10.1103/PhysRevLett.134.071403} {\bibfield  {journal} {\bibinfo
  {journal} {Phys. Rev. Lett.}\ }\textbf {\bibinfo {volume} {134}},\ \bibinfo
  {pages} {071403} (\bibinfo {year} {2025})},\ \Eprint
  {http://arxiv.org/abs/2407.08407} {arXiv:2407.08407 [astro-ph.HE]}
  \BibitemShut {NoStop}%
\bibitem [{\citenamefont {Abbott}\ \emph
  {et~al.}(2021{\natexlab{c}})\citenamefont {Abbott} \emph
  {et~al.}}]{GW200105_GW200115}%
  \BibitemOpen
  \bibfield  {author} {\bibinfo {author} {\bibfnamefont {R.}~\bibnamefont
  {Abbott}} \emph {et~al.} (\bibinfo {collaboration} {LIGO Scientific, KAGRA,
  VIRGO}),\ }\href {\doibase 10.3847/2041-8213/ac082e} {\bibfield  {journal}
  {\bibinfo  {journal} {Astrophys. J. Lett.}\ }\textbf {\bibinfo {volume}
  {915}},\ \bibinfo {pages} {L5} (\bibinfo {year} {2021}{\natexlab{c}})},\
  \Eprint {http://arxiv.org/abs/2106.15163} {arXiv:2106.15163 [astro-ph.HE]}
  \BibitemShut {NoStop}%
\bibitem [{\citenamefont {Abbott}\ \emph {et~al.}(2023)\citenamefont {Abbott}
  \emph {et~al.}}]{GWTC-3-pop}%
  \BibitemOpen
  \bibfield  {author} {\bibinfo {author} {\bibfnamefont {R.}~\bibnamefont
  {Abbott}} \emph {et~al.} (\bibinfo {collaboration} {KAGRA, VIRGO, LIGO
  Scientific}),\ }\href {\doibase 10.1103/PhysRevX.13.011048} {\bibfield
  {journal} {\bibinfo  {journal} {Phys. Rev. X}\ }\textbf {\bibinfo {volume}
  {13}},\ \bibinfo {pages} {011048} (\bibinfo {year} {2023})},\ \Eprint
  {http://arxiv.org/abs/2111.03634} {arXiv:2111.03634 [astro-ph.HE]}
  \BibitemShut {NoStop}%
\bibitem [{\citenamefont {Landry}\ and\ \citenamefont
  {Read}(2021)}]{Landry:2021hvl}%
  \BibitemOpen
  \bibfield  {author} {\bibinfo {author} {\bibfnamefont {P.}~\bibnamefont
  {Landry}}\ and\ \bibinfo {author} {\bibfnamefont {J.~S.}\ \bibnamefont
  {Read}},\ }\href {\doibase 10.3847/2041-8213/ac2f3e} {\bibfield  {journal}
  {\bibinfo  {journal} {Astrophys. J. Lett.}\ }\textbf {\bibinfo {volume}
  {921}},\ \bibinfo {pages} {L25} (\bibinfo {year} {2021})},\ \Eprint
  {http://arxiv.org/abs/2107.04559} {arXiv:2107.04559 [astro-ph.HE]}
  \BibitemShut {NoStop}%
\bibitem [{\citenamefont {Antonini}\ and\ \citenamefont
  {Rasio}(2016)}]{Antonini:2016gqe}%
  \BibitemOpen
  \bibfield  {author} {\bibinfo {author} {\bibfnamefont {F.}~\bibnamefont
  {Antonini}}\ and\ \bibinfo {author} {\bibfnamefont {F.~A.}\ \bibnamefont
  {Rasio}},\ }\href {\doibase 10.3847/0004-637X/831/2/187} {\bibfield
  {journal} {\bibinfo  {journal} {Astrophys. J.}\ }\textbf {\bibinfo {volume}
  {831}},\ \bibinfo {pages} {187} (\bibinfo {year} {2016})},\ \Eprint
  {http://arxiv.org/abs/1606.04889} {arXiv:1606.04889 [astro-ph.HE]}
  \BibitemShut {NoStop}%
\bibitem [{\citenamefont {Mahapatra}\ \emph {et~al.}(2021)\citenamefont
  {Mahapatra}, \citenamefont {Gupta}, \citenamefont {Favata}, \citenamefont
  {Arun},\ and\ \citenamefont {Sathyaprakash}}]{Mahapatra:2021hme}%
  \BibitemOpen
  \bibfield  {author} {\bibinfo {author} {\bibfnamefont {P.}~\bibnamefont
  {Mahapatra}}, \bibinfo {author} {\bibfnamefont {A.}~\bibnamefont {Gupta}},
  \bibinfo {author} {\bibfnamefont {M.}~\bibnamefont {Favata}}, \bibinfo
  {author} {\bibfnamefont {K.~G.}\ \bibnamefont {Arun}}, \ and\ \bibinfo
  {author} {\bibfnamefont {B.~S.}\ \bibnamefont {Sathyaprakash}},\ }\href
  {\doibase 10.3847/2041-8213/ac20db} {\bibfield  {journal} {\bibinfo
  {journal} {Astrophys. J. Lett.}\ }\textbf {\bibinfo {volume} {918}},\
  \bibinfo {pages} {L31} (\bibinfo {year} {2021})},\ \Eprint
  {http://arxiv.org/abs/2106.07179} {arXiv:2106.07179 [astro-ph.HE]}
  \BibitemShut {NoStop}%
\bibitem [{\citenamefont {Abbott}\ \emph {et~al.}(2019)\citenamefont {Abbott}
  \emph {et~al.}}]{LIGOScientific:2018hze}%
  \BibitemOpen
  \bibfield  {author} {\bibinfo {author} {\bibfnamefont {B.~P.}\ \bibnamefont
  {Abbott}} \emph {et~al.} (\bibinfo {collaboration} {LIGO Scientific,
  Virgo}),\ }\href {\doibase 10.1103/PhysRevX.9.011001} {\bibfield  {journal}
  {\bibinfo  {journal} {Phys. Rev. X}\ }\textbf {\bibinfo {volume} {9}},\
  \bibinfo {pages} {011001} (\bibinfo {year} {2019})},\ \Eprint
  {http://arxiv.org/abs/1805.11579} {arXiv:1805.11579 [gr-qc]} \BibitemShut
  {NoStop}%
\bibitem [{\citenamefont {Abbott}\ \emph {et~al.}(2018)\citenamefont {Abbott}
  \emph {et~al.}}]{LIGOScientific:2018cki}%
  \BibitemOpen
  \bibfield  {author} {\bibinfo {author} {\bibfnamefont {B.~P.}\ \bibnamefont
  {Abbott}} \emph {et~al.} (\bibinfo {collaboration} {LIGO Scientific,
  Virgo}),\ }\href {\doibase 10.1103/PhysRevLett.121.161101} {\bibfield
  {journal} {\bibinfo  {journal} {Phys. Rev. Lett.}\ }\textbf {\bibinfo
  {volume} {121}},\ \bibinfo {pages} {161101} (\bibinfo {year} {2018})},\
  \Eprint {http://arxiv.org/abs/1805.11581} {arXiv:1805.11581 [gr-qc]}
  \BibitemShut {NoStop}%
\bibitem [{\citenamefont {De}\ \emph {et~al.}(2018)\citenamefont {De},
  \citenamefont {Finstad}, \citenamefont {Lattimer}, \citenamefont {Brown},
  \citenamefont {Berger},\ and\ \citenamefont {Biwer}}]{De:2018uhw}%
  \BibitemOpen
  \bibfield  {author} {\bibinfo {author} {\bibfnamefont {S.}~\bibnamefont
  {De}}, \bibinfo {author} {\bibfnamefont {D.}~\bibnamefont {Finstad}},
  \bibinfo {author} {\bibfnamefont {J.~M.}\ \bibnamefont {Lattimer}}, \bibinfo
  {author} {\bibfnamefont {D.~A.}\ \bibnamefont {Brown}}, \bibinfo {author}
  {\bibfnamefont {E.}~\bibnamefont {Berger}}, \ and\ \bibinfo {author}
  {\bibfnamefont {C.~M.}\ \bibnamefont {Biwer}},\ }\href {\doibase
  10.1103/PhysRevLett.121.091102} {\bibfield  {journal} {\bibinfo  {journal}
  {Phys. Rev. Lett.}\ }\textbf {\bibinfo {volume} {121}},\ \bibinfo {pages}
  {091102} (\bibinfo {year} {2018})},\ \bibinfo {note} {[Erratum:
  Phys.Rev.Lett. 121, 259902 (2018)]},\ \Eprint
  {http://arxiv.org/abs/1804.08583} {arXiv:1804.08583 [astro-ph.HE]}
  \BibitemShut {NoStop}%
\bibitem [{\citenamefont {{Radice}}\ \emph {et~al.}(2018)\citenamefont
  {{Radice}}, \citenamefont {{Perego}}, \citenamefont {{Zappa}},\ and\
  \citenamefont {{Bernuzzi}}}]{Radice2018ApJL}%
  \BibitemOpen
  \bibfield  {author} {\bibinfo {author} {\bibfnamefont {D.}~\bibnamefont
  {{Radice}}}, \bibinfo {author} {\bibfnamefont {A.}~\bibnamefont {{Perego}}},
  \bibinfo {author} {\bibfnamefont {F.}~\bibnamefont {{Zappa}}}, \ and\
  \bibinfo {author} {\bibfnamefont {S.}~\bibnamefont {{Bernuzzi}}},\ }\href
  {\doibase 10.3847/2041-8213/aaa402} {\bibfield  {journal} {\bibinfo
  {journal} {"Astrophys. J. Lett."}\ }\textbf {\bibinfo {volume} {852}},\
  \bibinfo {eid} {L29} (\bibinfo {year} {2018})},\ \Eprint
  {http://arxiv.org/abs/1711.03647} {arXiv:1711.03647 [astro-ph.HE]}
  \BibitemShut {NoStop}%
\bibitem [{\citenamefont {Radice}\ and\ \citenamefont
  {Dai}(2019)}]{Radice:2018ozg}%
  \BibitemOpen
  \bibfield  {author} {\bibinfo {author} {\bibfnamefont {D.}~\bibnamefont
  {Radice}}\ and\ \bibinfo {author} {\bibfnamefont {L.}~\bibnamefont {Dai}},\
  }\href {\doibase 10.1140/epja/i2019-12716-4} {\bibfield  {journal} {\bibinfo
  {journal} {Eur. Phys. J. A}\ }\textbf {\bibinfo {volume} {55}},\ \bibinfo
  {pages} {50} (\bibinfo {year} {2019})},\ \Eprint
  {http://arxiv.org/abs/1810.12917} {arXiv:1810.12917 [astro-ph.HE]}
  \BibitemShut {NoStop}%
\bibitem [{\citenamefont {Radice}\ \emph {et~al.}(2018)\citenamefont {Radice},
  \citenamefont {Perego}, \citenamefont {Hotokezaka}, \citenamefont {Fromm},
  \citenamefont {Bernuzzi},\ and\ \citenamefont {Roberts}}]{Radice:2018pdn}%
  \BibitemOpen
  \bibfield  {author} {\bibinfo {author} {\bibfnamefont {D.}~\bibnamefont
  {Radice}}, \bibinfo {author} {\bibfnamefont {A.}~\bibnamefont {Perego}},
  \bibinfo {author} {\bibfnamefont {K.}~\bibnamefont {Hotokezaka}}, \bibinfo
  {author} {\bibfnamefont {S.~A.}\ \bibnamefont {Fromm}}, \bibinfo {author}
  {\bibfnamefont {S.}~\bibnamefont {Bernuzzi}}, \ and\ \bibinfo {author}
  {\bibfnamefont {L.~F.}\ \bibnamefont {Roberts}},\ }\href {\doibase
  10.3847/1538-4357/aaf054} {\bibfield  {journal} {\bibinfo  {journal}
  {Astrophys. J.}\ }\textbf {\bibinfo {volume} {869}},\ \bibinfo {pages} {130}
  (\bibinfo {year} {2018})},\ \Eprint {http://arxiv.org/abs/1809.11161}
  {arXiv:1809.11161 [astro-ph.HE]} \BibitemShut {NoStop}%
\bibitem [{\citenamefont {Coughlin}\ \emph {et~al.}(2018)\citenamefont
  {Coughlin} \emph {et~al.}}]{Coughlin:2018miv}%
  \BibitemOpen
  \bibfield  {author} {\bibinfo {author} {\bibfnamefont {M.~W.}\ \bibnamefont
  {Coughlin}} \emph {et~al.},\ }\href {\doibase 10.1093/mnras/sty2174}
  {\bibfield  {journal} {\bibinfo  {journal} {Mon. Not. Roy. Astron. Soc.}\
  }\textbf {\bibinfo {volume} {480}},\ \bibinfo {pages} {3871} (\bibinfo {year}
  {2018})},\ \Eprint {http://arxiv.org/abs/1805.09371} {arXiv:1805.09371
  [astro-ph.HE]} \BibitemShut {NoStop}%
\bibitem [{\citenamefont {Flanagan}\ and\ \citenamefont
  {Hinderer}(2008)}]{Flanagan:2007ix}%
  \BibitemOpen
  \bibfield  {author} {\bibinfo {author} {\bibfnamefont {E.~E.}\ \bibnamefont
  {Flanagan}}\ and\ \bibinfo {author} {\bibfnamefont {T.}~\bibnamefont
  {Hinderer}},\ }\href {\doibase 10.1103/PhysRevD.77.021502} {\bibfield
  {journal} {\bibinfo  {journal} {Phys. Rev. D}\ }\textbf {\bibinfo {volume}
  {77}},\ \bibinfo {pages} {021502} (\bibinfo {year} {2008})},\ \Eprint
  {http://arxiv.org/abs/0709.1915} {arXiv:0709.1915 [astro-ph]} \BibitemShut
  {NoStop}%
\bibitem [{\citenamefont {Favata}(2014)}]{Favata:2013rwa}%
  \BibitemOpen
  \bibfield  {author} {\bibinfo {author} {\bibfnamefont {M.}~\bibnamefont
  {Favata}},\ }\href {\doibase 10.1103/PhysRevLett.112.101101} {\bibfield
  {journal} {\bibinfo  {journal} {Phys. Rev. Lett.}\ }\textbf {\bibinfo
  {volume} {112}},\ \bibinfo {pages} {101101} (\bibinfo {year} {2014})},\
  \Eprint {http://arxiv.org/abs/1310.8288} {arXiv:1310.8288 [gr-qc]}
  \BibitemShut {NoStop}%
\bibitem [{\citenamefont {Kulkarni}\ \emph {et~al.}(2023)\citenamefont
  {Kulkarni}, \citenamefont {Padamata}, \citenamefont {Gupta}, \citenamefont
  {Radice},\ and\ \citenamefont {Kashyap}}]{Kulkarni:2023tex}%
  \BibitemOpen
  \bibfield  {author} {\bibinfo {author} {\bibfnamefont {S.}~\bibnamefont
  {Kulkarni}}, \bibinfo {author} {\bibfnamefont {S.}~\bibnamefont {Padamata}},
  \bibinfo {author} {\bibfnamefont {A.}~\bibnamefont {Gupta}}, \bibinfo
  {author} {\bibfnamefont {D.}~\bibnamefont {Radice}}, \ and\ \bibinfo {author}
  {\bibfnamefont {R.}~\bibnamefont {Kashyap}},\ }\href {\doibase
  10.1103/PhysRevD.108.103023} {\bibfield  {journal} {\bibinfo  {journal}
  {Phys. Rev. D}\ }\textbf {\bibinfo {volume} {108}},\ \bibinfo {pages}
  {103023} (\bibinfo {year} {2023})},\ \Eprint
  {http://arxiv.org/abs/2308.03955} {arXiv:2308.03955 [astro-ph.HE]}
  \BibitemShut {NoStop}%
\bibitem [{\citenamefont {{Fitchett}}(1983)}]{Fitchett83}%
  \BibitemOpen
  \bibfield  {author} {\bibinfo {author} {\bibfnamefont {M.~J.}\ \bibnamefont
  {{Fitchett}}},\ }\href@noop {} {\bibfield  {journal} {\bibinfo  {journal}
  {Mon. Not. R. Astron. Soc.}\ }\textbf {\bibinfo {volume} {203}},\ \bibinfo
  {pages} {1049} (\bibinfo {year} {1983})}\BibitemShut {NoStop}%
\bibitem [{\citenamefont {Favata}\ \emph {et~al.}(2004)\citenamefont {Favata},
  \citenamefont {Hughes},\ and\ \citenamefont {Holz}}]{Favata:2004wz}%
  \BibitemOpen
  \bibfield  {author} {\bibinfo {author} {\bibfnamefont {M.}~\bibnamefont
  {Favata}}, \bibinfo {author} {\bibfnamefont {S.~A.}\ \bibnamefont {Hughes}},
  \ and\ \bibinfo {author} {\bibfnamefont {D.~E.}\ \bibnamefont {Holz}},\
  }\href {\doibase 10.1086/421552} {\bibfield  {journal} {\bibinfo  {journal}
  {Astrophys. J. Lett.}\ }\textbf {\bibinfo {volume} {607}},\ \bibinfo {pages}
  {L5} (\bibinfo {year} {2004})},\ \Eprint
  {http://arxiv.org/abs/astro-ph/0402056} {arXiv:astro-ph/0402056} \BibitemShut
  {NoStop}%
\bibitem [{\citenamefont {Nedora}\ \emph {et~al.}(2022)\citenamefont {Nedora},
  \citenamefont {Schianchi}, \citenamefont {Bernuzzi}, \citenamefont {Radice},
  \citenamefont {Daszuta}, \citenamefont {Endrizzi}, \citenamefont {Perego},
  \citenamefont {Prakash},\ and\ \citenamefont {Zappa}}]{Nedora:2020qtd}%
  \BibitemOpen
  \bibfield  {author} {\bibinfo {author} {\bibfnamefont {V.}~\bibnamefont
  {Nedora}}, \bibinfo {author} {\bibfnamefont {F.}~\bibnamefont {Schianchi}},
  \bibinfo {author} {\bibfnamefont {S.}~\bibnamefont {Bernuzzi}}, \bibinfo
  {author} {\bibfnamefont {D.}~\bibnamefont {Radice}}, \bibinfo {author}
  {\bibfnamefont {B.}~\bibnamefont {Daszuta}}, \bibinfo {author} {\bibfnamefont
  {A.}~\bibnamefont {Endrizzi}}, \bibinfo {author} {\bibfnamefont
  {A.}~\bibnamefont {Perego}}, \bibinfo {author} {\bibfnamefont
  {A.}~\bibnamefont {Prakash}}, \ and\ \bibinfo {author} {\bibfnamefont
  {F.}~\bibnamefont {Zappa}},\ }\href {\doibase 10.1088/1361-6382/ac35a8}
  {\bibfield  {journal} {\bibinfo  {journal} {Class. Quant. Grav.}\ }\textbf
  {\bibinfo {volume} {39}},\ \bibinfo {pages} {015008} (\bibinfo {year}
  {2022})},\ \Eprint {http://arxiv.org/abs/2011.11110} {arXiv:2011.11110
  [astro-ph.HE]} \BibitemShut {NoStop}%
\bibitem [{\citenamefont {Mahapatra}\ \emph {et~al.}(2025)\citenamefont
  {Mahapatra}, \citenamefont {Chattopadhyay}, \citenamefont {Gupta},
  \citenamefont {Favata}, \citenamefont {Sathyaprakash},\ and\ \citenamefont
  {Arun}}]{Mahapatra:2022ngs}%
  \BibitemOpen
  \bibfield  {author} {\bibinfo {author} {\bibfnamefont {P.}~\bibnamefont
  {Mahapatra}}, \bibinfo {author} {\bibfnamefont {D.}~\bibnamefont
  {Chattopadhyay}}, \bibinfo {author} {\bibfnamefont {A.}~\bibnamefont
  {Gupta}}, \bibinfo {author} {\bibfnamefont {M.}~\bibnamefont {Favata}},
  \bibinfo {author} {\bibfnamefont {B.~S.}\ \bibnamefont {Sathyaprakash}}, \
  and\ \bibinfo {author} {\bibfnamefont {K.~G.}\ \bibnamefont {Arun}},\ }\href
  {\doibase 10.1103/PhysRevD.111.023013} {\bibfield  {journal} {\bibinfo
  {journal} {Phys. Rev. D}\ }\textbf {\bibinfo {volume} {111}},\ \bibinfo
  {pages} {023013} (\bibinfo {year} {2025})},\ \Eprint
  {http://arxiv.org/abs/2209.05766} {arXiv:2209.05766 [astro-ph.HE]}
  \BibitemShut {NoStop}%
\bibitem [{\citenamefont {{Webb}}\ and\ \citenamefont
  {{Leigh}}(2015)}]{Webb2015MNRAS}%
  \BibitemOpen
  \bibfield  {author} {\bibinfo {author} {\bibfnamefont {J.~J.}\ \bibnamefont
  {{Webb}}}\ and\ \bibinfo {author} {\bibfnamefont {N.~W.~C.}\ \bibnamefont
  {{Leigh}}},\ }\href {\doibase 10.1093/mnras/stv1780} {\bibfield  {journal}
  {\bibinfo  {journal} {Mon. Not. Roy. Astron. Soc.}\ }\textbf {\bibinfo
  {volume} {453}},\ \bibinfo {pages} {3278} (\bibinfo {year} {2015})},\ \Eprint
  {http://arxiv.org/abs/1508.00577} {arXiv:1508.00577 [astro-ph.GA]}
  \BibitemShut {NoStop}%
\bibitem [{\citenamefont {{Carretta}}\ \emph {et~al.}(2011)\citenamefont
  {{Carretta}}, \citenamefont {{Lucatello}}, \citenamefont {{Gratton}},
  \citenamefont {{Bragaglia}},\ and\ \citenamefont {{D'Orazi}}}]{Carretta2011}%
  \BibitemOpen
  \bibfield  {author} {\bibinfo {author} {\bibfnamefont {E.}~\bibnamefont
  {{Carretta}}}, \bibinfo {author} {\bibfnamefont {S.}~\bibnamefont
  {{Lucatello}}}, \bibinfo {author} {\bibfnamefont {R.~G.}\ \bibnamefont
  {{Gratton}}}, \bibinfo {author} {\bibfnamefont {A.}~\bibnamefont
  {{Bragaglia}}}, \ and\ \bibinfo {author} {\bibfnamefont {V.}~\bibnamefont
  {{D'Orazi}}},\ }\href {\doibase 10.1051/0004-6361/201117269} {\bibfield
  {journal} {\bibinfo  {journal} {Astronomy \& Astrophysics}\ }\textbf
  {\bibinfo {volume} {533}},\ \bibinfo {eid} {A69} (\bibinfo {year} {2011})},\
  \Eprint {http://arxiv.org/abs/1106.3174} {arXiv:1106.3174 [astro-ph.SR]}
  \BibitemShut {NoStop}%
\bibitem [{\citenamefont {{Carballo-Bello}}\ \emph {et~al.}(2018)\citenamefont
  {{Carballo-Bello}}, \citenamefont {{Mart{\'\i}nez-Delgado}}, \citenamefont
  {{Navarrete}}, \citenamefont {{Catelan}}, \citenamefont {{Mu{\~n}oz}},
  \citenamefont {{Antoja}},\ and\ \citenamefont
  {{Sollima}}}]{CarballoBello2018}%
  \BibitemOpen
  \bibfield  {author} {\bibinfo {author} {\bibfnamefont {J.~A.}\ \bibnamefont
  {{Carballo-Bello}}}, \bibinfo {author} {\bibfnamefont {D.}~\bibnamefont
  {{Mart{\'\i}nez-Delgado}}}, \bibinfo {author} {\bibfnamefont
  {C.}~\bibnamefont {{Navarrete}}}, \bibinfo {author} {\bibfnamefont
  {M.}~\bibnamefont {{Catelan}}}, \bibinfo {author} {\bibfnamefont {R.~R.}\
  \bibnamefont {{Mu{\~n}oz}}}, \bibinfo {author} {\bibfnamefont
  {T.}~\bibnamefont {{Antoja}}}, \ and\ \bibinfo {author} {\bibfnamefont
  {A.}~\bibnamefont {{Sollima}}},\ }\href {\doibase 10.1093/mnras/stx2767}
  {\bibfield  {journal} {\bibinfo  {journal} {Mon. Not. Roy. Astron. Soc.}\
  }\textbf {\bibinfo {volume} {474}},\ \bibinfo {pages} {683} (\bibinfo {year}
  {2018})},\ \Eprint {http://arxiv.org/abs/1710.08927} {arXiv:1710.08927
  [astro-ph.GA]} \BibitemShut {NoStop}%
\bibitem [{\citenamefont {{Portegies Zwart}}\ \emph {et~al.}(2011)\citenamefont
  {{Portegies Zwart}}, \citenamefont {{van den Heuvel}}, \citenamefont {{van
  Leeuwen}},\ and\ \citenamefont {{Nelemans}}}]{TriplePortegies2011}%
  \BibitemOpen
  \bibfield  {author} {\bibinfo {author} {\bibfnamefont {S.}~\bibnamefont
  {{Portegies Zwart}}}, \bibinfo {author} {\bibfnamefont {E.~P.~J.}\
  \bibnamefont {{van den Heuvel}}}, \bibinfo {author} {\bibfnamefont
  {J.}~\bibnamefont {{van Leeuwen}}}, \ and\ \bibinfo {author} {\bibfnamefont
  {G.}~\bibnamefont {{Nelemans}}},\ }\href {\doibase
  10.1088/0004-637X/734/1/55} {\bibfield  {journal} {\bibinfo  {journal}
  {\apj}\ }\textbf {\bibinfo {volume} {734}},\ \bibinfo {eid} {55} (\bibinfo
  {year} {2011})},\ \Eprint {http://arxiv.org/abs/1103.2375} {arXiv:1103.2375
  [astro-ph.SR]} \BibitemShut {NoStop}%
\bibitem [{\citenamefont {{Pijloo}}\ \emph {et~al.}(2012)\citenamefont
  {{Pijloo}}, \citenamefont {{Caputo}},\ and\ \citenamefont {{Portegies
  Zwart}}}]{TriplePijloo2012}%
  \BibitemOpen
  \bibfield  {author} {\bibinfo {author} {\bibfnamefont {J.~T.}\ \bibnamefont
  {{Pijloo}}}, \bibinfo {author} {\bibfnamefont {D.~P.}\ \bibnamefont
  {{Caputo}}}, \ and\ \bibinfo {author} {\bibfnamefont {S.~F.}\ \bibnamefont
  {{Portegies Zwart}}},\ }\href {\doibase 10.1111/j.1365-2966.2012.21431.x}
  {\bibfield  {journal} {\bibinfo  {journal} {Mon. Not. Roy. Astron. Soc.}\
  }\textbf {\bibinfo {volume} {424}},\ \bibinfo {pages} {2914} (\bibinfo {year}
  {2012})},\ \Eprint {http://arxiv.org/abs/1207.0009} {arXiv:1207.0009
  [astro-ph.GA]} \BibitemShut {NoStop}%
\bibitem [{\citenamefont {{Serylak}}\ \emph {et~al.}(2022)\citenamefont
  {{Serylak}}, \citenamefont {{Venkatraman Krishnan}}, \citenamefont
  {{Freire}}, \citenamefont {{Tauris}}, \citenamefont {{Kramer}}, \citenamefont
  {{Geyer}}, \citenamefont {{Parthasarathy}}, \citenamefont {{Bailes}},
  \citenamefont {{Bernadich}}, \citenamefont {{Buchner}}, \citenamefont
  {{Burgay}}, \citenamefont {{Camilo}}, \citenamefont {{Karastergiou}},
  \citenamefont {{Lower}}, \citenamefont {{Possenti}}, \citenamefont
  {{Reardon}}, \citenamefont {{Shannon}}, \citenamefont {{Spiewak}},
  \citenamefont {{Stairs}},\ and\ \citenamefont {{van
  Straten}}}]{TripleSerylak2022}%
  \BibitemOpen
  \bibfield  {author} {\bibinfo {author} {\bibfnamefont {M.}~\bibnamefont
  {{Serylak}}}, \bibinfo {author} {\bibfnamefont {V.}~\bibnamefont
  {{Venkatraman Krishnan}}}, \bibinfo {author} {\bibfnamefont {P.~C.~C.}\
  \bibnamefont {{Freire}}}, \bibinfo {author} {\bibfnamefont {T.~M.}\
  \bibnamefont {{Tauris}}}, \bibinfo {author} {\bibfnamefont {M.}~\bibnamefont
  {{Kramer}}}, \bibinfo {author} {\bibfnamefont {M.}~\bibnamefont {{Geyer}}},
  \bibinfo {author} {\bibfnamefont {A.}~\bibnamefont {{Parthasarathy}}},
  \bibinfo {author} {\bibfnamefont {M.}~\bibnamefont {{Bailes}}}, \bibinfo
  {author} {\bibfnamefont {M.~C.~i.}\ \bibnamefont {{Bernadich}}}, \bibinfo
  {author} {\bibfnamefont {S.}~\bibnamefont {{Buchner}}}, \bibinfo {author}
  {\bibfnamefont {M.}~\bibnamefont {{Burgay}}}, \bibinfo {author}
  {\bibfnamefont {F.}~\bibnamefont {{Camilo}}}, \bibinfo {author}
  {\bibfnamefont {A.}~\bibnamefont {{Karastergiou}}}, \bibinfo {author}
  {\bibfnamefont {M.~E.}\ \bibnamefont {{Lower}}}, \bibinfo {author}
  {\bibfnamefont {A.}~\bibnamefont {{Possenti}}}, \bibinfo {author}
  {\bibfnamefont {D.~J.}\ \bibnamefont {{Reardon}}}, \bibinfo {author}
  {\bibfnamefont {R.~M.}\ \bibnamefont {{Shannon}}}, \bibinfo {author}
  {\bibfnamefont {R.}~\bibnamefont {{Spiewak}}}, \bibinfo {author}
  {\bibfnamefont {I.~H.}\ \bibnamefont {{Stairs}}}, \ and\ \bibinfo {author}
  {\bibfnamefont {W.}~\bibnamefont {{van Straten}}},\ }\href {\doibase
  10.1051/0004-6361/202142670} {\bibfield  {journal} {\bibinfo  {journal}
  {Astronomy \& Astrophysics}\ }\textbf {\bibinfo {volume} {665}},\ \bibinfo
  {eid} {A53} (\bibinfo {year} {2022})},\ \Eprint
  {http://arxiv.org/abs/2203.00607} {arXiv:2203.00607 [astro-ph.HE]}
  \BibitemShut {NoStop}%
\bibitem [{\citenamefont {{Wex}}\ and\ \citenamefont
  {{Kopeikin}}(1999)}]{Wex1999}%
  \BibitemOpen
  \bibfield  {author} {\bibinfo {author} {\bibfnamefont {N.}~\bibnamefont
  {{Wex}}}\ and\ \bibinfo {author} {\bibfnamefont {S.~M.}\ \bibnamefont
  {{Kopeikin}}},\ }\href {\doibase 10.1086/306933} {\bibfield  {journal}
  {\bibinfo  {journal} {\apj}\ }\textbf {\bibinfo {volume} {514}},\ \bibinfo
  {pages} {388} (\bibinfo {year} {1999})},\ \Eprint
  {http://arxiv.org/abs/astro-ph/9811052} {arXiv:astro-ph/9811052 [astro-ph]}
  \BibitemShut {NoStop}%
\bibitem [{\citenamefont {Morras}\ \emph {et~al.}(2025)\citenamefont {Morras},
  \citenamefont {Pratten},\ and\ \citenamefont {Schmidt}}]{Morras:2025xfu}%
  \BibitemOpen
  \bibfield  {author} {\bibinfo {author} {\bibfnamefont {G.}~\bibnamefont
  {Morras}}, \bibinfo {author} {\bibfnamefont {G.}~\bibnamefont {Pratten}}, \
  and\ \bibinfo {author} {\bibfnamefont {P.}~\bibnamefont {Schmidt}},\
  }\href@noop {} {\  (\bibinfo {year} {2025})},\ \Eprint
  {http://arxiv.org/abs/2503.15393} {arXiv:2503.15393 [astro-ph.HE]}
  \BibitemShut {NoStop}%
\bibitem [{\citenamefont {Aasi}\ \emph {et~al.}(2015)\citenamefont {Aasi} \emph
  {et~al.}}]{ALIGO}%
  \BibitemOpen
  \bibfield  {author} {\bibinfo {author} {\bibfnamefont {J.}~\bibnamefont
  {Aasi}} \emph {et~al.} (\bibinfo {collaboration} {LIGO Scientific}),\ }\href
  {\doibase 10.1088/0264-9381/32/7/074001} {\bibfield  {journal} {\bibinfo
  {journal} {Class. Quant. Grav.}\ }\textbf {\bibinfo {volume} {32}},\ \bibinfo
  {pages} {074001} (\bibinfo {year} {2015})},\ \Eprint
  {http://arxiv.org/abs/1411.4547} {arXiv:1411.4547 [gr-qc]} \BibitemShut
  {NoStop}%
\bibitem [{\citenamefont {Acernese}\ \emph {et~al.}(2015)\citenamefont
  {Acernese} \emph {et~al.}}]{AVirgo}%
  \BibitemOpen
  \bibfield  {author} {\bibinfo {author} {\bibfnamefont {F.}~\bibnamefont
  {Acernese}} \emph {et~al.} (\bibinfo {collaboration} {VIRGO}),\ }\href
  {\doibase 10.1088/0264-9381/32/2/024001} {\bibfield  {journal} {\bibinfo
  {journal} {Class. Quant. Grav.}\ }\textbf {\bibinfo {volume} {32}},\ \bibinfo
  {pages} {024001} (\bibinfo {year} {2015})},\ \Eprint
  {http://arxiv.org/abs/1408.3978} {arXiv:1408.3978 [gr-qc]} \BibitemShut
  {NoStop}%
\bibitem [{\citenamefont {Reitze}\ \emph {et~al.}(2019)\citenamefont {Reitze}
  \emph {et~al.}}]{CE:2019iox}%
  \BibitemOpen
  \bibfield  {author} {\bibinfo {author} {\bibfnamefont {D.}~\bibnamefont
  {Reitze}} \emph {et~al.},\ }\href@noop {} {\bibfield  {journal} {\bibinfo
  {journal} {Bull. Am. Astron. Soc.}\ }\textbf {\bibinfo {volume} {51}},\
  \bibinfo {pages} {035} (\bibinfo {year} {2019})},\ \Eprint
  {http://arxiv.org/abs/arXiv:1907.04833 [astro-ph.IM]} {arXiv:arXiv:1907.04833
  [astro-ph.IM] [astro-ph.IM]} \BibitemShut {NoStop}%
\bibitem [{\citenamefont {Sathyaprakash}\ \emph {et~al.}(2012)\citenamefont
  {Sathyaprakash}, \citenamefont {Abernathy}, \citenamefont {Acernese},
  \citenamefont {N.}, \citenamefont {Arun} \emph {et~al.}}]{ETScience11}%
  \BibitemOpen
  \bibfield  {author} {\bibinfo {author} {\bibfnamefont {B.}~\bibnamefont
  {Sathyaprakash}}, \bibinfo {author} {\bibfnamefont {M.}~\bibnamefont
  {Abernathy}}, \bibinfo {author} {\bibfnamefont {A.}~\bibnamefont {Acernese},
  \bibfnamefont {F.}}, \bibinfo {author} {\bibfnamefont {P.-S.}\ \bibnamefont
  {N.}}, \bibinfo {author} {\bibfnamefont {K.}~\bibnamefont {Arun}},  \emph
  {et~al.},\ }\href@noop {} {\bibfield  {journal} {\bibinfo  {journal}
  {Class.Quant.Grav.}\ }\textbf {\bibinfo {volume} {29}},\ \bibinfo {pages}
  {124013} (\bibinfo {year} {2012})},\ \Eprint {http://arxiv.org/abs/1108.1423}
  {arXiv:1108.1423 [gr-qc]} \BibitemShut {NoStop}%
\bibitem [{\citenamefont {Abbott}\ \emph
  {et~al.}(2017{\natexlab{b}})\citenamefont {Abbott} \emph
  {et~al.}}]{ET:2016wof}%
  \BibitemOpen
  \bibfield  {author} {\bibinfo {author} {\bibfnamefont {B.~P.}\ \bibnamefont
  {Abbott}} \emph {et~al.} (\bibinfo {collaboration} {LIGO Scientific
  Collaboration}),\ }\href {\doibase 10.1088/1361-6382/aa51f4} {\bibfield
  {journal} {\bibinfo  {journal} {Classical Quantum Gravity}\ }\textbf
  {\bibinfo {volume} {34}},\ \bibinfo {pages} {044001} (\bibinfo {year}
  {2017}{\natexlab{b}})},\ \Eprint {http://arxiv.org/abs/arXiv:1607.08697
  [astro-ph.IM]} {arXiv:arXiv:1607.08697 [astro-ph.IM] [astro-ph.IM]}
  \BibitemShut {NoStop}%
\bibitem [{\citenamefont {Dietrich}\ \emph {et~al.}(2018)\citenamefont
  {Dietrich}, \citenamefont {Radice}, \citenamefont {Bernuzzi}, \citenamefont
  {Zappa}, \citenamefont {Perego}, \citenamefont {Br\"ugmann}, \citenamefont
  {Chaurasia}, \citenamefont {Dudi}, \citenamefont {Tichy},\ and\ \citenamefont
  {Ujevic}}]{Dietrich:2018phi}%
  \BibitemOpen
  \bibfield  {author} {\bibinfo {author} {\bibfnamefont {T.}~\bibnamefont
  {Dietrich}}, \bibinfo {author} {\bibfnamefont {D.}~\bibnamefont {Radice}},
  \bibinfo {author} {\bibfnamefont {S.}~\bibnamefont {Bernuzzi}}, \bibinfo
  {author} {\bibfnamefont {F.}~\bibnamefont {Zappa}}, \bibinfo {author}
  {\bibfnamefont {A.}~\bibnamefont {Perego}}, \bibinfo {author} {\bibfnamefont
  {B.}~\bibnamefont {Br\"ugmann}}, \bibinfo {author} {\bibfnamefont {S.~V.}\
  \bibnamefont {Chaurasia}}, \bibinfo {author} {\bibfnamefont {R.}~\bibnamefont
  {Dudi}}, \bibinfo {author} {\bibfnamefont {W.}~\bibnamefont {Tichy}}, \ and\
  \bibinfo {author} {\bibfnamefont {M.}~\bibnamefont {Ujevic}},\ }\href
  {\doibase 10.1088/1361-6382/aaebc0} {\bibfield  {journal} {\bibinfo
  {journal} {Class. Quant. Grav.}\ }\textbf {\bibinfo {volume} {35}},\ \bibinfo
  {pages} {24LT01} (\bibinfo {year} {2018})},\ \Eprint
  {http://arxiv.org/abs/1806.01625} {arXiv:1806.01625 [gr-qc]} \BibitemShut
  {NoStop}%
\bibitem [{\citenamefont {Bernuzzi}\ \emph {et~al.}(2016)\citenamefont
  {Bernuzzi}, \citenamefont {Radice}, \citenamefont {Ott}, \citenamefont
  {Roberts}, \citenamefont {Moesta},\ and\ \citenamefont
  {Galeazzi}}]{Bernuzzi:2015opx}%
  \BibitemOpen
  \bibfield  {author} {\bibinfo {author} {\bibfnamefont {S.}~\bibnamefont
  {Bernuzzi}}, \bibinfo {author} {\bibfnamefont {D.}~\bibnamefont {Radice}},
  \bibinfo {author} {\bibfnamefont {C.~D.}\ \bibnamefont {Ott}}, \bibinfo
  {author} {\bibfnamefont {L.~F.}\ \bibnamefont {Roberts}}, \bibinfo {author}
  {\bibfnamefont {P.}~\bibnamefont {Moesta}}, \ and\ \bibinfo {author}
  {\bibfnamefont {F.}~\bibnamefont {Galeazzi}},\ }\href {\doibase
  10.1103/PhysRevD.94.024023} {\bibfield  {journal} {\bibinfo  {journal} {Phys.
  Rev. D}\ }\textbf {\bibinfo {volume} {94}},\ \bibinfo {pages} {024023}
  (\bibinfo {year} {2016})},\ \Eprint {http://arxiv.org/abs/1512.06397}
  {arXiv:1512.06397 [gr-qc]} \BibitemShut {NoStop}%
\bibitem [{\citenamefont {Gonzalez}\ \emph
  {et~al.}(2023{\natexlab{b}})\citenamefont {Gonzalez} \emph
  {et~al.}}]{Gonzalez:2022mgo}%
  \BibitemOpen
  \bibfield  {author} {\bibinfo {author} {\bibfnamefont {A.}~\bibnamefont
  {Gonzalez}} \emph {et~al.},\ }\href {\doibase 10.1088/1361-6382/acc231}
  {\bibfield  {journal} {\bibinfo  {journal} {Class. Quant. Grav.}\ }\textbf
  {\bibinfo {volume} {40}},\ \bibinfo {pages} {085011} (\bibinfo {year}
  {2023}{\natexlab{b}})},\ \Eprint {http://arxiv.org/abs/2210.16366}
  {arXiv:2210.16366 [gr-qc]} \BibitemShut {NoStop}%
\end{thebibliography}%

\clearpage


\onecolumngrid
\appendix*
\section{Binary Neutron Star Fitting Formulas}\label{sec:NRfit-appendix}
Here, we summarize the numerical relativity (NR) fits employed in our study to model the remnant mass, spin, and kick magnitude of a binary neutron star (BNS) system. We adopt the NR fitting formulas for the remnant mass ($m_{f}^{\rm NR}$) and spin ($\chi_{f}^{\rm NR}$) of non-spinning BNSs given in Ref.~\cite{Coughlin:2018fis}. These fits are expressed in terms of the following parameters:
\begin{eqnarray}
\nu &=& \frac{Q}{(1+Q)^{2}}\,,\\ \label{eq:nu}
M &=& m_1+m_2\,,\\ \label{eq:mass_tot}
\tilde{\Lambda} &=& \frac{16}{13} \frac{\Lambda_1 Q^5 + 12 \Lambda_1 Q^4 + 12 \Lambda_2 Q + \Lambda_2}{(1+Q)^5}\,,\label{eq:tidal_eff}
\end{eqnarray}
where, $m_1$ and $m_2$ are the masses of the primary (more massive) and secondary (less massive) components of the BNS system, respectively, expressed in solar mass units ($M_{\odot}$). The parameters $\Lambda_1$ and $\Lambda_2$ are their corresponding dimensionless tidal deformabilities, and $Q=\tfrac{m_1}{m_2}$ denotes the inverse mass ratio.
The NR fits for the remnant mass and spin are given by the following expressions [see Eqs.~(3) and (4), and Appendix D of Ref.~\cite{Coughlin:2018fis}]:
\begin{eqnarray}
m_{f}^{\rm NR} &=& a_m \left(\frac{\nu}{0.25}\right)^{2} \left(M + b_m \frac{\tilde{\Lambda}}{400}\right)\,,\\ \label{eq:rem_mass}
\chi_{f}^{\rm NR} &=& \Bigg| \tanh{\Big[ a_{\chi} \nu^{2} \left(M + b_{\chi} \tilde{\Lambda}\right) + c_{\chi} \Big]} \Bigg|\,. \label{eq:rem_spin}
\end{eqnarray}
The values of the different numerical fitting coefficients are $a_M=0.980$, $b_M=-0.093$, $a_{\chi}=0.537$, $b_{\chi}=-0.185$, and $c_{\chi}=-0.514$. The use of the \(\tanh\) function and the absolute value in the ansatz for \(\chi_{f}^{\rm NR}\) ensures that the final spin remains within the physically valid range, i.e., \( 0 \leq \chi_{f}^{\rm NR} \leq 1\). These phenomenological fits were derived in Ref.~\cite{Coughlin:2018fis} based on the set of NR data publicly released in the CoRe catalog~\cite{Dietrich:2018phi} together with results published in Ref.~\cite{Bernuzzi:2015opx}.

Using the dataset of 200 NR simulations of BNS mergers from the updated CoRe database~\cite{Gonzalez:2022mgo}, Ref.~\cite{Kulkarni:2023tex} derived a fitting formula for kick magnitude ($V_{\rm kick}$) imparted to the BNS merger remnants as a function of dynamical ejecta mass ($M_{\rm ej}$) [see Eq.~(13) of Ref.~\cite{Kulkarni:2023tex}]:
\begin{equation}
\label{eq:v_rem_fit}
    \ln V_{\rm kick} = 
    \begin{cases}
        1.075 & \text{if } \ln{(5\times10^{-6})}\leq \ln{M_{\rm ej}} < -8\,,\\ \\
        1.01 \ln{M_{\rm ej}} + 9.16 & \text{if } -8 \leq \ln{M_{\rm ej}} \leq \ln{(2\times10^{-2})}\,,
    \end{cases}
\end{equation}
where, $\ln{}$ denotes the natural logarithm (logarithm to the base $e$), and $V_{\rm kick}$ and $M_{\rm ej}$ are expressed in km/s and $M_\odot$, respectively. This fit is calibrated for ejected masses in the range $5 \times 10^{-6} M_{\odot} \lesssim M_{\rm ej} \lesssim 0.02 M_{\odot}$.
Furthermore, Ref.~\cite{Nedora:2020qtd} developed fitting formulas for the dynamical ejecta mass in BNS mergers based on NR simulations incorporating microphysical nuclear equations of state and neutrino transport, as well as results using polytropic equations of state. They present the fitting formula for the dynamical ejecta of BNS mergers as a second-order polynomial in Q and $\tilde{\Lambda}$ (See Eq.~(6) of Ref.~\cite{Nedora:2020qtd}):
\begin{equation}
\label{eq:m_ej_fit}
    {\rm log}_{10} (M_{\rm ej}) = b_0 + b_1 Q + b_2 \tilde{\Lambda} + b_3 Q^{2} + b_4 Q \tilde{\Lambda} + b_5 \tilde{\Lambda}^{2}\,.
\end{equation}
The values of the different numerical fitting coefficients are $b_0=-1.32$, $b_1=-0.382$, $b_2=-4.47\times10^{-3}$, $b_3=-0.339$, $b_4=3.21\times10^{-3}$, $b_5=4.31\times10^{-7}$ (See Table IV of Ref.~\cite{Nedora:2020qtd} for more details). By substituting Eq.~(\ref{eq:m_ej_fit}) into Eq.~(\ref{eq:v_rem_fit}), we can express the kick magnitude as a function of the masses and dimensionaless tidal parameters of the BNS. Note that while estimating $V_{\rm kick}$, we discarded posterior/prior samples of BNS parameters that result in $M_{\rm ej}$ values outside the calibration range. If a relatively large number of samples needss to be removed, the results should not be considered reliable. 
In our case, all posterior samples of parent BNS parameters resulted in $M_{\rm ej}$ values within the calibration region. However, we had to discard nearly half of the total prior samples of parent BNS parameters. As a result, the prior distributions on kick are not reliable and may not accurately represent the true prior distributions.

\end{document}